\documentclass[structabstract]{aa}  
\usepackage{graphicx}
\usepackage{lscape}
\usepackage{longtable}
\usepackage[latin1]{inputenc}
\usepackage{tikz}
\usetikzlibrary{shapes,arrows}
\usepackage[varg]{txfonts}

\begin{document}

\title{Automated measurements of diffuse interstellar bands in early-type star spectra}
\subtitle{Correlations with the Color Excess}

\author{L.Puspitarini, R.Lallement, H.-C. Chen
 }

\institute{GEPI Observatoire de Paris, CNRS, Universit\'e Paris Diderot, Place Jules Janssen  92190 Meudon, France\\
\email{rosine.lallement@obspm.fr, \\lucky.puspitarini@obspm.fr, hui-chen.chen@obspm.fr}
}

\date{}

\abstract
{}
{Stellar spectroscopic surveys may bring useful statistical information on the links between Diffuse Interstellar Bands (DIBs) and interstellar environment.  
DIB databases can also be used as a complementary tool for locating interstellar (IS) clouds. Our goal is to develop fully automated methods of DIB measurements to be applied to extensive data from stellar surveys.}
{We present a method appropriate for early-type nearby stars, its application to high-resolution spectra of $\sim$130 targets recorded with ESO FEROS spectrograph, and comparisons with other determinations. Using a DIB average profile deduced from the most reddened stars, we performed an automated fitting of a combination of a smooth stellar continuum, the DIB profile, and, when necessary, a synthetic telluric transmission.}
{Measurements are presented for 16 DIBs in the optical domain that could be extracted automatically: 
4726.8, 4762.6, 4963.9, 5780.4, 5797.1, 5849.8, 6089.8, 6196.0, 6203.0-6204.5, 6269.8, 6283.8, 6379.3, 6445.3, 6613.6, 6660.7, and 6699.3 \AA.
Two approaches to equivalent width determination (fitted equivalent width and continuum-integrated equivalent width) were tested and the two determinations were found to agree within the uncertainties, except for the 6203-6204 multiple band, and compatible with previous non-automated measurements.  Errors arising from statistical noise and continuum placement  were estimated conservatively by means of the "sliding window" method.  We derived the mean relationship between the DIB equivalent width and the color excess. Parameters of the linear correlation relationships are positively compared with published values, and most correlation coefficients were found higher then those based on earlier-type, UV bright target stars, confirming previous results on the influence of the radiation field. 
We provide the first correlation parameters for the 5849.8, 6089.8, 6269.8, 6445.3, 6660.7 and 6699.3 \AA\  bands in the Galaxy. The coefficients are within the same range as for other optical bands.
}
 {Automated DIB equivalent width extractions can be performed for nearby early-type stars without significant losses in the accuracy. This opens the perspective of building extended DIB catalogs  based on high resolution spectroscopic surveys. We discuss the limitations of this method in the case of distant stars and potential improvements.
 }

\authorrunning{Puspitarini et al}
\titlerunning{Automated DIB measurements}
\maketitle

\section{Introduction}

Identifying the carriers of the diffuse interstellar bands (DIBs) is a long standing  challenge in astronomical spectroscopy. 
Large molecules or their ions appear as likely candidates, however identifications are still actively debated (see e.g. reviews by \cite{1995ARA&A..33...19H}, \cite{sarre2006}, \cite{2011ApJ...727...33F} and references therein). 
On the other hand, an increasing number of bands were discovered in the optical domain (\cite{JD94}, \cite{krelo95}, \cite{tuairisg00}, \cite{weselak10}, \cite{gala00}), about 500 DIBs now identified between 4400 and 8600 {\AA} (\cite{hobbs2008}, \cite{2009ApJ...705...32H}, \cite{mccall2013}).

As discussed by \cite{2011ApJ...727...33F} and references therein, attempts to identify DIBs can be divided in four fields: (1) searches for signatures in the laboratory, (2) theoretical modeling of the detected profiles, (3) searches for correlations of DIBs with other interstellar parameters, and (4) searches for correlations between individual pairs of DIBs. 
Extensive datasets, in particular from stellar spectroscopic surveys, can be useful in the two latter perspectives. 
However, analyses of  large data sets require automated methods of DIB extraction. As part of this work we built and tested such methods.

Another advantage of extensive DIB measurements from spectroscopic surveys is their potential use as a tracer of the IS matter to build three-dimensional (3D) maps, by means of the derived spatial gradients of their strengths. 
This is made possible by the link between the DIBs and the amounts of intervening IS matter. 
Such a link has been extensively studied for extinction (or color excess) and atomic or molecular tracers.  Many of them are significantly or weakly correlated with interstellar integrated quantities, namely the HI, H$_2$ column or the color excess, as shown by numerous studies (e.g. \cite{2011ApJ...727...33F}, \cite{vos11}). In particular, the correlation of the DIB strength with the extinction or the color excess has been investigated for a large number of bands, and it has been shown that the degree of correlation is highly variable among the DIBs (see e.g. \cite{moutou99} and references in Table \ref{tabledibparam}). 
For some of them  (e.g. \cite{2011ApJ...727...33F}, \cite{Ordaz2011}), the correlation is moderate to good, and sometimes improved when restricting to regions which are not exposed by a strong UV radiation field (\cite{raimond12}). As a matter of fact, there is a strong and complex influence of the UV radiation field, which now is studied in details (e.g. \cite{kre94}, \cite{salama96}, \cite{sonnentrucker97}, \cite{cordiner08}, \cite{vos11}). In particular, two distinct types of sight-lines with respect to DIB properties and UV extinction law have been identified (\cite{kre94}, \cite{krelo95}): the $\zeta$-type sight-lines correspond to UV-shielded cloud cores, while the $\sigma$-type sight-lines probe external regions of the clouds, that are partially ionized by UV radiation field. The DIBs behave differently in response to those conditions, and there are families of DIBs that respond in a similar way (e.g. \cite{krelo87}).
Even if their degree of correlation with the dust or gas column is poor, DIBs can be used to locate IS clouds in the same way interstellar lines are used, simply by detecting gradients that mark the cloud boundary. 
E.g., neutral sodium is very poorly correlated with HI, nevertheless abrupt jumps in neutral sodium columns clearly mark the boundary of atomic or partially-ionized clouds. 

DIBs would be complementary to other tracers that have been used for IS mapping, such as 
atomic lines or color excesses (\cite{welsh10}, \cite{vergely010}).  
The easily observed IS lines such as the NaI D1-D2 and the CaII H-K lines saturate for moderate and high reddening so that the uncertainties associated to the derived gas columns are large.  
This is a problem when one uses spatial gradients of the absorbing columns to locate the IS clouds. 
At variance with those lines, most of the DIBS in the visible do not suffer from this saturation effects and can potentially be used for larger distances. 
With the advent of spectroscopic surveys such as the ESO-Gaia Spectroscopic Survey (\cite{2012Msngr.147...25G}), large numbers of DIB absorption measurements are expected, that can feed databases for 3D mapping. 
Thanks to GAIA mission, parallax distances should become available for billions of stars and allow to build 3D interstellar matter (ISM) maps by means of inversion methods (\cite{vergely01}, \cite{vergely010}). 
Despite the fact that assigning distances to clouds by means of absorption data does not require a proportionality between the measured quantity (column of a gaseous species, reddening, or DIB strength) and the column of IS gas or dust, the existence of an established average relationship is useful 
to estimate average volume opacities of 3D dust clouds, in addition to their locations. 
Because the better the DIB correlation is, the more meaningful the
estimation of the dust (or gas) column that will be based on the DIB strength, it is interesting to study in details all correlation coefficients. 
In another respect, the possibility of estimating the extinction based on a DIB strength and the correlation law may be extremely valuable in case photometric or spectrophotometric data are absent or ambiguous. 
As part of the present work, we extend previous correlative studies and provide new average relationships for several DIBs.

In a first paper (\cite{raimond12}), we extracted two of the strongest and well studied interstellar bands  ($\lambda\lambda$ 
5780.4, 6283.8)
from about 150 high quality, high-resolution early-type spectra of nearby stars and performed a study of their correlations with the color excess. 
We found an overall agreement with previous studies, in particular the work of \cite{2011ApJ...727...33F}. 
This comparison showed that the dispersion around the average relationship is diminished when using later type target stars, and it was argued that this is due to the decrease of radiation field influences. As a matter of fact, short sightline may be dominated by dense clouds located immediately in front of the target stars and strongly irradiated. 
In this previous study, the measurement of the DIB equivalent width (EW) was performed in a classical, non-automated way similar to the methods used by several other groups, i.e. by means of a preliminary fit of the continuum around the DIB to 
a polynomial, a subsequent normalization of the spectrum and the computation of the absorption area below the normalized continuum. 
At variance with this earlier work, here we performed automated DIB measurements and we compared their results with those from the previous interactive method. 
As said above, automated methods are mandatory in the case of spectroscopic surveys.  Here we have been testing methods based on template DIB profiles, in the same way \cite{cordiner08} proceeded in the case of M31. We have developed an automated program and make use here of the same database to test it and extract a series of measurable DIBs, namely the $\lambda\lambda$ 
4726.83, 4762.61, 4963.88, 5780.38, 5797.06, 5849.81, 6089.85, 6195.98, 6203.05-6204.49, 6269.85, 6283.84, 6379.32, 6445.28, 6613.62, 6660.71, and 6699.32 
DIBs. The central wavelengths here are the values listed in the recent catalog of Hobbs et al (2008). 
An exception is the 6204 \AA\ complex, which has been shown to be made of two distinct absorptions centered at 6203.05 and 6204.49 \AA\ resp. and corresponding to two distinct carriers (see e.g. \cite{porceddu91}). 
For simplicity here we have chosen to derive the equivalent width of the ensemble of those two absorptions, with an approximate centering at 6204 \AA. For the two strong DIBs of our first paper, we compare the derived equivalent widths with our previous non-automated analysis results. 
We finally compare the DIB strengths with the color-excess values, and the derived relationships with previously established results, when they exist.

Section 2 describes the data, while section 3 presents the spectral analysis and the error estimates. Section 4 presents the results, and describes the relationships with the color excess, as well as comparisons with previous works. 
Section 5 discusses the limitations of the automated method and potential improvements.

\section{Data}

We use high-resolution (R $\approx$ 48000), high signal-to-noise (S/N $\geq$ 100) spectra of nearby, early-type (B to A5), fast rotating stars that are mostly located within 400 pc.
Observations were done with the FEROS spectrograph at the ESO/Max Planck 2.2 m telescope in La Silla and were part of the LP179.C-0197 program. FEROS provides a broad spectral range: $\sim$ 3530 to $\sim$ 9210  {\AA}. This dataset has been previously used for interstellar (IS) neutral sodium and ionized calcium measurements that,  after merging to an existing database, have been inverted to produce 3D maps of the nearby IS matter (\cite{welsh10,vergely010}). They have also been used to study the two strong bands at 5780.38 and 6283.84 \AA\  (\cite{raimond12}). Such high-quality and wide range spectra are well suited for the study of diffuse bands, including weak ones.

For the purpose of comparisons with the reddening, we use an homogeneous dataset of color-excess data based on the seven colors Geneva photometric system (\cite{cram99}, \cite{burcram12}). Those data, that have also been used by \cite{raimond12}, were all determined based on the same calibration method. The homogeneity avoids additional dispersion and biases in the correlation computation. We thus have restricted this study to target stars possessing a color excess determination from this single source. This reduced the number of useful targets from about 400 to 150. 
The spectra were also preliminarily analyzed in the search for narrow lines due to cold companions. As a matter of fact, narrow and weak stellar features may contaminate the DIBs and make the continuum fitting difficult or biased.  This resulted in the exclusion of  a significant fraction of binary stars. The final number of targets available for the fully automated DIB extraction is 129. 

The Geneva E(B-V) color excess values were converted to Str\"omgen E(b-y) color excesses, using all Str\"omgen data compiled by \cite{vergely010} for which both determinations do exist. 
A scaling factor was derived from the cross-correlation and a similar cross-correlation was then performed  to convert to  the  commonly used Johnson system. 
The relationship with the Johnson system is E(B-V)(Geneva)= 1.703 E(B-V) (Johnson). 
Finally, the error on E(B-V) was estimated conservatively to be 0.03, as discussed in \cite{raimond12}. 

\section{Data analysis}

 \subsection{Choice of DIBs.}
 
We started with the recent and extended DIB catalog of \cite{hobbs2008} and initially considered all potential bands. 
DIBs were first selected when they were strong enough to be detected in more than about 20 target stars, in such a way a correlation with the color excess can be established with some statistical significance.
Because our targets are nearby and slightly reddened, this resulted in a drastic selection of the strongest bands, despite the high signal-to-noise. 
In a second step, we excluded the strong but very broad and shallow bands such as the 4426 \AA\ band. 
As a matter of fact, in the case of our early-type stars those bands have widths that become comparable to the stellar line widths and are difficult to disentangle from those stellar lines in an automated way  (those DIBs are paradoxically easier to measure for late-type stars).
For the selection  we used a threshold of $\gtrsim$ 0.05 for the ratio between the DIB equivalent width and the full width at half maximum (FWHM). 
For those broad and shallow bands, methods based on stellar spectral models (\cite{chen12}) would be more appropriate and will be the subject of future work. 
Table \ref{tabledibparam} shows  our 16 selected DIBs and lists their parameters as they have been determined by \cite{hobbs2008}. 
They have all been detected in several objects, including in the Magellanic clouds, M31, M33, and for nine of them previous determinations of the relationship with the reddening do exist for Milky Way targets and are listed in the table. 

\begin{table}[h!]
	\caption{List of the selected DIBs, their parameters and relevant studies}
	\label{tabledibparam}
	\begin{center}
	\begin{tiny}
		\begin{tabular}{|c|c|c|c| l | }
		\hline
		$\lambda_c^*$ 	&  FWHM  	&  EW  & $\Delta$ EW & Ref.  \\
		(\AA) 			& (\AA)  			   &  (m\AA)  & rms err. &  \\\hline \hline
		4726.83 &  2.74	&  283.8	& 	4.2 						& 4,13	 												\\ \hline
		4762.61 &  1.00 	&  50.8	&	1.6 						& 4,7,13													 \\ \hline	
		4963.88 &  0.62	&  53.4	&	1.0 						& 4,7,13 	 												\\ \hline
		5780.38 &  2.11	&  257.0	&	3.0 						& 1,2,3,4,6,7,8,9$^{nG}$,10,$11^{nG}$,		  \\  
		 &  	&  	& 						&  12$^{nG}$,14,15,18		  \\  \hline
		5797.06 &  0.77	&  199.0	&	1.1 						& 1,2,3,4,5,6,7,8,9$^{nG}$,10,$11^{nG}$,  		  \\ 
		 &  	&  	&	 						&  12$^{nG}$,14,17,18 		  \\ \hline
		5849.81 &  0.82	&  95.6	&	1.2 						& 1,4 													\\ \hline
		6089.85 &  0.54	&  28.0	&	0.8 						& 4 														\\ \hline
		6195.98 &  0.42	&  37.8	&	0.6 						& 2,4,14,16,17,18,19					 					\\ \hline
		6203.05; 	& 1.21;	&  57.1; 	& 1.7;	 & 2,4,7, 9$^{nG}$,$11^{nG}$,12$^{nG}$,14,16,17				 \\ 
		6204.49	&  4.87	&  71.6	&  6.9			 & 			 \\ \hline
		6269.85 &  1.17	&  77.0	&	1.7 						&  2,4,7,$11^{nG}$,12$^{nG}$								 \\ \hline
		6283.84 &  4.77	&  459.7	&	6.9 						&  2,4,7,9$^{nG}$,$11^{nG}$,12$^{nG}$,14,15,16,17			\\  \hline
		6379.32 &  0.58	&  94.9	&	0.8 						&  4,5,7,$11^{nG}$,12$^{nG}$,18,19								 \\ \hline
		6445.28 &  0.46 	&  35.7	&	0.7 						&  4														 \\ \hline
		6613.62 &  0.93	&  165.1	&	1.4 						&  4,5,7,8,9$^{nG}$,$11^{nG}$,12$^{nG}$,14,17,18				 \\ \hline
		6660.71 &  0.58	&  33.0	&	0.9 						&  4,$11^{nG}$							 				\\ \hline
		6699.32 &  0.63	&  21.6	&  	0.9 						&  4, 17													 \\	
		\hline
		\end{tabular}
	\end{tiny}
	\end{center}
\tablefoottext{*}{$\lambda_c$, FWHM, EW, and $\Delta$ EW from \cite{hobbs2008} (measurements towards HD204827)}\\
\tablefoottext{nG}{extragalactic target: M31, M33, etc}\\
\tablebib{
(1){~\cite{josa87} (reflection nebulae targets)}; 
(2){~\cite{benvenuti89} (link to the 2175 \AA\ feature)};  
(3){~\cite{herbig93} (correlations with gaseous lines and extinction)};  
(4){~\cite{JD94} (DIBs survey with EW/E(B-V))}; 
(5){~\cite{sonnentrucker97} (Ionization properties)}; 
(6){~\cite{krelo99} (relation with CH and CH+)};  
(7){~\cite{thor03} (relation with $C_2$ molecules)}; 
(8){~\cite{2007ApJ...664..909D} (cool stars)}; 
(9){~\cite{cordiner08} (M31)}; 
(10){~\cite{vanloon09} (Tarantula survey)}; 
(11){~\cite{cordiner11} (M31)}; 
(12){~\cite{2011AAS...21711204C} (M31, M33) }; 
(13){~\cite{Ordaz2011} (O stars)}; 
(14){~\cite{2011ApJ...727...33F} (Correlations with gaseous lines sand color excess)}; 
(15){~\cite{raimond12} (survey: correlation with color excess)}; 
(16){~\cite{chen12} (cool giant bulge stars)} ; 
(17){~\cite{xiang12} (compilation of correlations}); 
(18){~\cite{vos11} (study of environmental effects)}; 
(18){~\cite{walker01} (study of profiles)} 
}
\end{table}

\subsection{Fitting method}

Our goal is to measure DIB equivalent widths (EWs) in series of stellar spectra in an entirely automated way. 
There are several steps in this process. First, prior to the actual spectral analysis, we establish an empirical DIB shape model. Alternatively, we use a pre-existing model from prior studies. 
This is the case for 6283.84 DIB profile, which is taken from \cite{raimond12} and is already adapted to the resolution and instrumental response.
When weak telluric lines are present, we use a telluric transmission model that is fitted in airmass and Doppler shift to the data and is used to eliminate those features. 
Then, we fit a pre-defined spectral interval around the DIB to a combination of a smooth continuum, represented by a polynomial function, here of third to seventh-degree, and the DIB model. The DIB model is allowed to vary in strength and velocity shift, and the polynomial coefficients are all free to vary.
The spectral interval over which the fit is done is pre-defined, it contains the DIB and broad enough intervals both short-ward and long-ward of the DIB  to ensure a good fit 
of the stellar continuum.  When strong telluric lines are present, instead of the combination of a DIB model and a smooth continuum, we fit the data to a triple combination of  telluric transmission model (also fit in strength and shift), DIB model and smooth continuum.
In all cases (non-existent, weak or strong telluric lines), by using such a global fitting, we avoid any manual selection of intervals for the continuum fit. 

The DIB EW is derived in two separate ways. The first way assumes that the continuum is reasonably fitted during the fit 
and also that the actual DIB absorption has a shape that is similar to the model. In this case the EW is simply the product of the model EW by the fit coefficient (or DIB's strength or power index of the absorption), since for our sightlines the linear regime prevails. 
The second estimate is solely based on the first, less restrictive assumption that the continuum has been correctly fitted during the DIB+continuum(+telluric) global fit, but does NOT make use of the DIB model nor of the fit coefficient. It simply uses the area between the spectrum and the normalized continuum. Following \cite{cordiner08}, we call the two measurements respectively as the fitted EW (which is based on the DIB model and power index) and continuum-integrated EW (which is independent of the DIB model).

The comparison between the two EW values allows to detect potential discrepancies between the observed and modeled DIB shapes, including in particular some DIB broadening due to a velocity shift spread of the interstellar clouds, or DIB shape variations with sightline.
As a matter of fact, using a unique DIB shape is equivalent to assuming that sub-structures in the DIB profile vary in the same proportions from one sightline to the other. This is only true at first order, and it is known from high resolution spectral analyses that the shape may vary in some cases (e.g. \cite{gala2008}). It is also equivalent to assuming that the DIB is much broader than the velocity shift between the intervening clouds. 
This is in principle justified when the DIBs are relatively broad, and especially broad in comparison with the radial velocity dispersion of the IS clouds. 
Conversely, if for a target the two estimates of the EW are in good agreement, this implies that the actual DIB shape in the stellar spectrum  is very similar to the template shape. This also implies that the the continuum has been realistically fitted and the DIB EW well estimated, and constitutes a validity check of the automated method. We will come back to this point in the last section.

The DIB model was obtained empirically during the preliminary phase in the following way. 
We selected a small number (about 3 to 5) of targets such as HD172488, HD147932, HD176853, HD151884 and HD171957 that have high color excess values and excellent spectra, extracted the DIB in a classic way and scaled them to the same EW. Then, we simply averaged them and used the mean as a template.
In the case of the weakest DIBs or the 6204 \AA\ blend, the model has to be used carefully. 
As we mentioned before, the comparison of  the two measurements, the fitted and continuum-integrated EWs, will allow to check its reliability. 

\begin{figure}[h!]
\centering
	\includegraphics[width=\linewidth]{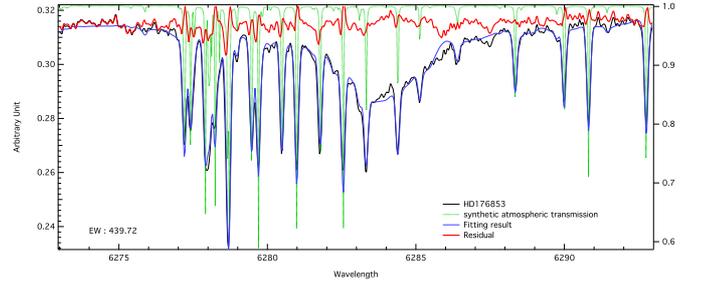}
\caption{Illustration of telluric line corrections using an automated fitting method that take into account stellar continuum, telluric lines, and also DIB. The black line is the original spectrum of HD176853, the green line is the synthetic telluric transmission, the blue line is the fitting result, and the red line is the residual. The very strong telluric lines are not entirely removed, however the residuals do not impact significantly on the DIB equivalent width.}
\label{telluriccorr}%
\end{figure}

\begin{figure}[h!]
\centering
	\includegraphics[width=\linewidth]{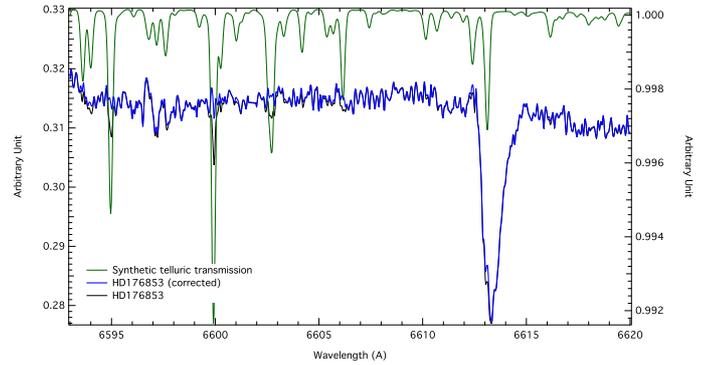}
%
\caption{Example of a preliminarily correction for weak telluric lines. The black line is the spectrum of HD176853, the green line is the  synthetic telluric transmission, and the blue line is the corrected spectrum.}
\label{telluriccorr2}%
\end{figure}

Atmospheric lines were modeled by means of a synthetic telluric transmission computed using the Line-By-Line Radiative Transfert Model (LBLRTM) code and the HIgh-Resolution-TRANsmission molecular absorption (HITRAN) spectroscopic database, here for a standard atmosphere profile and for the altitude of La Silla (\cite{lblrtm05}, \cite{hitran2008}). 
For the 6283.8 DIB, telluric lines of diatomic oxygen are particularly strong. For this DIB, and this DIB only, the correction was not done in a preliminary phase, but instead we fitted the data to a triple combination of stellar, DIB and telluric models as described above. During the fit, both the DIB and the telluric transmission are free to vary in Doppler shift and strength, allowing for Earth velocity and airmass variations from one spectrum to the other. 
An illustration of the fitting can be found in Fig. \ref{telluriccorr}.
For the 5849.8, 6269.85, 6445.28, 6613.62,  and 6660.71 DIBs that are only slightly contaminated, we more simply used the synthetic telluric transmission to perform a correction prior to the DIB measurement (see Fig. \ref{telluriccorr2}).
Finally, in a few cases we also applied the classical corrections before the automated fit to correct %
for  cosmic rays,
by means of a classical interpolation in the contaminated area.
We also applied masks in the case of neighboring DIBs to separate their respective absorptions (e.g. for the 6195.98 DIB).  For some of the cooler targets, we also applied masks to avoid the contamination by narrow stellar lines 
as described in detail in \cite{raimond12}.

The consecutive steps of the EW measurement method can be found in Fig. \ref{DIBfitting}, and the illustration of the whole fitting process for the 6195.98 DIB can be found in Fig. \ref{DIBfit}. For the equivalent width measurements, we used the same limits of integration (DIB broadness) as \cite{2011ApJ...727...33F} for 5780.38, 6204, 6283.84, 6195.98, 6613.62, and 5797.06 {\AA}. Other DIB shapes, and integration limits can be seen in Fig. 
\ref{DIBmodel},\ref{DIBmodel2}, and \ref{DIBmodel3}.

\tikzstyle{decision} = [diamond, draw, fill=blue!20, 
    text width=6em, text badly centered, node distance=3cm, inner sep=0pt]
\tikzstyle{block} = [rectangle, draw, fill=blue!20, 
    text width=12em, text centered, rounded corners, minimum height=3em]
\tikzstyle{line} = [draw, -latex']
\tikzstyle{cloud} = [draw, ellipse,fill=red!20, node distance=3cm,
    minimum height=2em]

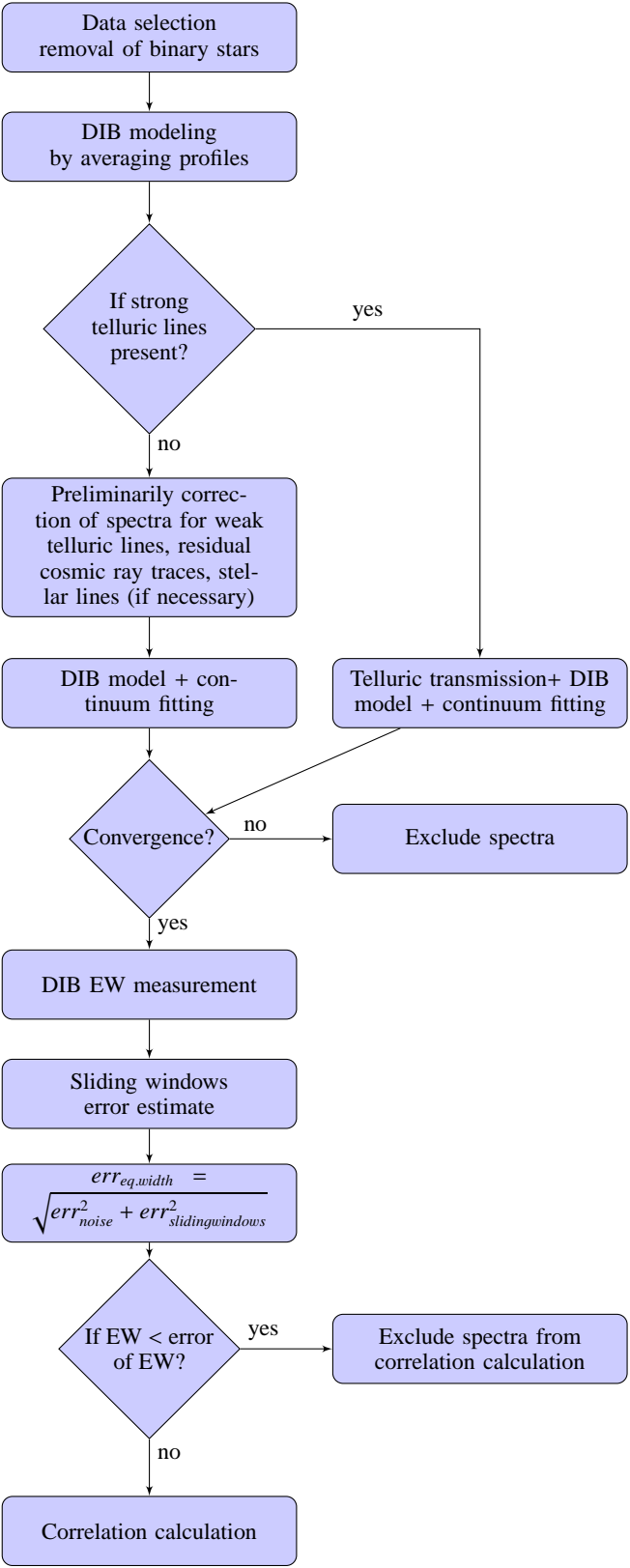
\begin{figure}[h!]
\begin{center}
\begin{tikzpicture}[node distance = 1cm, auto]
    
\node [block] (dataselection) {Data selection \\removal of binary stars};
\node [block,below of=dataselection, node distance=1.5cm] (dibmodel) {DIB modeling \\by averaging profiles};
\node [decision, below of=dibmodel, node distance=2.5cm] (decide) {If strong telluric lines present?};
\node [block, below of=decide, node distance=3cm] (telluric2) {Preliminarily correction of  spectra for weak telluric lines, residual cosmic ray traces, stellar lines (if necessary)};
\node [block, below of=telluric2, node distance=2cm] (contfit) {DIB model + continuum fitting }; 
\node [block, right of=contfit, node distance=4.5cm] (telluric) {Telluric transmission+ DIB model + continuum fitting}; 
\node [decision, below of=contfit, node distance=2cm] (decide2) {Convergence?};
\node [block, right of=decide2, node distance=4.5cm] (exclude) {Exclude spectra};
\node [block, below of=decide2, node distance=2cm] (dibew) {DIB EW measurement}; 
\node [block, below of=dibew, node distance=1.5cm] (diberr) {Sliding windows error estimate};
\node [block, below of=diberr, node distance=1.5cm] (comberr) {$err_{eq. width}=\sqrt{err_{noise}^2+err_{sliding windows}^2}$};
\node [decision, below of=comberr, node distance=2cm] (decide3) {If EW $<$ error of EW?};
\node [block, right of=decide3, node distance=4.5cm] (exclude2) {Exclude spectra from correlation calculation};
\node [block, below of=decide3, node distance=2.5cm] (corrcal) {Correlation calculation};
  %
  \path [line] (dataselection) -- (dibmodel);
  \path [line] (dibmodel) -- (decide);
  \path [line] (decide) -| node [near start] {yes} (telluric);
  \path [line] (decide) -- node [near start] {no} (telluric2);
  \path [line] (telluric2) -- (contfit);
  \path [line] (contfit) -- (decide2);
  \path [line] (telluric) -- (decide2);
  \path [line] (decide2) -- node [near start] {yes} (dibew);
  \path [line] (decide2) -- node [near start] {no} (exclude);  
  \path [line] (dibew) -- (diberr);
  \path [line] (diberr) -- (comberr);  
  \path [line] (comberr) -- (decide3);  
  \path [line] (decide3) -- node [near start] {yes} (exclude2);  
   \path [line] (decide3) -- node [near start] {no} (corrcal);
    

\end{tikzpicture}
\caption{DIB fitting process}
\label{DIBfitting}
\end{center}
\end{figure}


\begin{figure}[h!]
\centering

	\includegraphics[width=\linewidth]{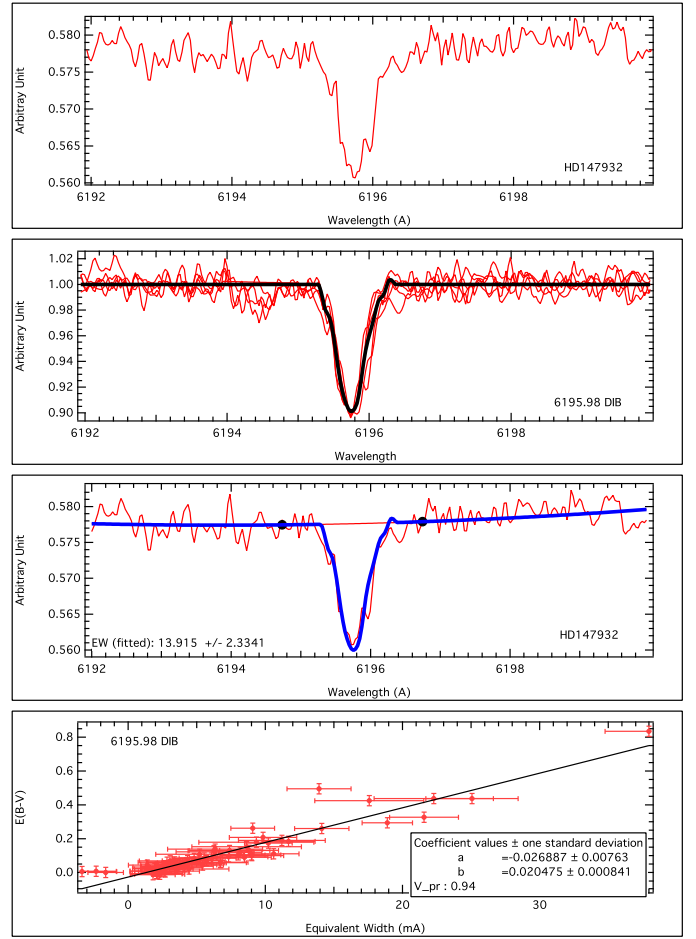}
\caption{Illustration of the automated DIB fitting method for the 6196.0 DIB. The first row image shows the spectrum of HD147932 (red line) with the DIB band at about 6195.98 \AA . The second row image shows an empirical DIB shape model (black line). It is an average profile deduced from profiles of the most reddened stars, scaled to the same EW. The third row image shows an example of automated DIB fitting (DIB model + continuum fitting) 
The red line is the spectrum of target star and the blue line is the fitting result. The last row shows the correlation between EWs and color excess values using OLS method.}
              \label{DIBfit}%
    \end{figure}

\subsection{Error estimates and results}

\subsubsection{Error estimates}

Errors on the equivalent width are of two kinds: first, the statistical noise in the spectrum, pixel to pixel, and second, the error due to uncertainties in defining the continuum. The latter corresponds to the departure between the extrapolated part of the fitted continuum and the true underlying  continuum at the DIB location. By definition it is impossible to know its exact shape, and those departures depend on the quality of the spectrum and of the properties of the target (mainly temperature and intrinsic variability). We estimated this potential departure in an empirical, statistical way  by means of the  \textit{sliding windows} method described in detail by  \cite{raimond12}.
In brief we applied the same continuum fitting, based on two spectral intervals of the same width and separated by the same gap, however this time by centering the gap NOT at the DIB location but at a large number of locations in the spectrum, in regions on both sides of the DIB that are free of IS absorption. To do so, we translate (or slide) the fitting intervals along the spectrum in a quasi-continuous way (hence the \textit{sliding window} name). The difference between the fitted continuum (which corresponds to the extrapolated part that would be used if  there was a DIB) and the actual spectrum provides an estimate of the error done during the continuum placement.
The entire set of locations provides a list of error values from which we can extract a standard deviation.
We then quadratically added this value to the error due to the measurement uncertainties $err_{noise}=\sigma \times \frac{\triangle\lambda}{N}$, where $\sigma$ is a standard deviation, $\triangle\lambda$ is the interval of DIB absorption, and $N$ is number of points in the interval of absorption.

\subsubsection{Results and comparison between the two determinations of the equivalent width}

Once the automated DIB fitting is performed 
it is possible to compute the equivalent width in two ways, as detailed in section 3.2: the first is based on the result of the continuum + DIB template fitting and is a simple scaling of the DIB template EW. The second consists in measuring directly the absorption area in the normalized spectrum, based on the fitted continuum. Differences between the two arise if the DIB shape in the spectrum differs significantly from the template, and as said above their similarity is a validity test of the assumptions and of the automated method. 
We have thus carefully compared the two equivalent widths for the various DIBs.

\begin{figure*}
\centering
	\includegraphics[width=0.2\linewidth]{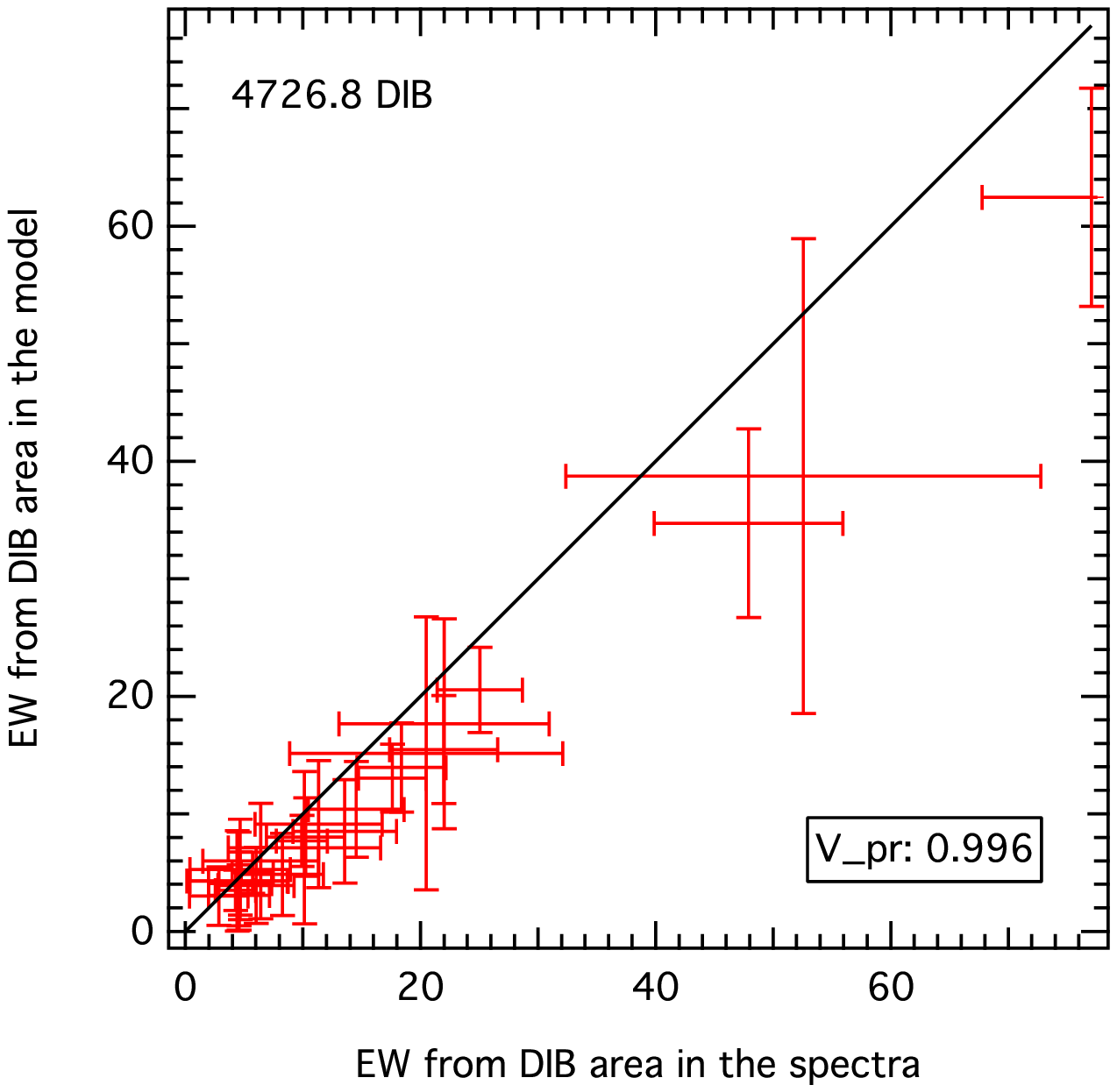}
	\includegraphics[width=0.2\linewidth]{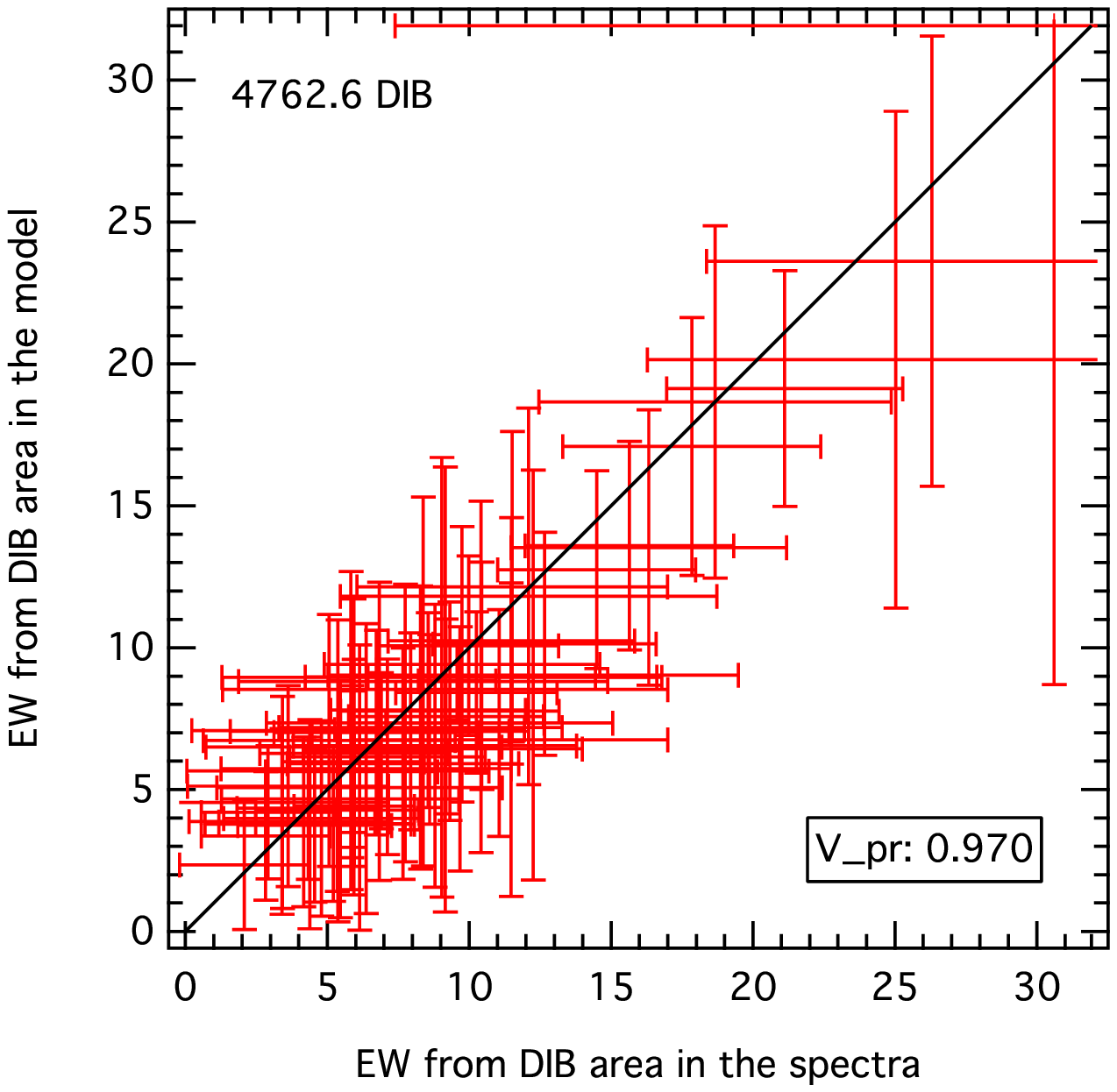}
	\includegraphics[width=0.2\linewidth]{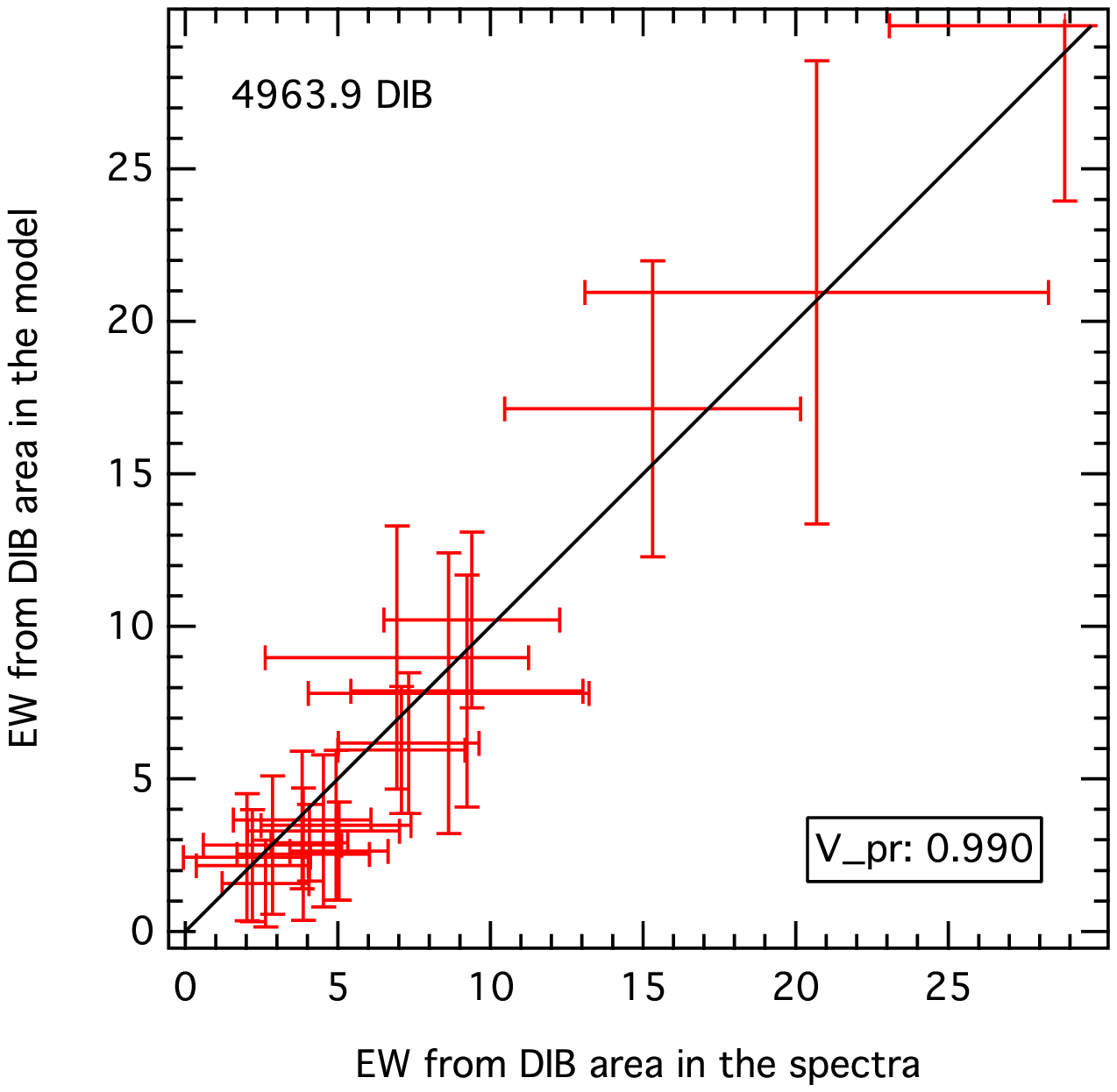}
	\includegraphics[width=0.2\linewidth]{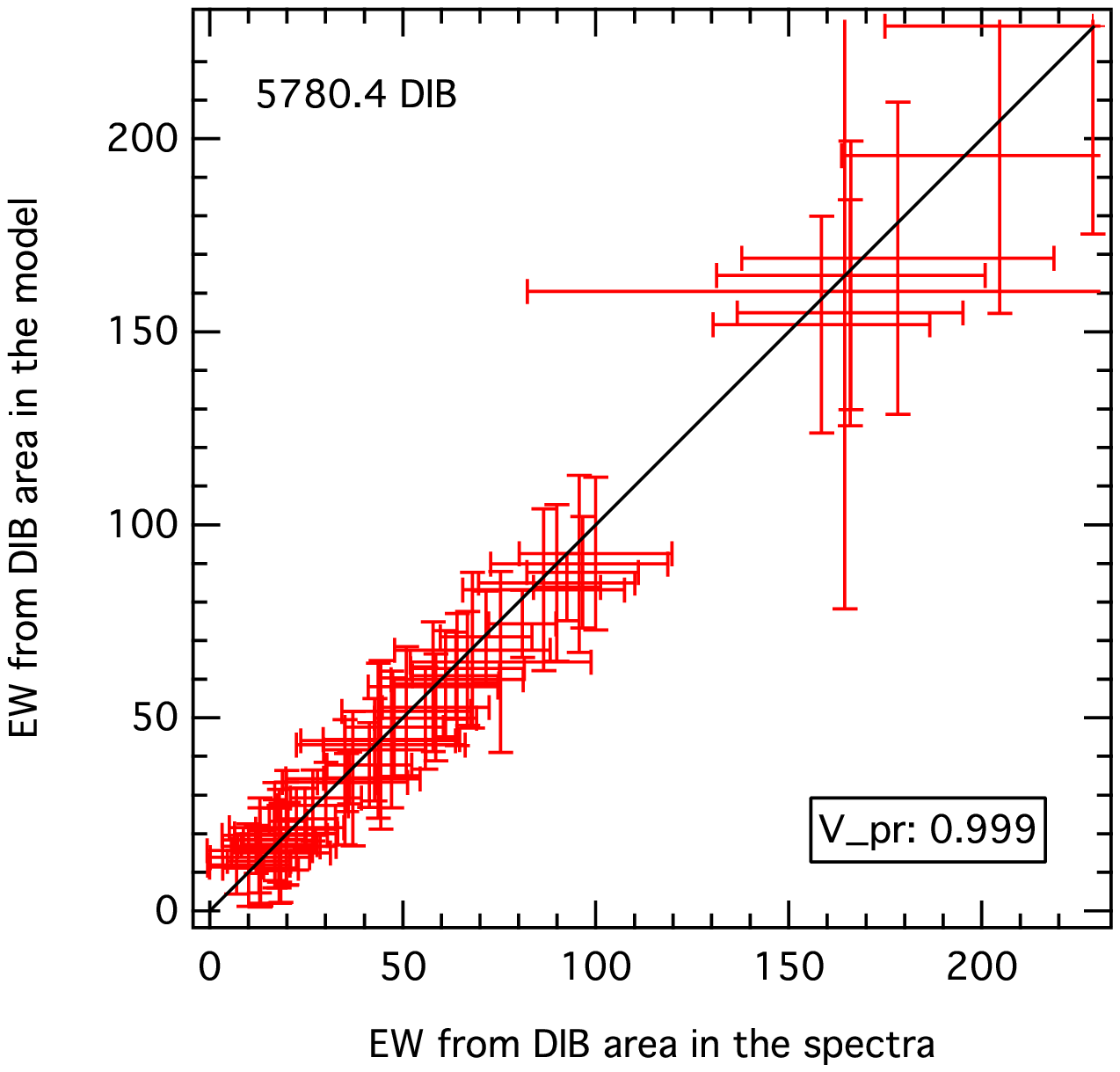}
	\includegraphics[width=0.2\linewidth]{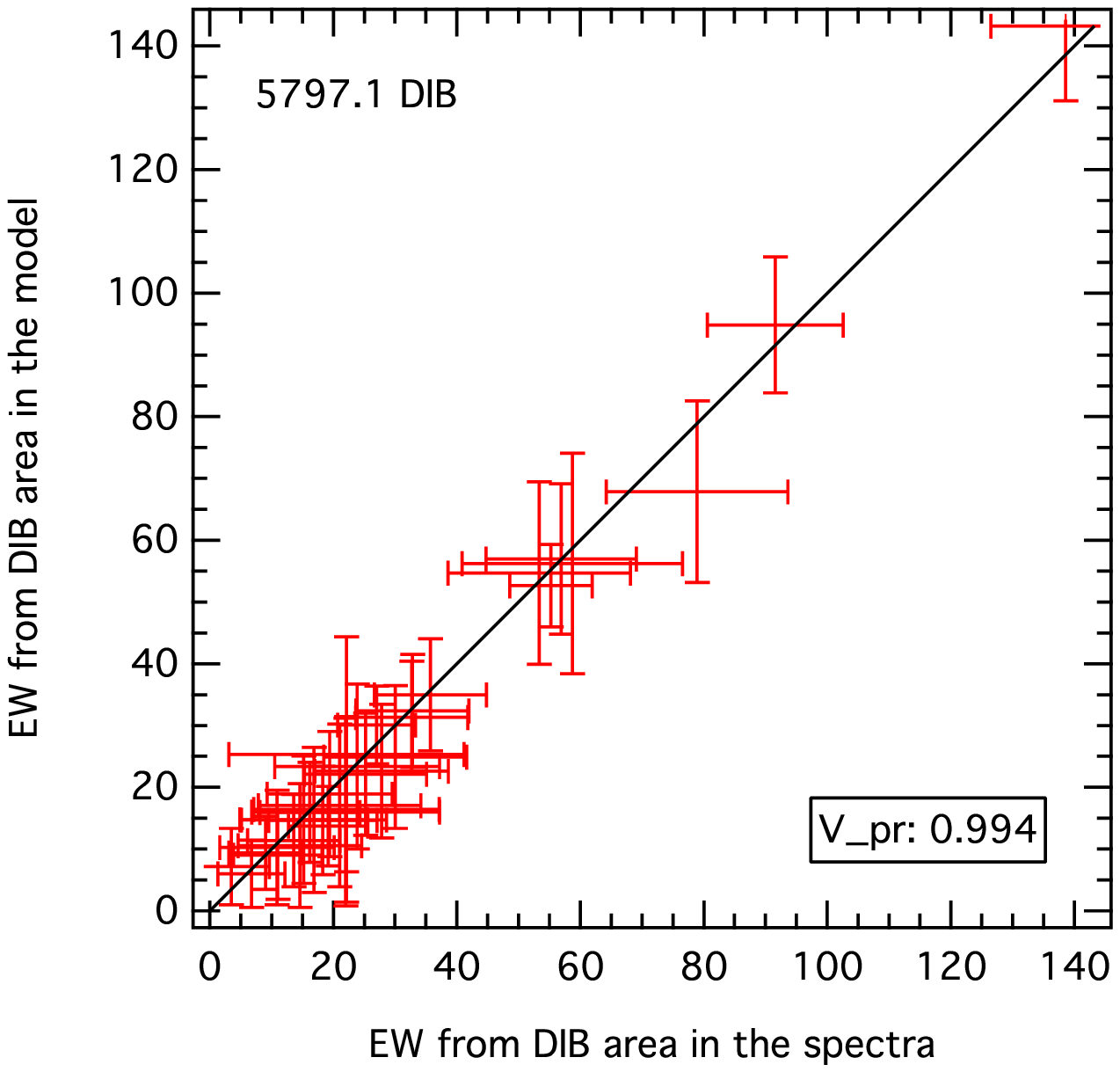}
	\includegraphics[width=0.2\linewidth]{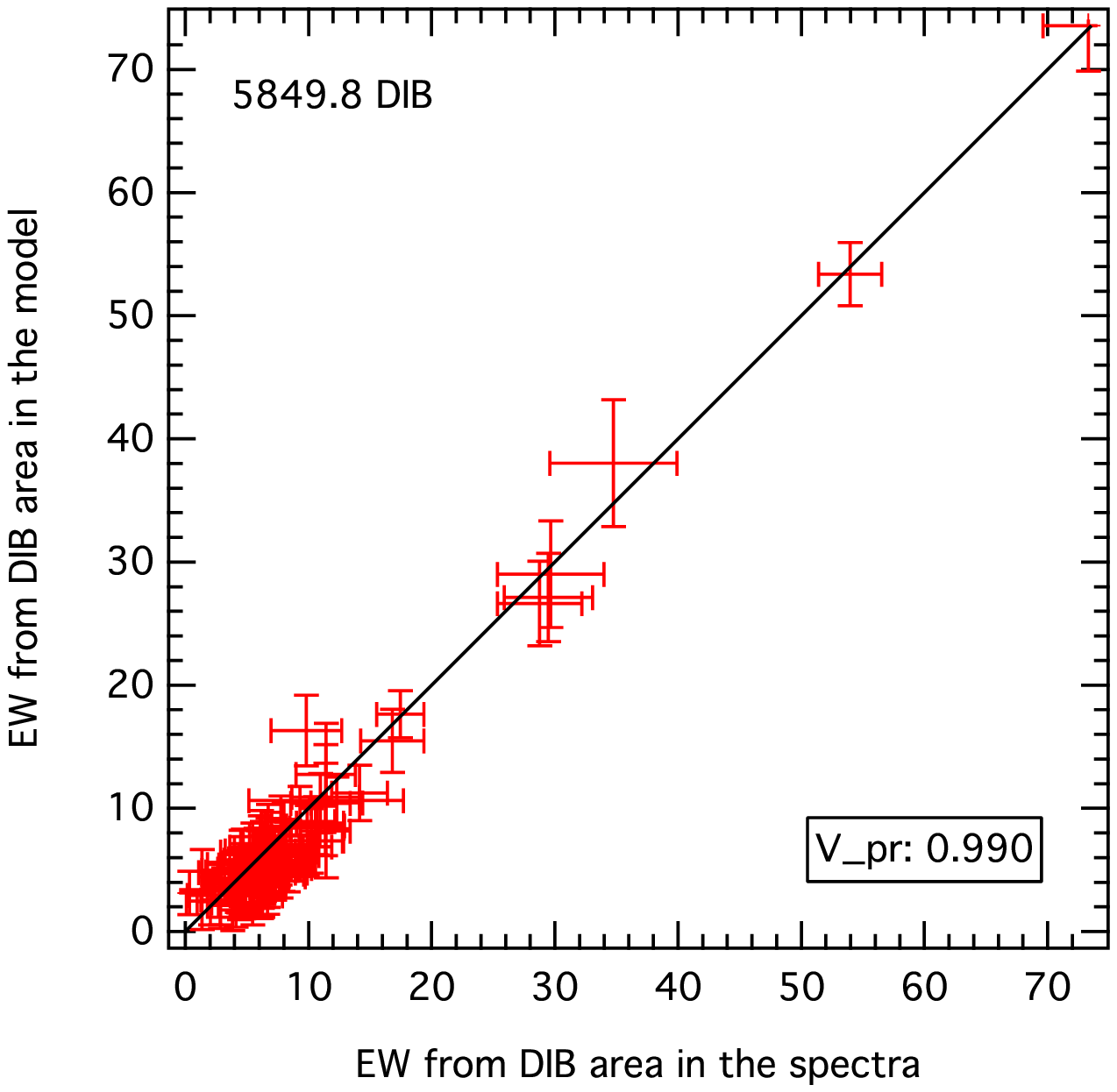}
	\includegraphics[width=0.2\linewidth]{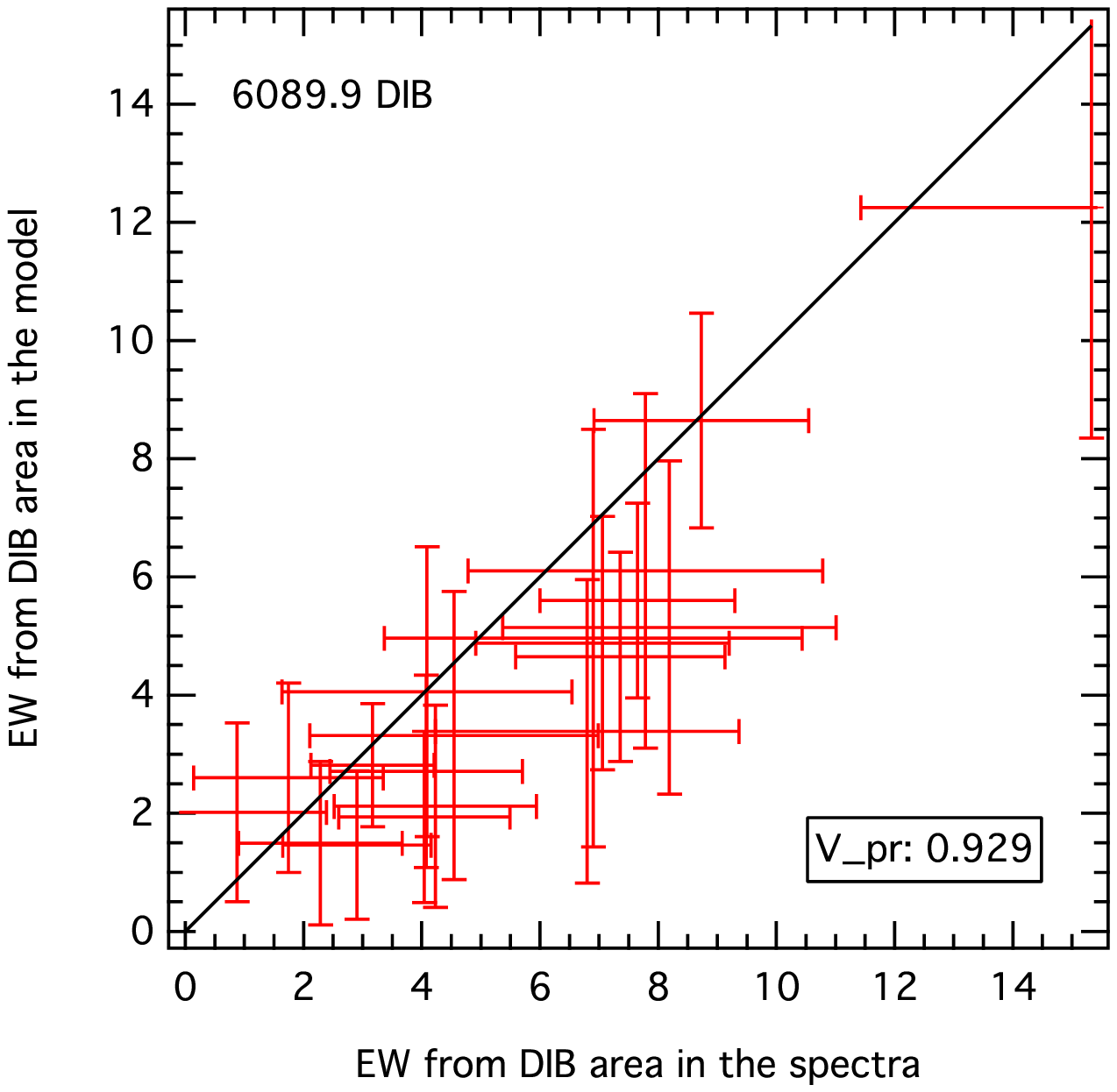}
	\includegraphics[width=0.2\linewidth]{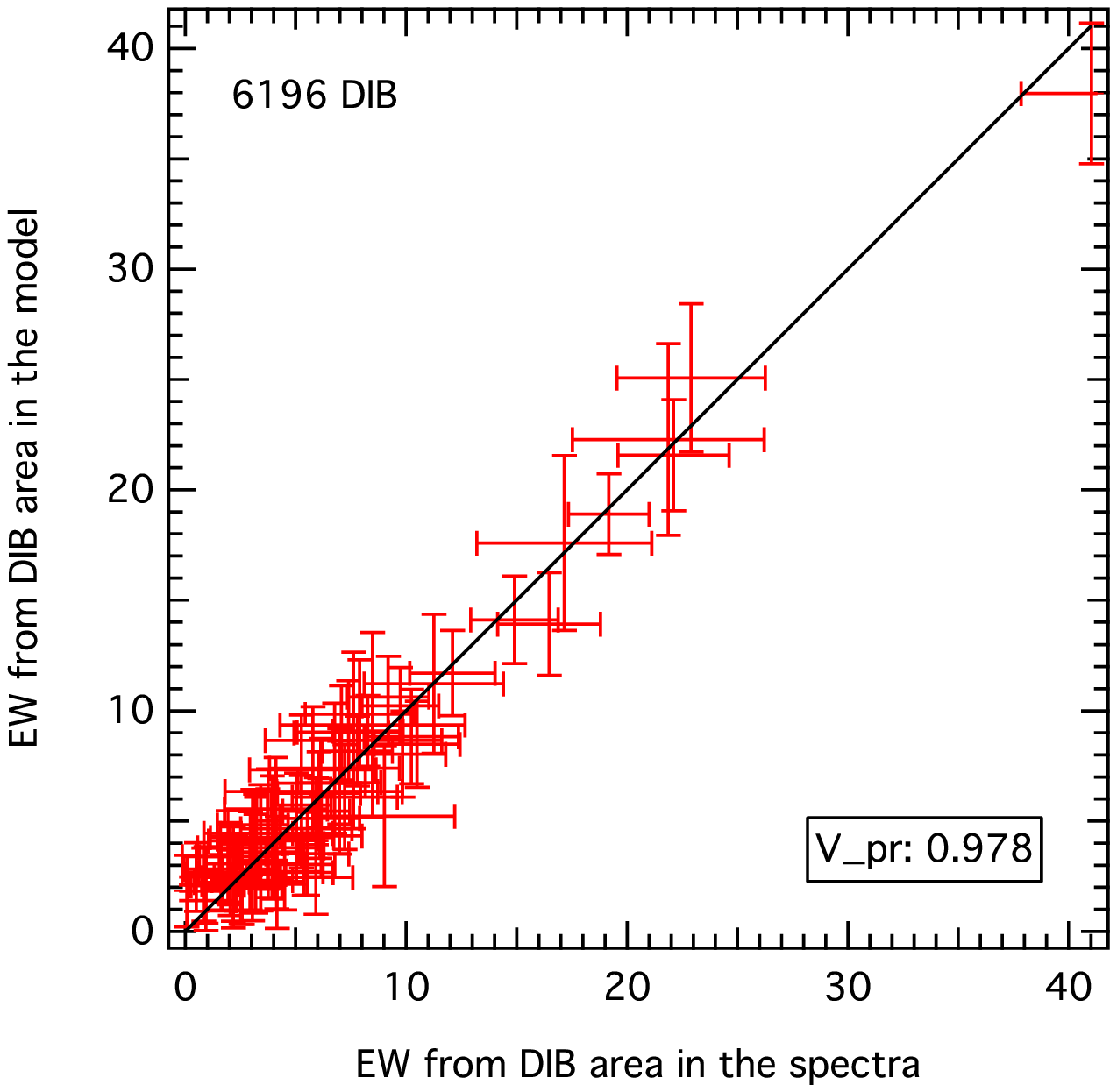}
	\includegraphics[width=0.2\linewidth]{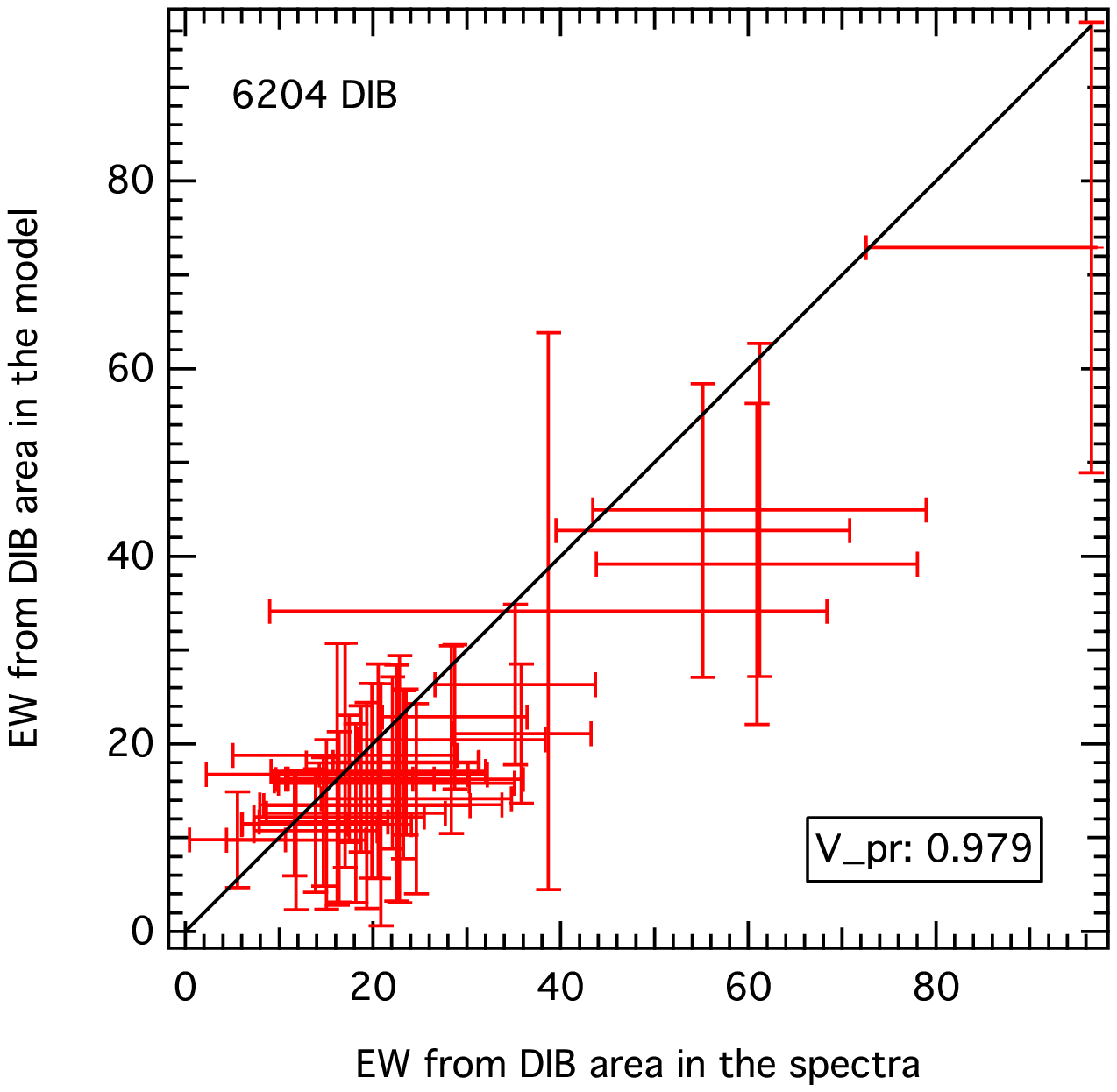}
	\includegraphics[width=0.2\linewidth]{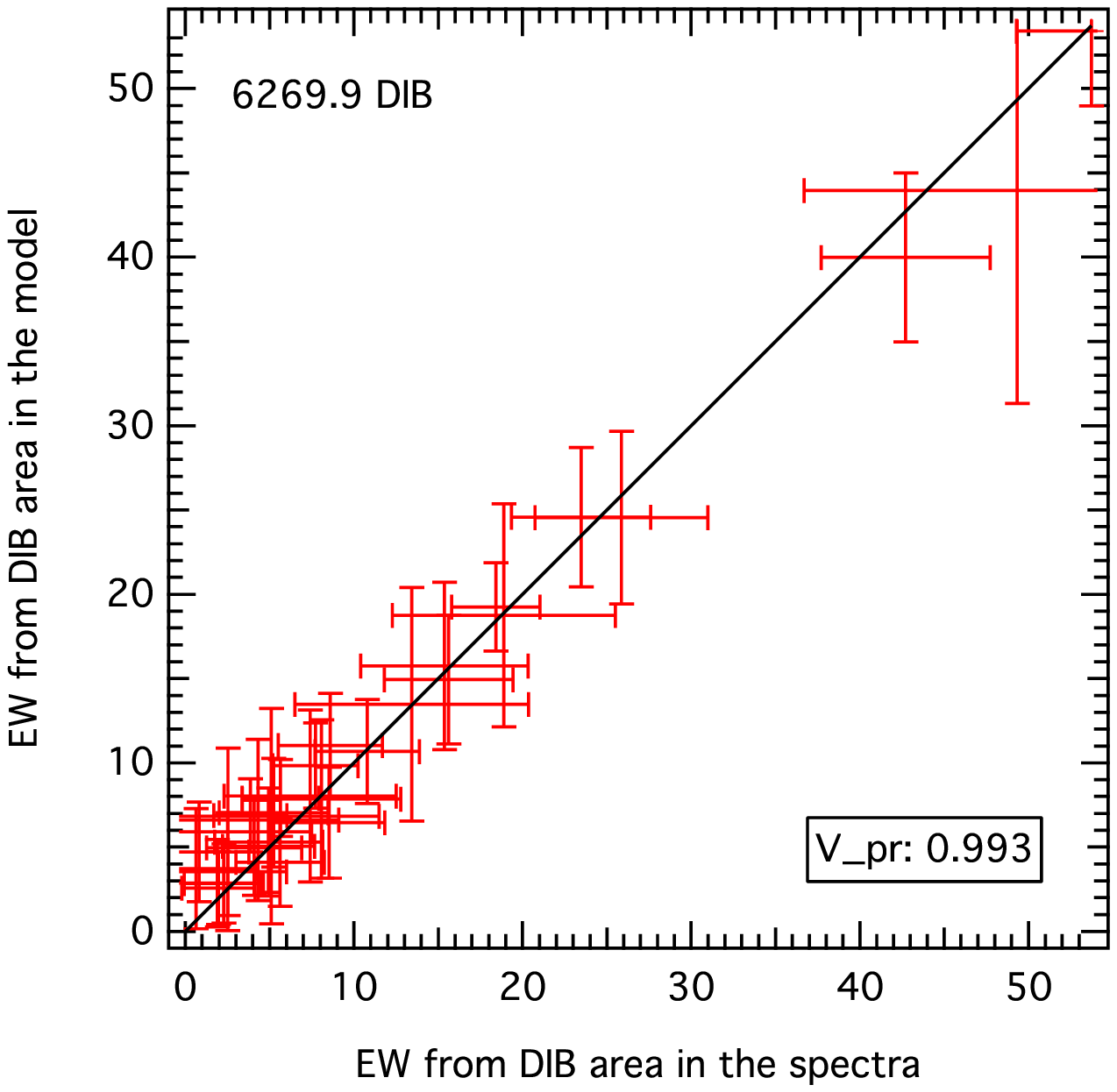}
	\includegraphics[width=0.2\linewidth]{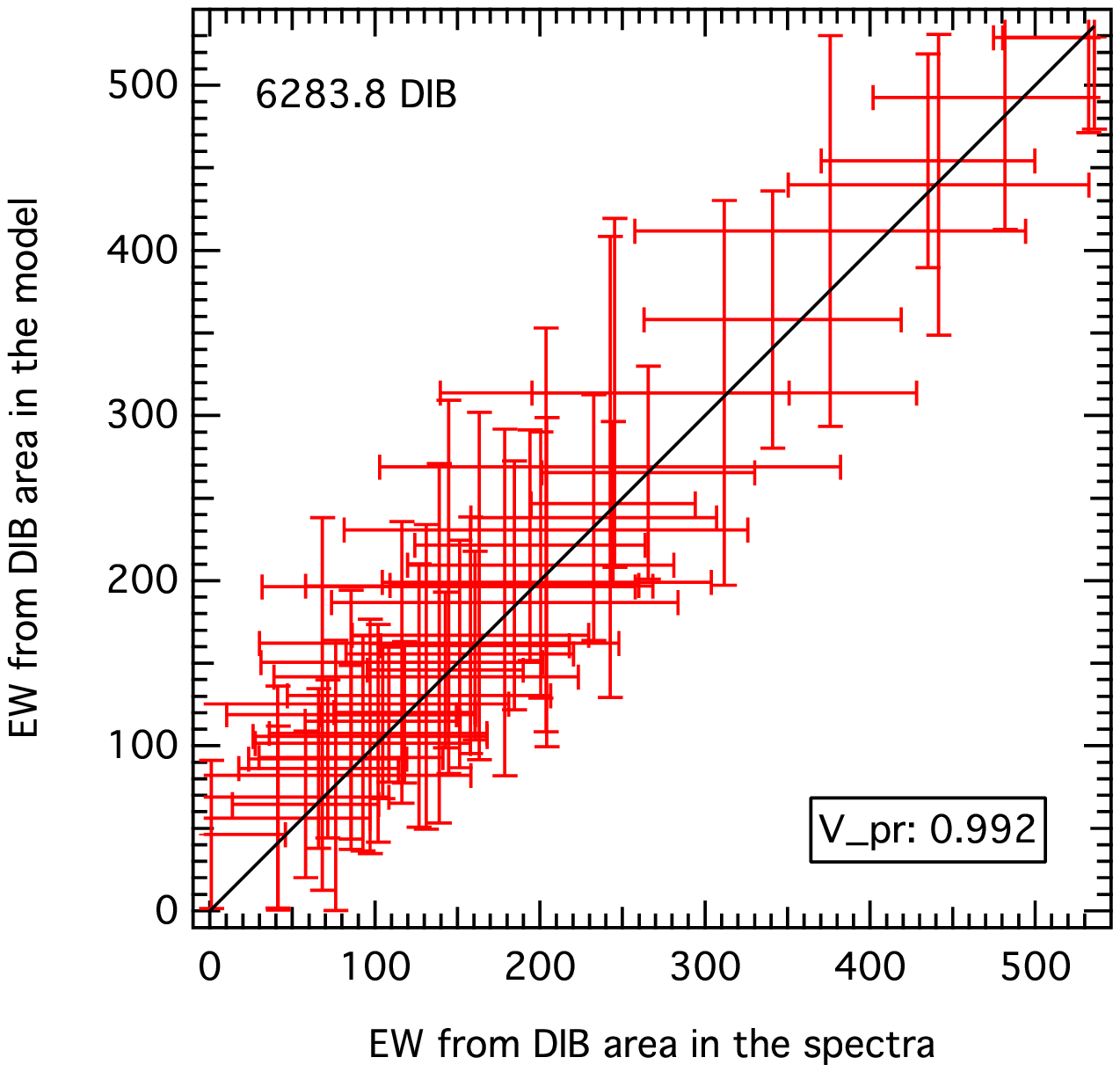}
	\includegraphics[width=0.2\linewidth]{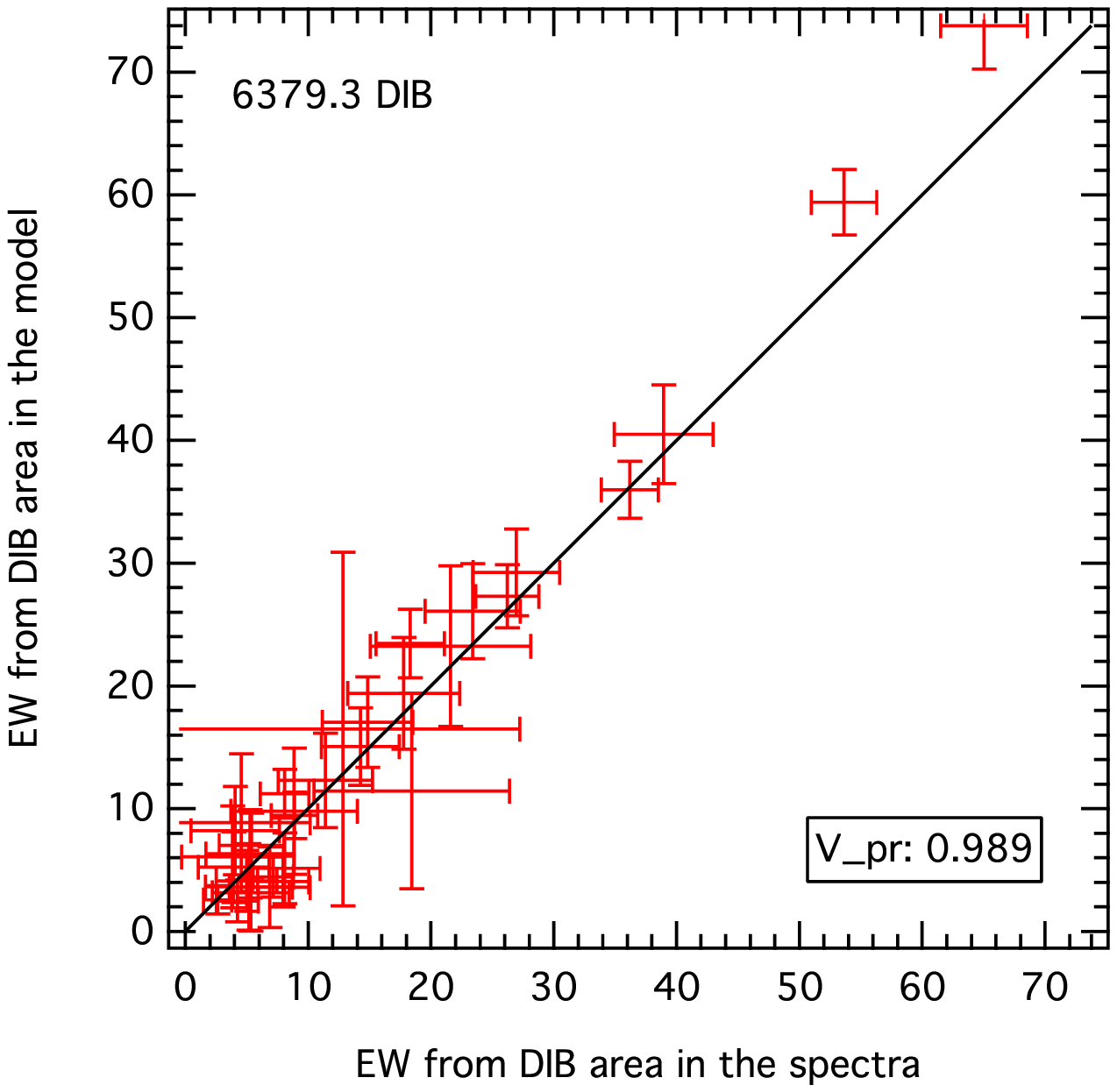}
	\includegraphics[width=0.2\linewidth]{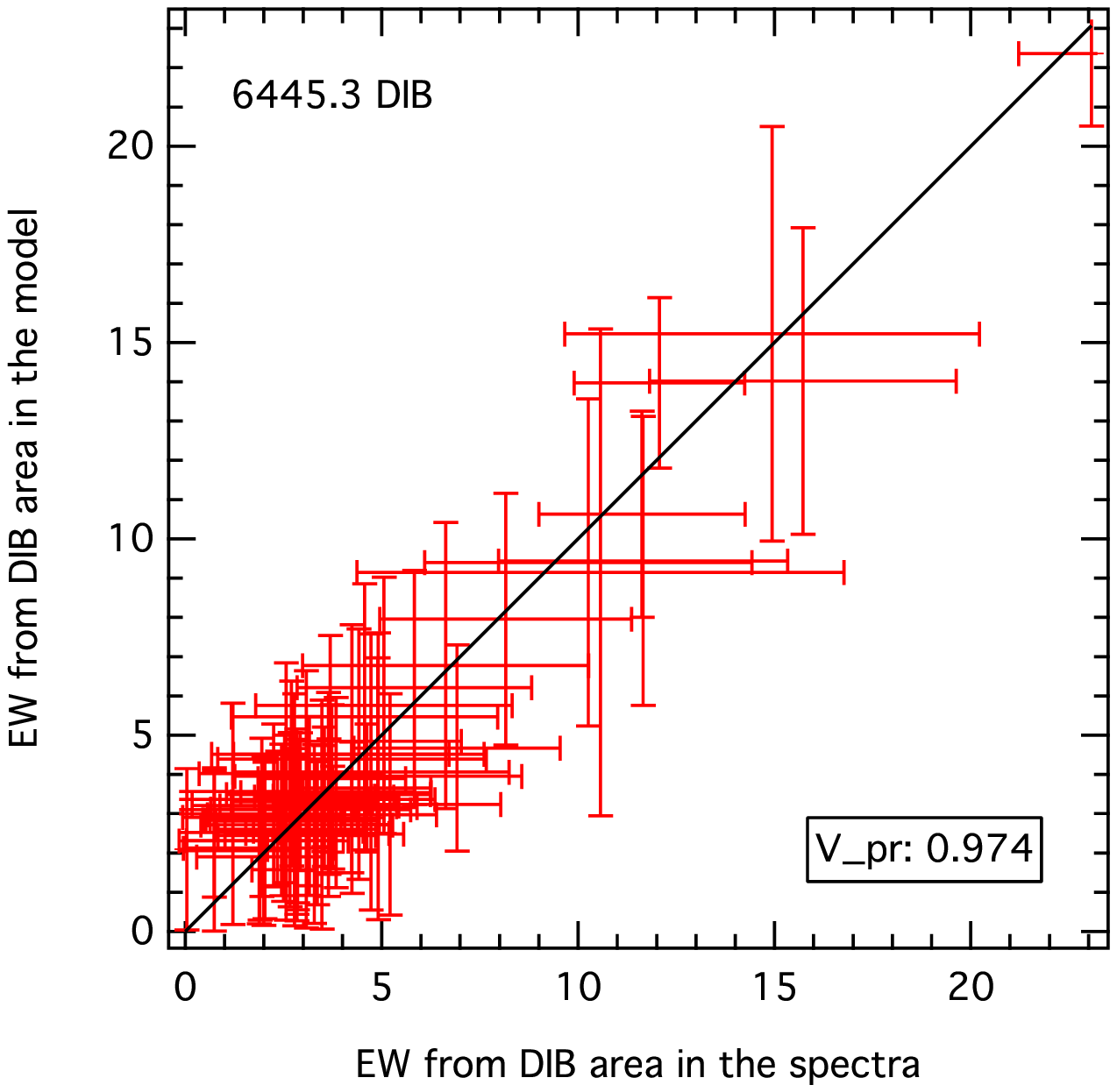}
	\includegraphics[width=0.2\linewidth]{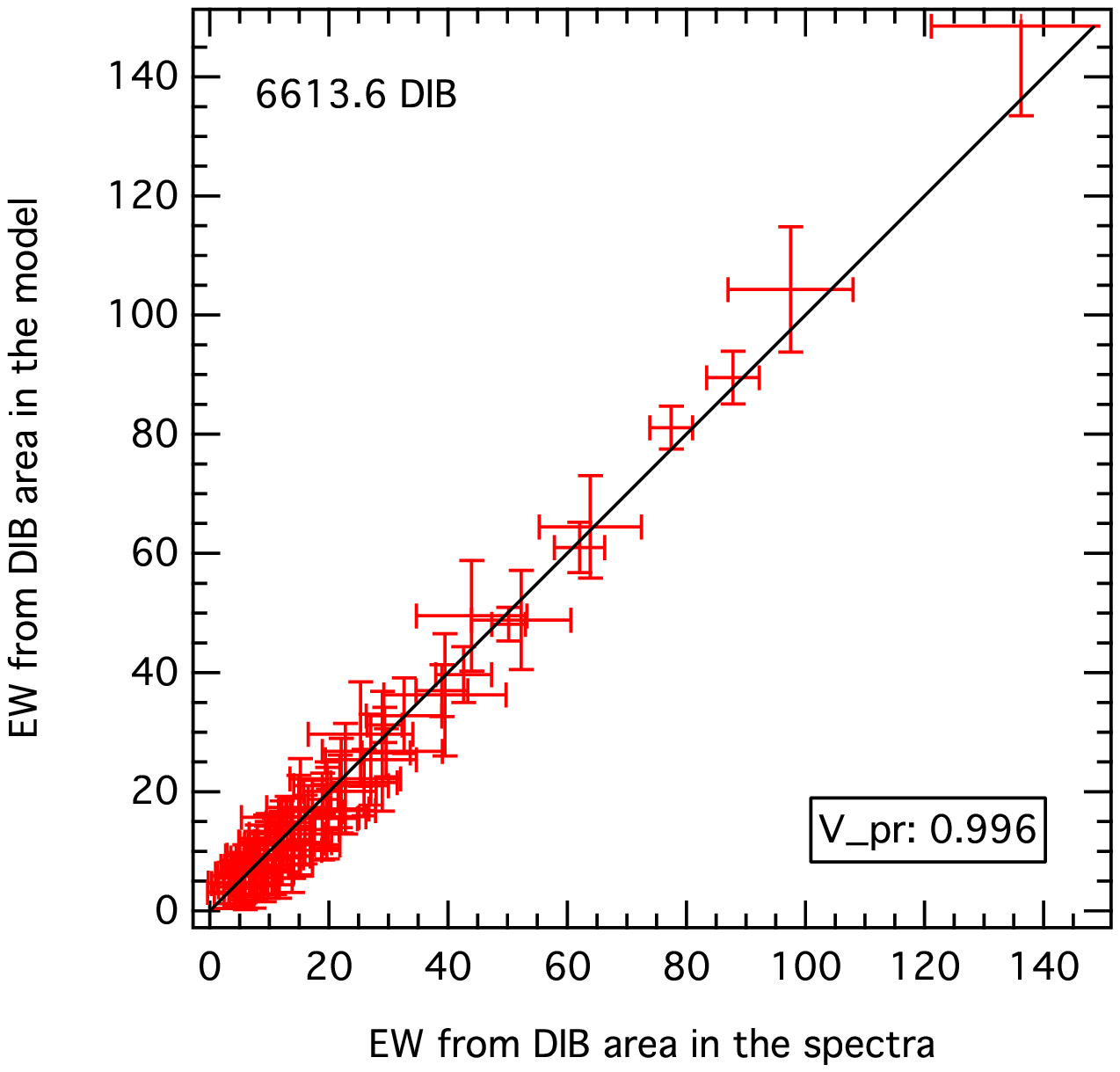}
	\includegraphics[width=0.2\linewidth]{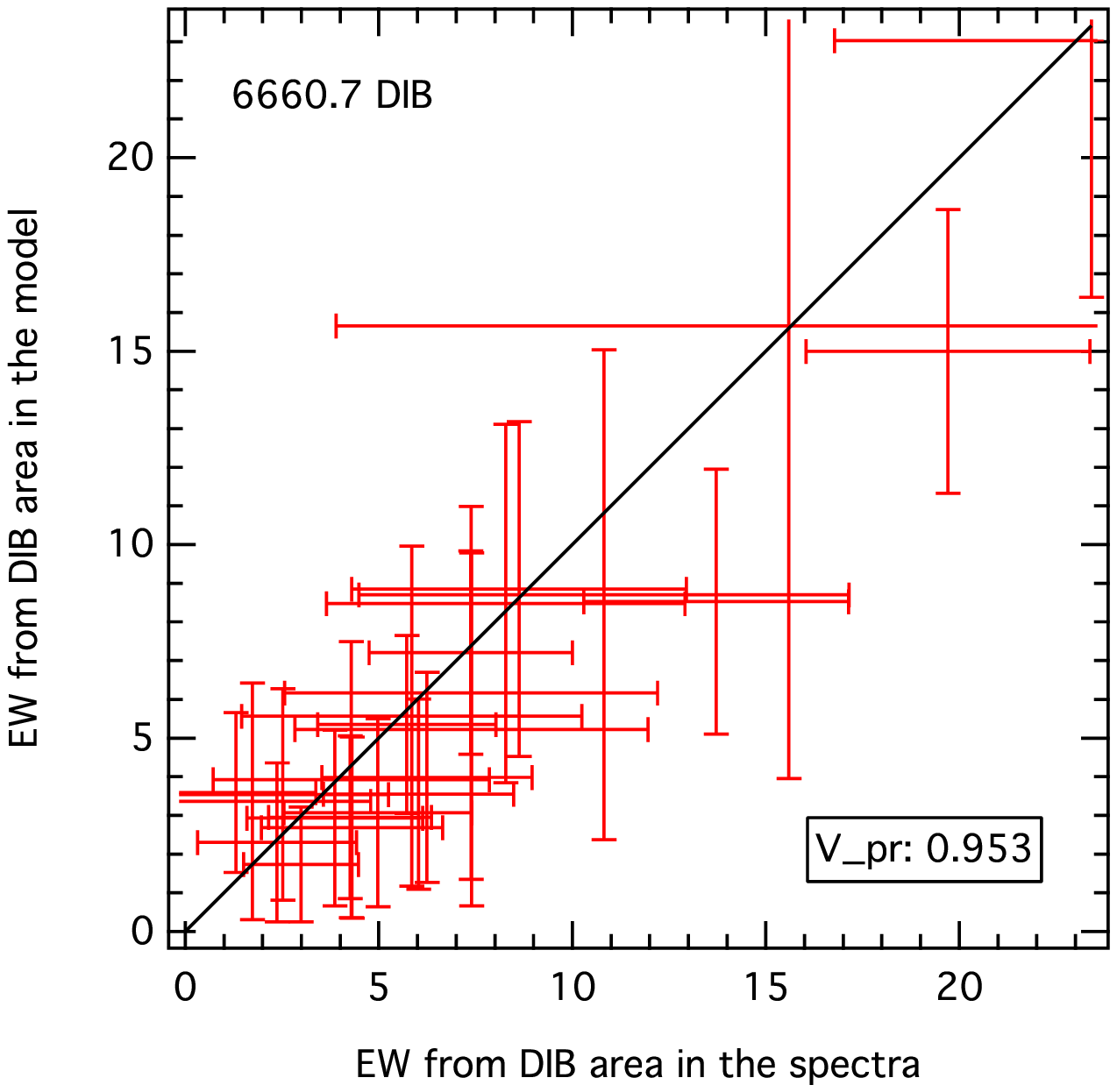}
	\includegraphics[width=0.2\linewidth]{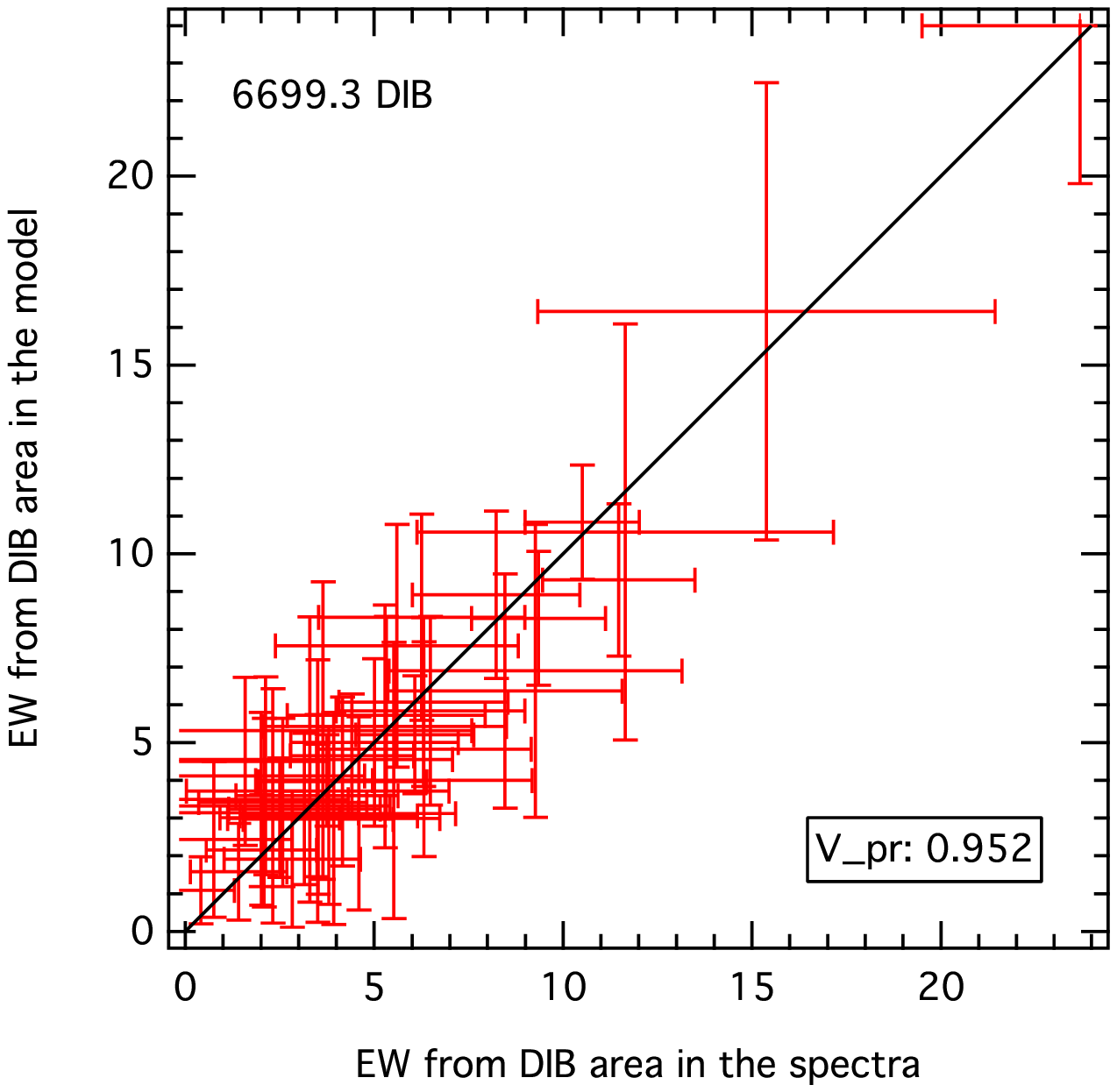}
\caption{Comparison of results from  the two methods of EW measurements, the fitted EWs vs. the continuum-integrated EWs.}%
\label{ewmodelvsspectra}%
\end{figure*}

Fig. \ref{ewmodelvsspectra} shows their comparison for the 16 DIBs. It can be seen that for almost all DIBs the differences are inferior to the uncertainties in the very large majority of measurements. As said above, DIB shape variabilities and velocity spread of the interstellar clouds must be reflected in the differences between the two EWs. The absence of significant differences may be explained here by the proximity of the target stars, which results in a small velocity dispersion. It also implies that the  DIB shape is representative enough of the majority of the absorptions in the nearby ISM. For that reason, we can use any of the EWs to study the correlations with other parameters.
The exception is the 6203-6204 \AA\ DIB, and this is certainly related to the fact that there are two components and that their relative strengths do vary from one target to the other. For this system it may be interesting to consider both the fitted EW and the continuum-integrated EW in the search for correlations (see the two results in table 
\ref{tablecorr}).

The online table contains the list of target stars, their coordinates, the color excess as determined from the Geneva photometry (see section 4) and  the measured equivalent width (the fitted EWs). 
The results show that the DIB absorptions are weak and that we are far from saturation, as expected for our nearby targets.

\begin{figure*}
\centering
\begin{minipage}[t]{0.5\linewidth}
\centering
	\includegraphics[width=0.8\linewidth]{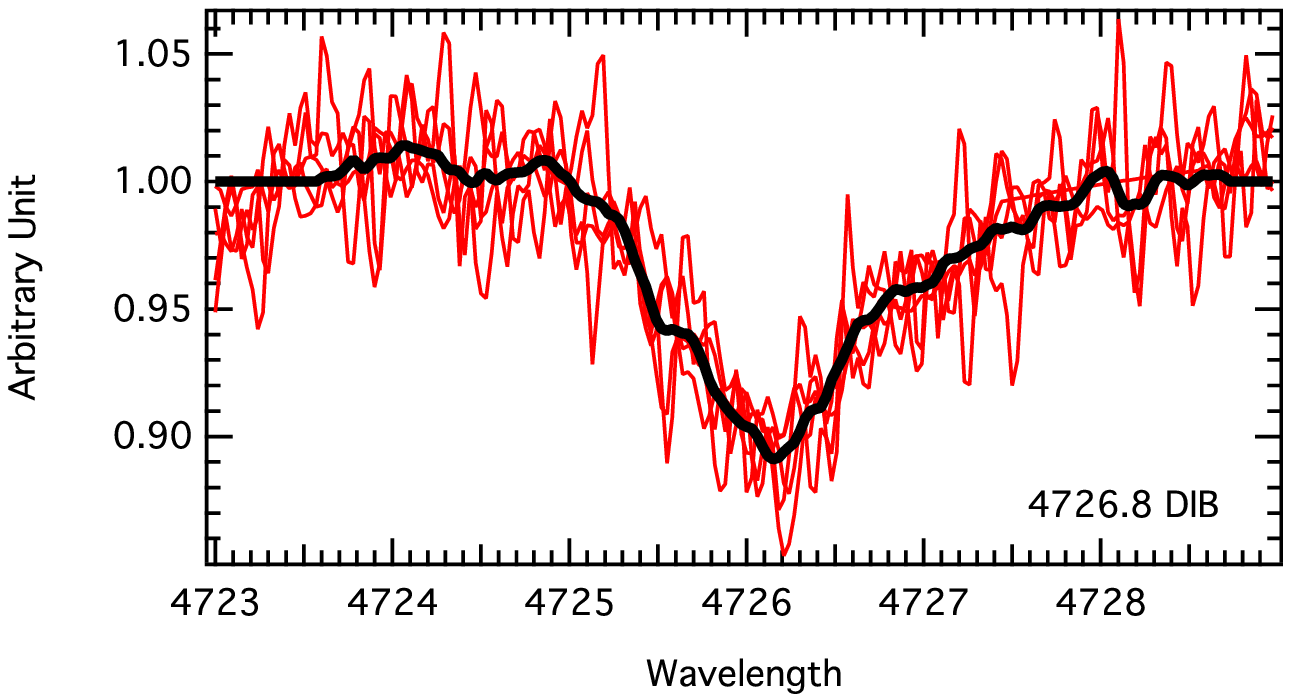} 
	\includegraphics[width=0.8\linewidth]{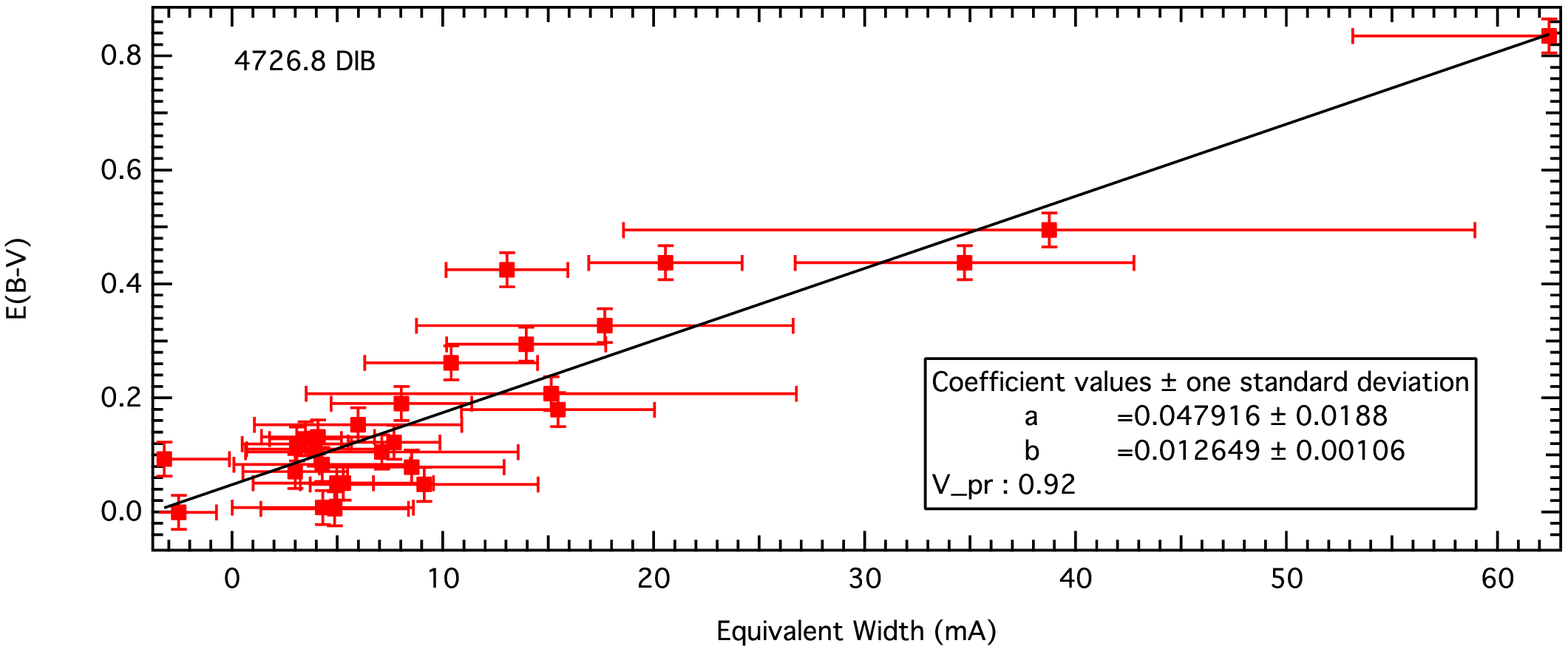}
\end{minipage}\hfill
\begin{minipage}[t]{0.5\linewidth}
\centering
	\includegraphics[width=0.8\linewidth]{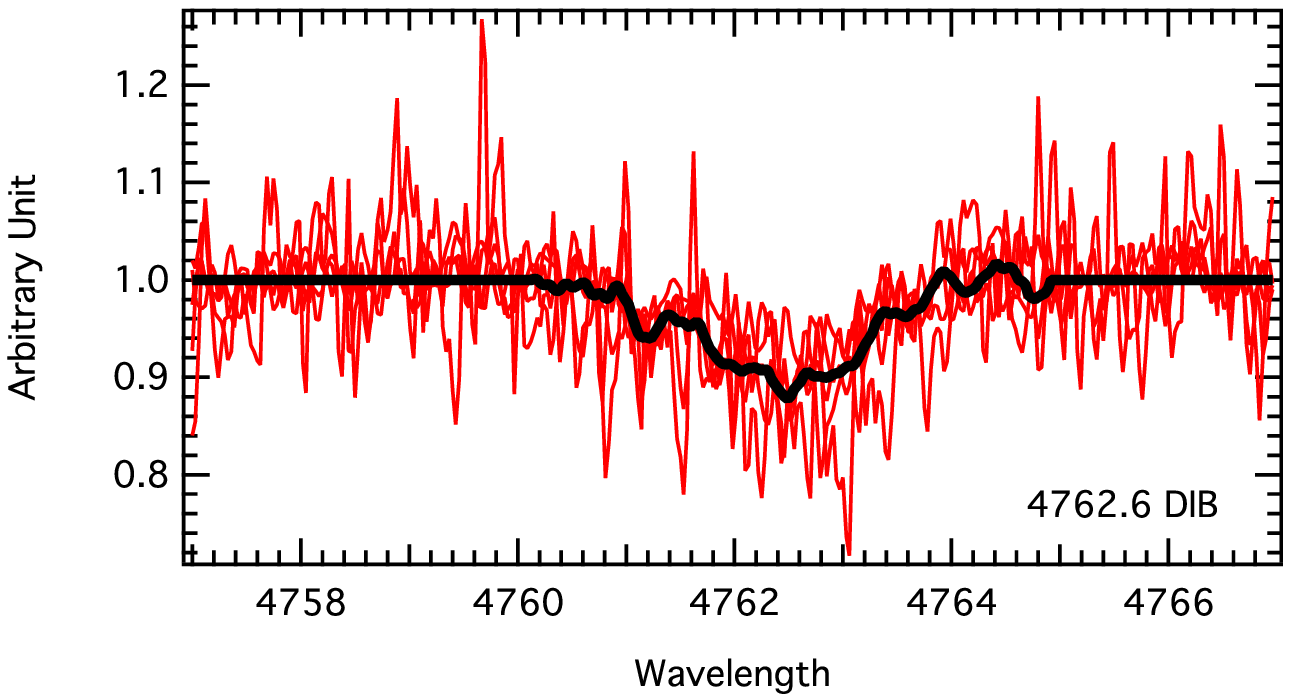} 
	\includegraphics[width=0.8\linewidth]{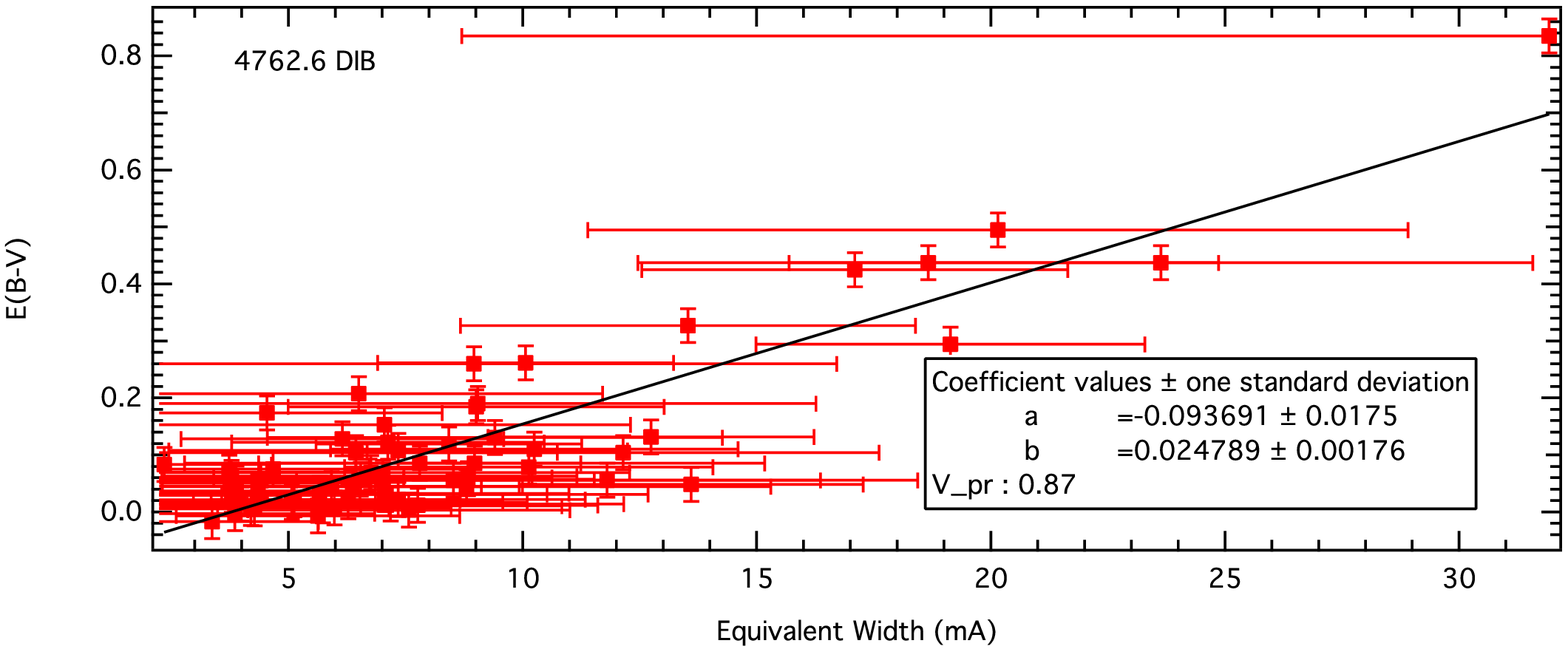}
\end{minipage}\hfill
\begin{minipage}[t]{0.5\linewidth}
\centering
	\includegraphics[width=0.8\linewidth]{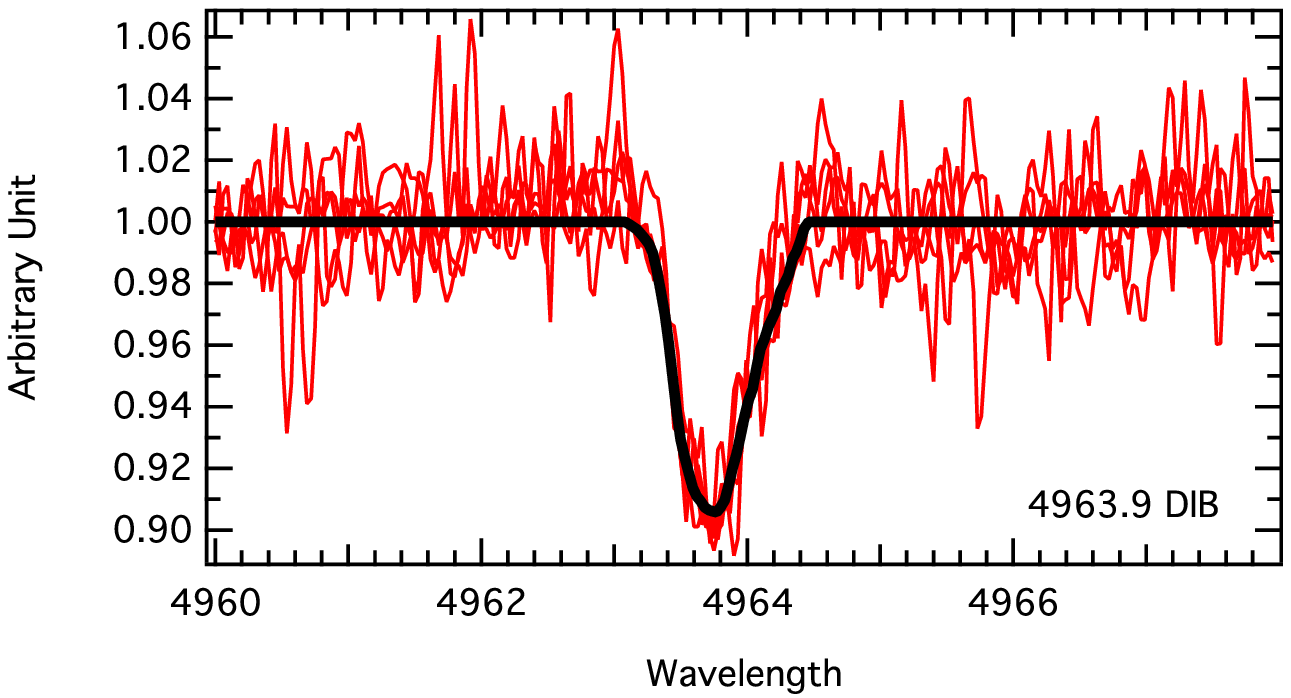} 
	\includegraphics[width=0.8\linewidth]{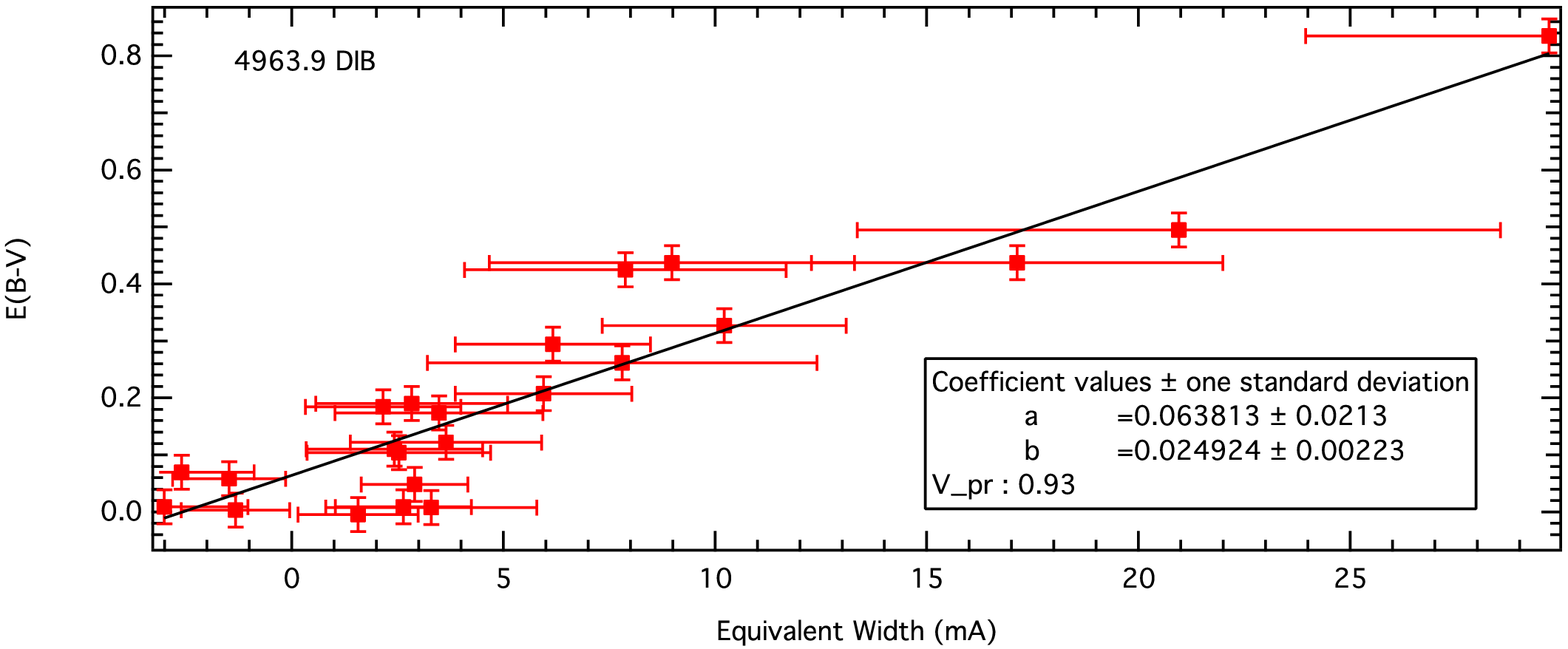}
\end{minipage}\hfill
\begin{minipage}[t]{0.5\linewidth}
\centering
	\includegraphics[width=0.8\linewidth]{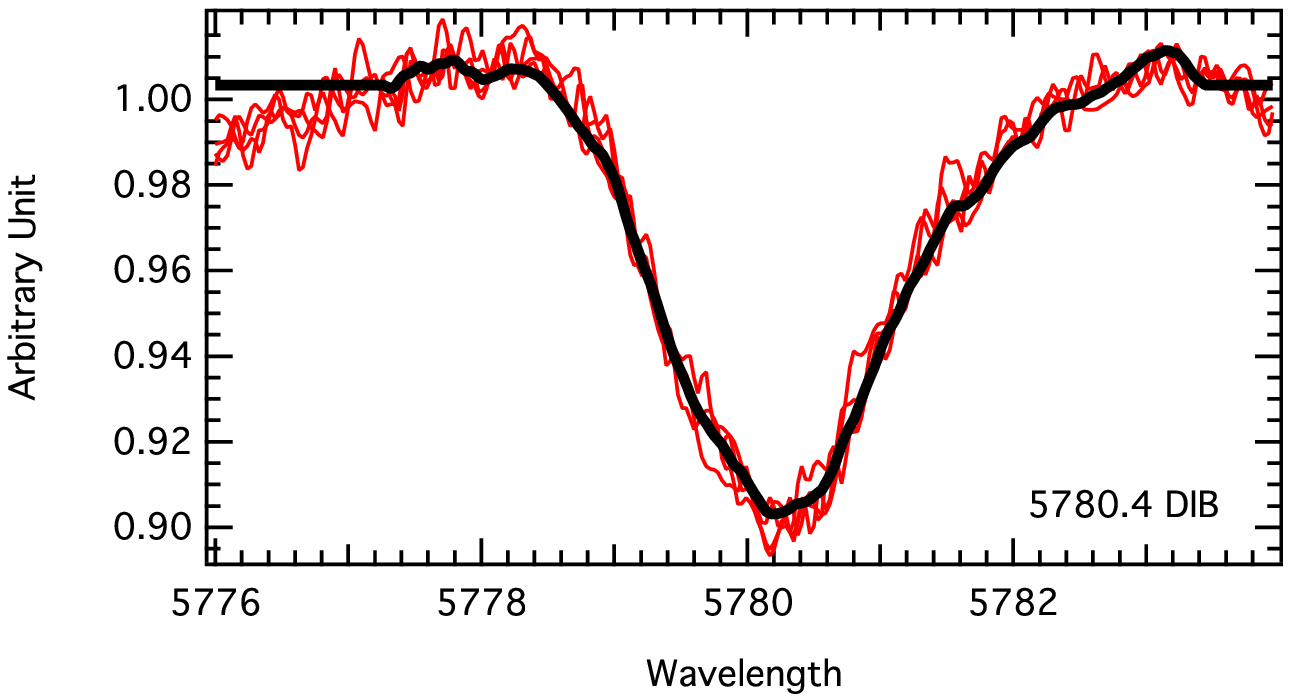} 
	\includegraphics[width=0.8\linewidth]{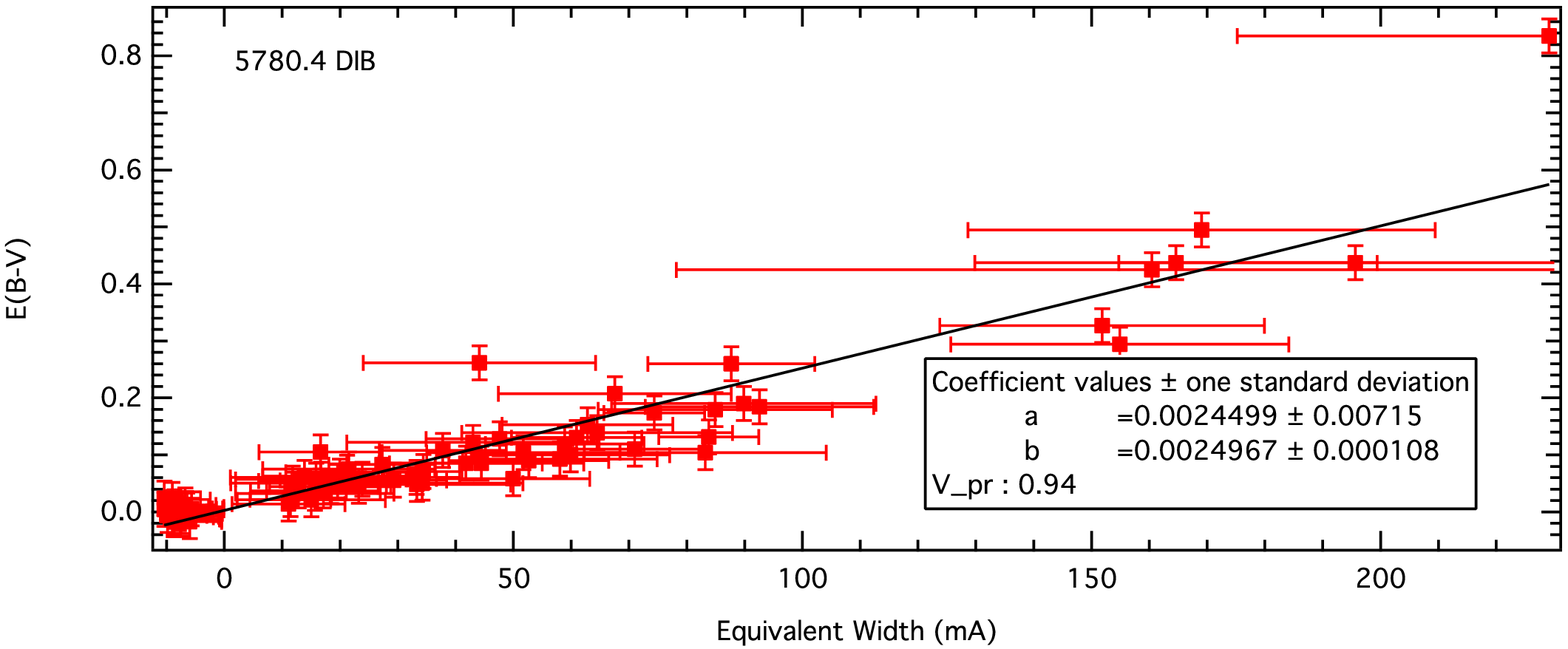}
\end{minipage}\hfill
\begin{minipage}[t]{0.5\linewidth}
\centering
	\includegraphics[width=0.8\linewidth]{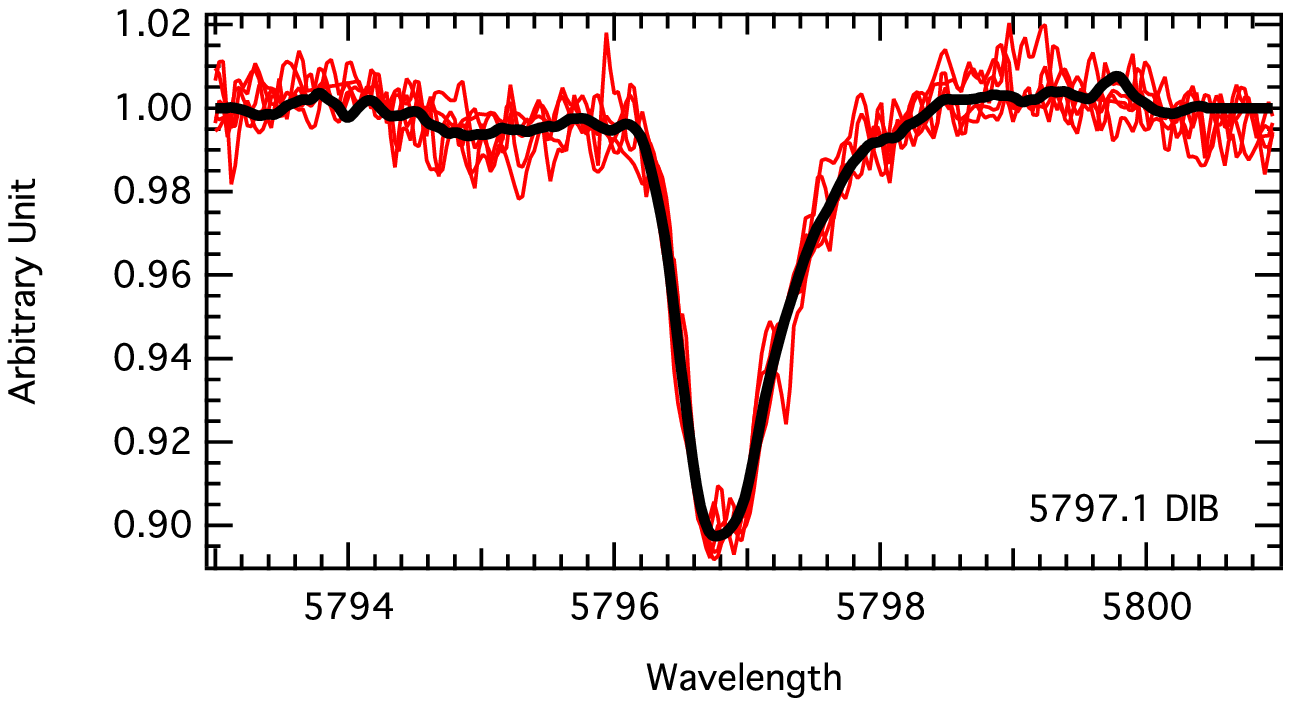} 
	\includegraphics[width=0.8\linewidth]{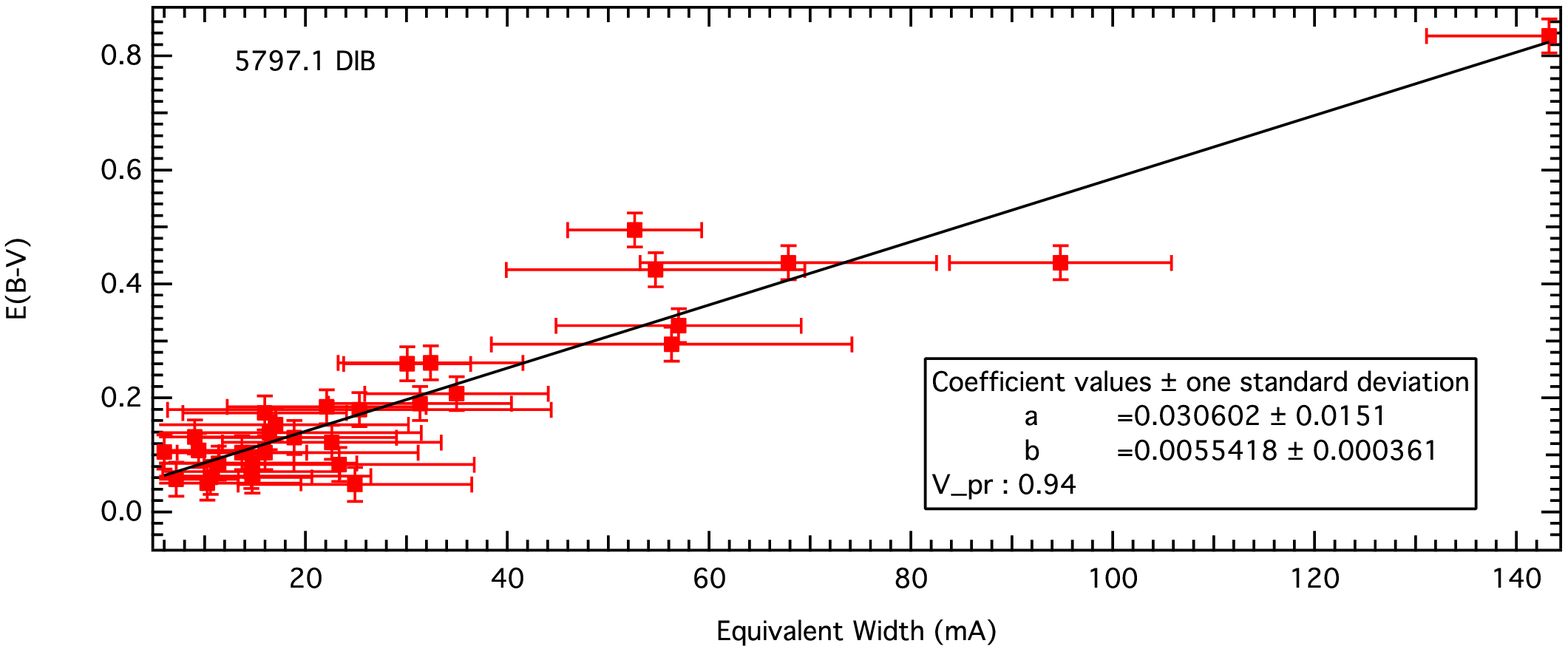}
\end{minipage}\hfill
\begin{minipage}[t]{0.5\linewidth}
\centering
	\includegraphics[width=0.8\linewidth]{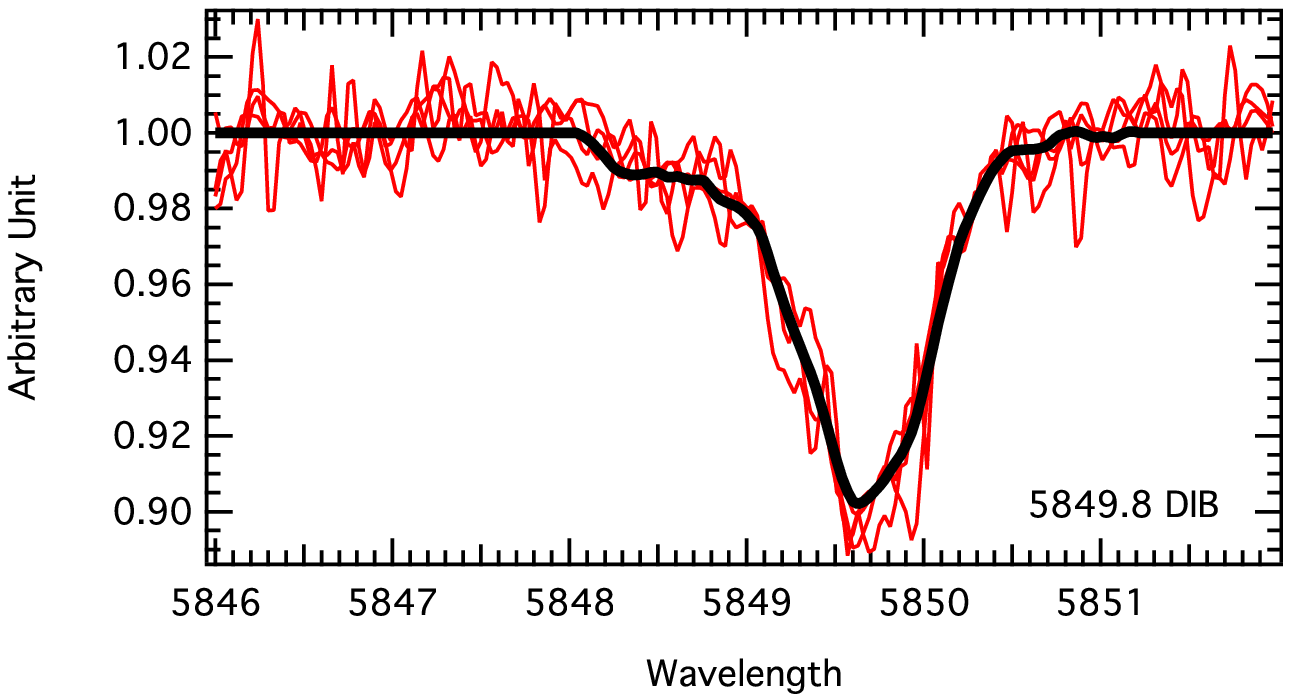} 
	\includegraphics[width=0.8\linewidth]{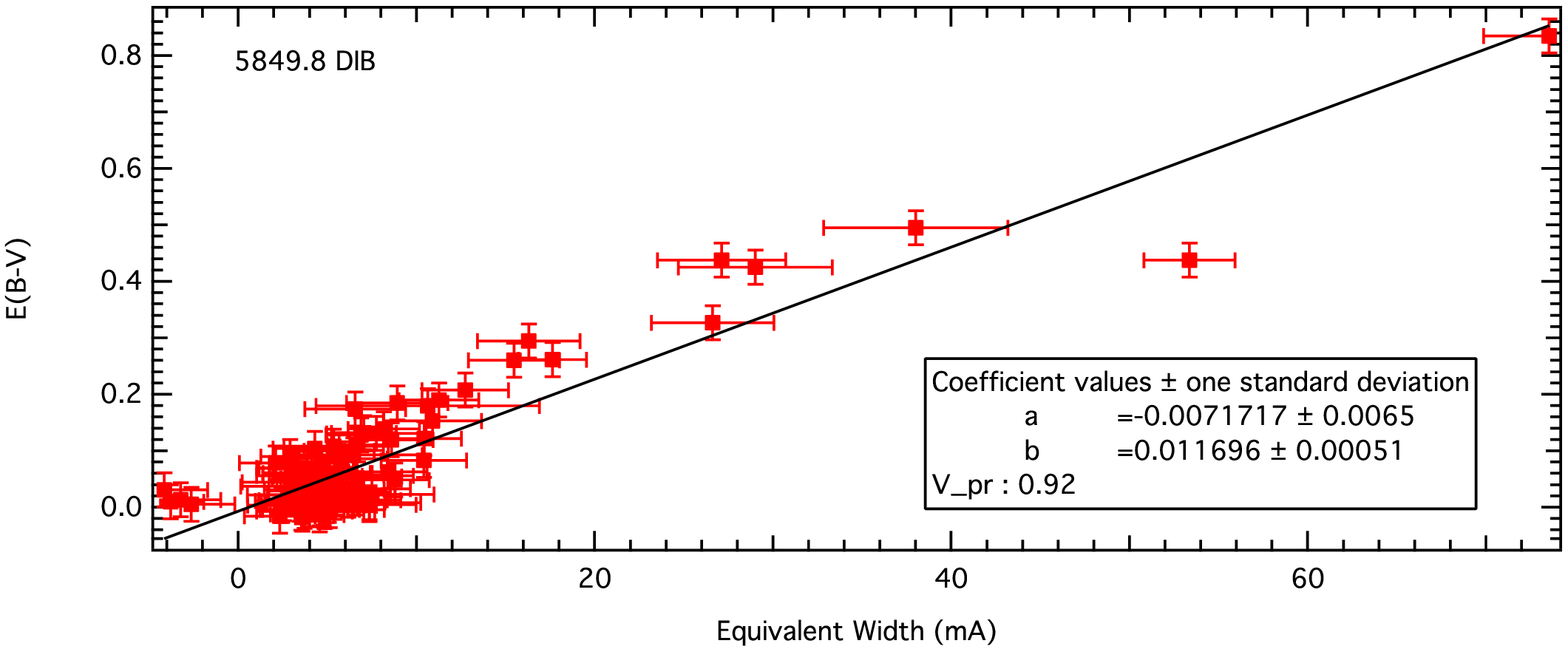}
\end{minipage}\hfill
\caption{DIB model and color excess-DIB equivalent width correlation. }
\label{DIBmodel}%
\end{figure*}

\begin{figure*}
\centering
\begin{minipage}[t]{0.5\linewidth}
\centering
	\includegraphics[width=0.8\linewidth]{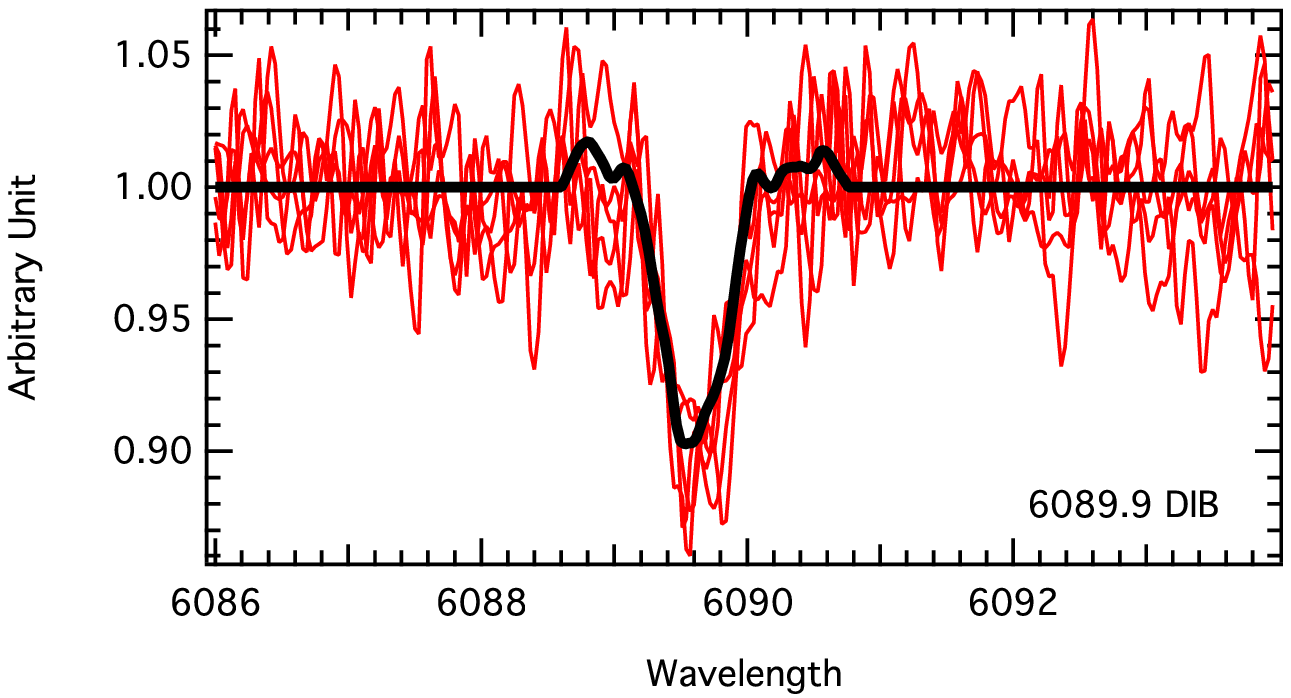} 
	\includegraphics[width=0.8\linewidth]{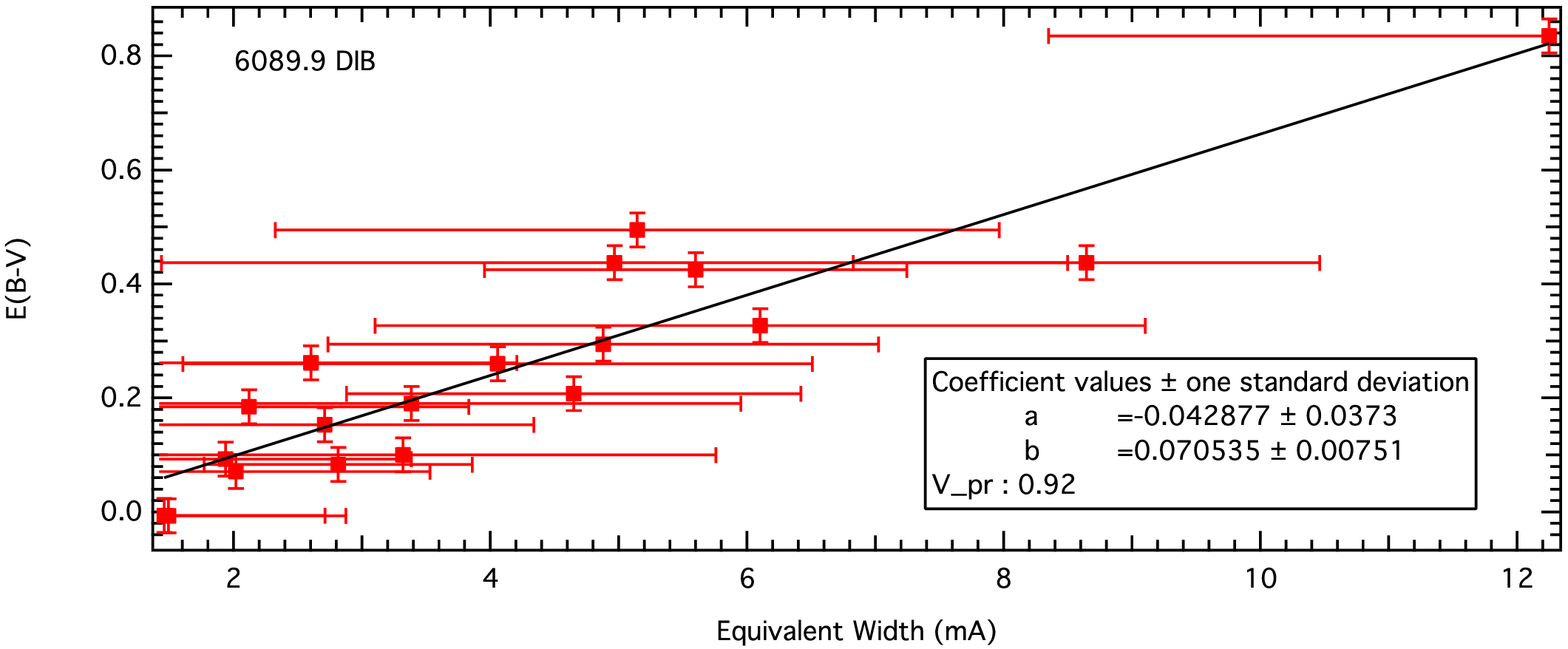}
\end{minipage}\hfill
\begin{minipage}[t]{0.5\linewidth}
\centering
	\includegraphics[width=0.8\linewidth]{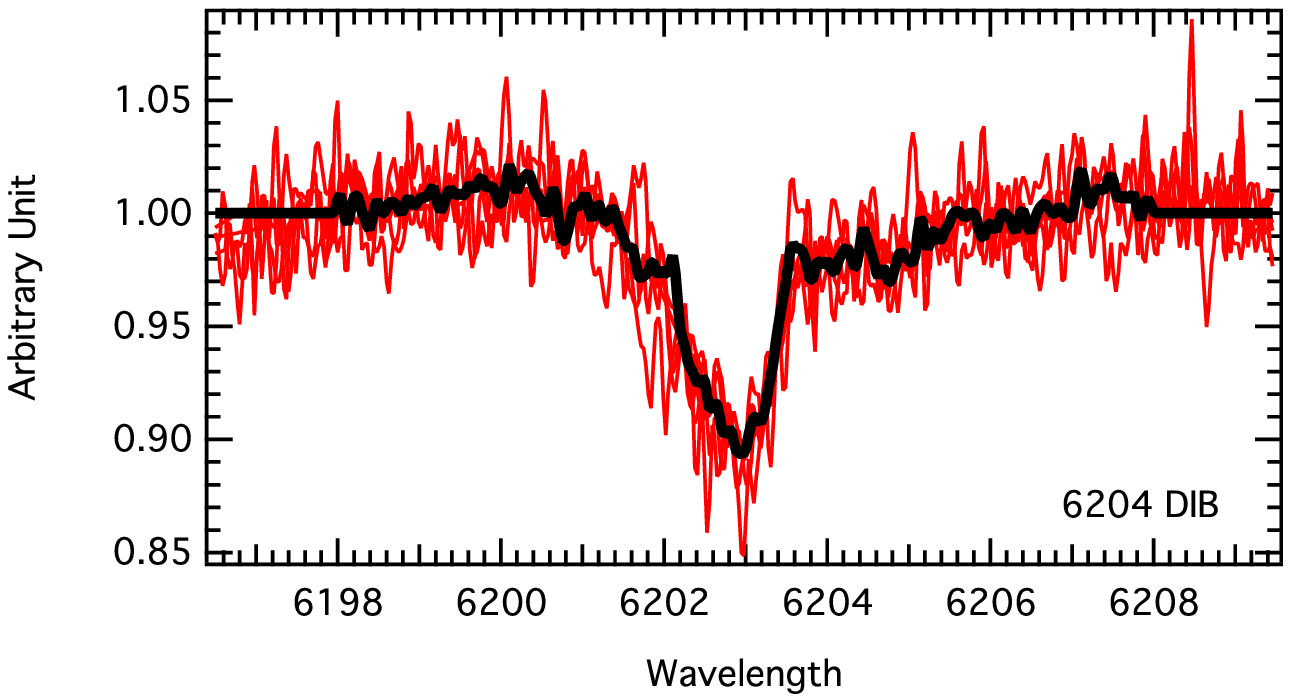} 
	\includegraphics[width=0.8\linewidth]{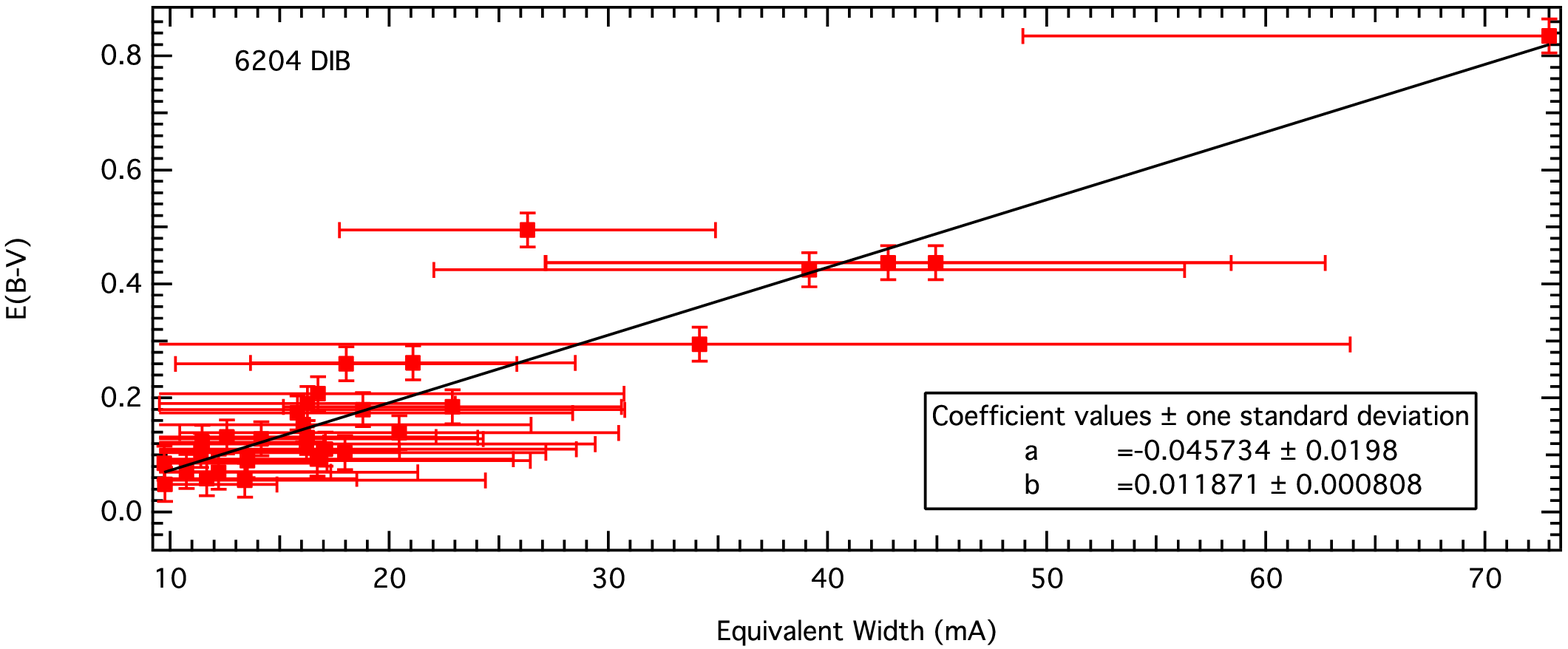}
\end{minipage}\hfill
\begin{minipage}[t]{0.5\linewidth}
\centering
	\includegraphics[width=0.8\linewidth]{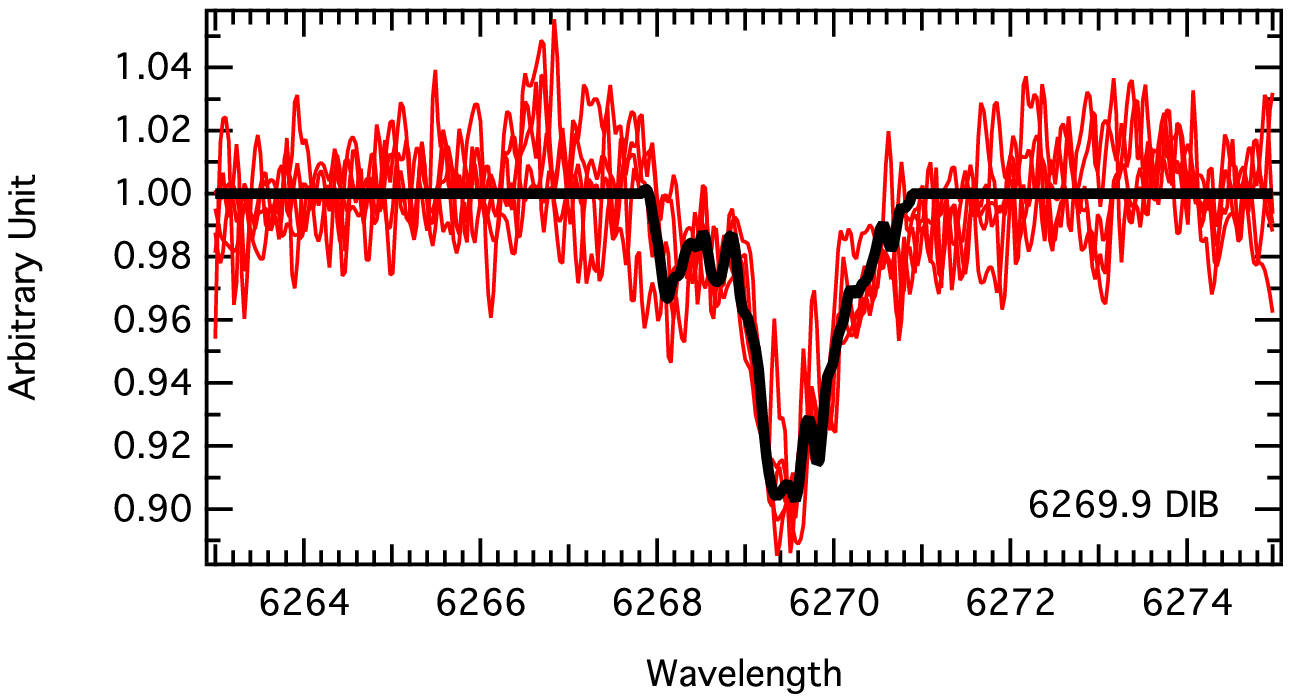} 
	\includegraphics[width=0.8\linewidth]{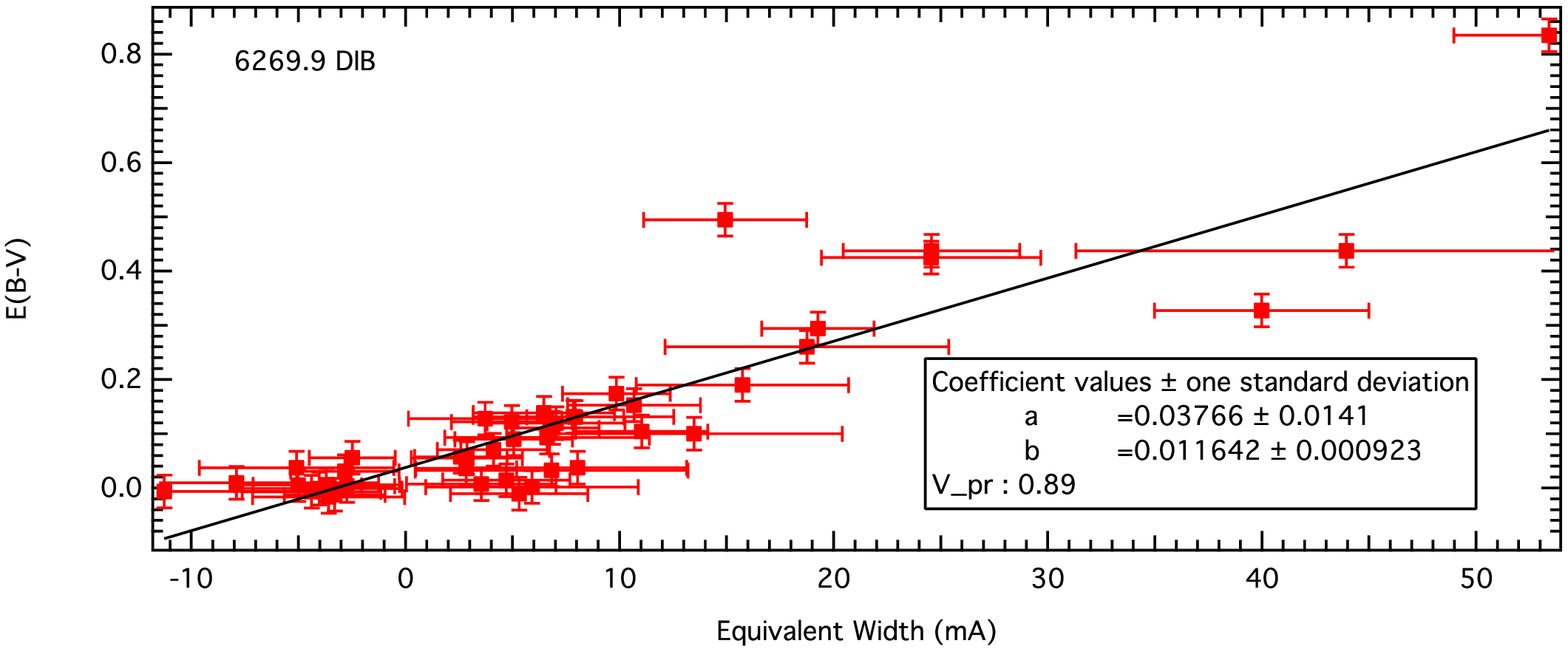}
\end{minipage}\hfill
\begin{minipage}[t]{0.5\linewidth}
\centering
	\includegraphics[width=0.8\linewidth]{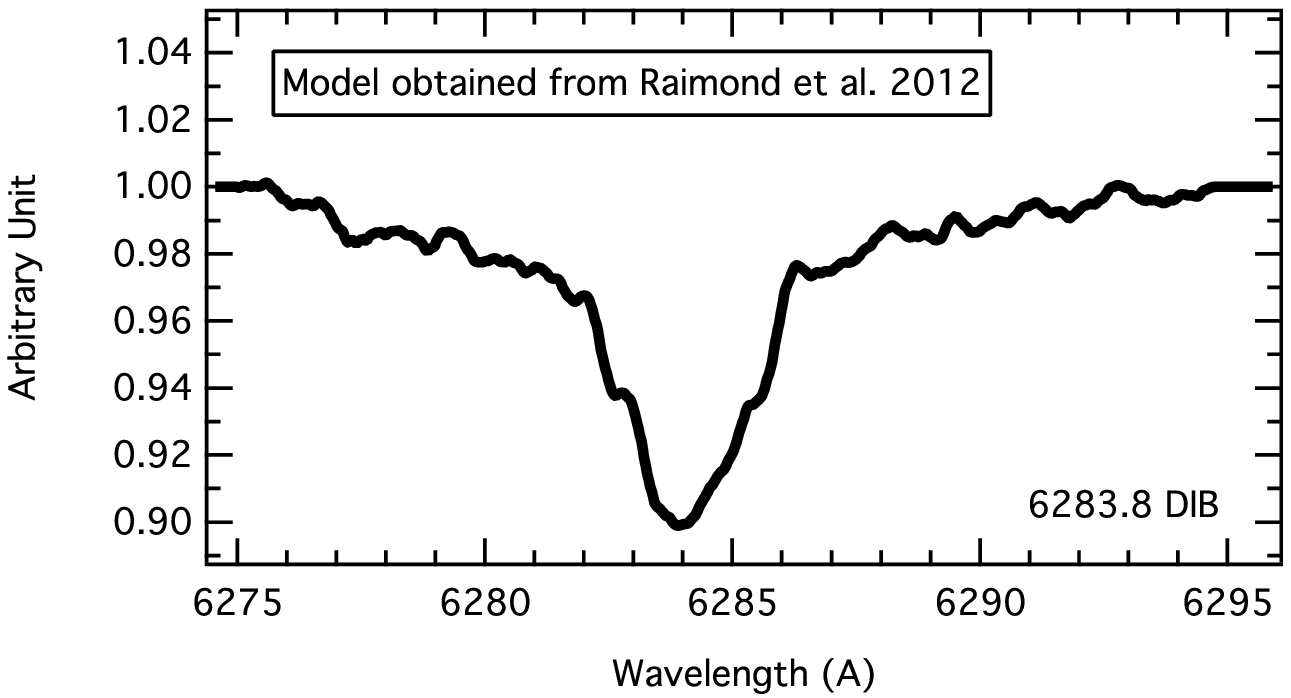} 
	\includegraphics[width=0.8\linewidth]{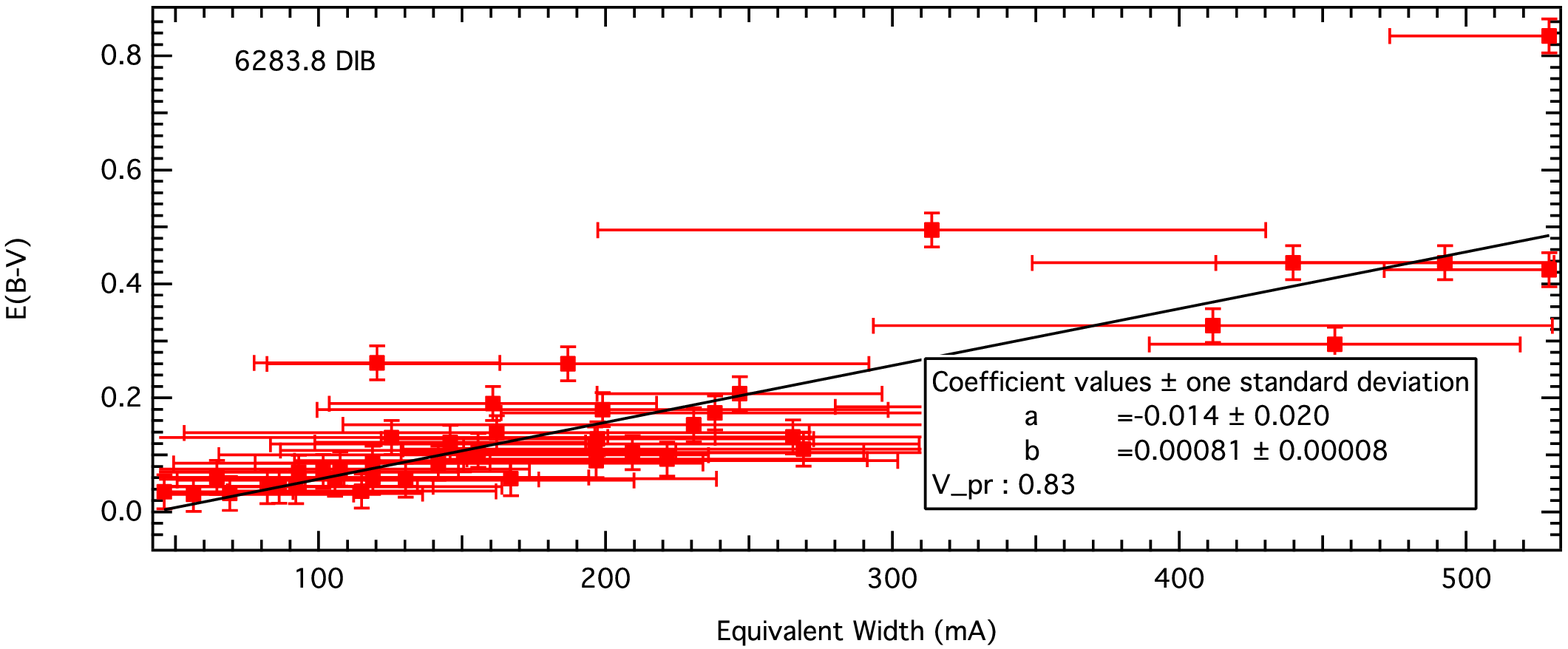}
\end{minipage}\hfill
\begin{minipage}[t]{0.5\linewidth}
\centering
	\includegraphics[width=0.8\linewidth]{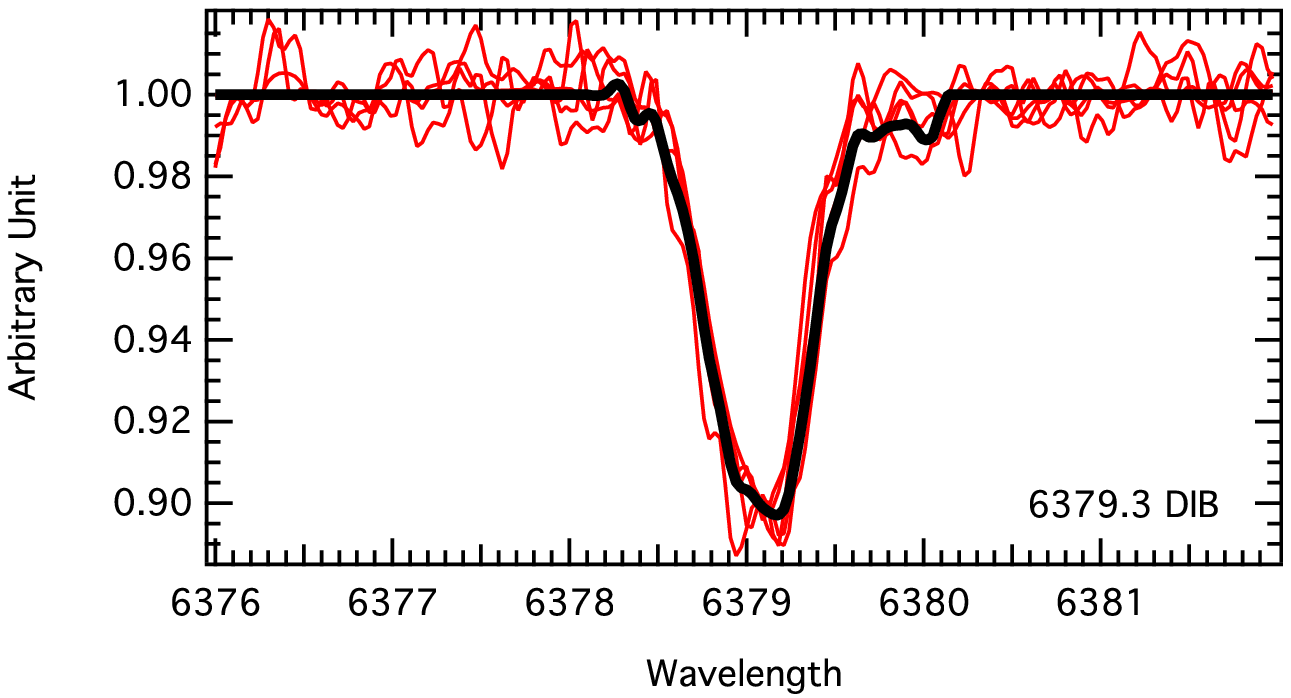} 
	\includegraphics[width=0.8\linewidth]{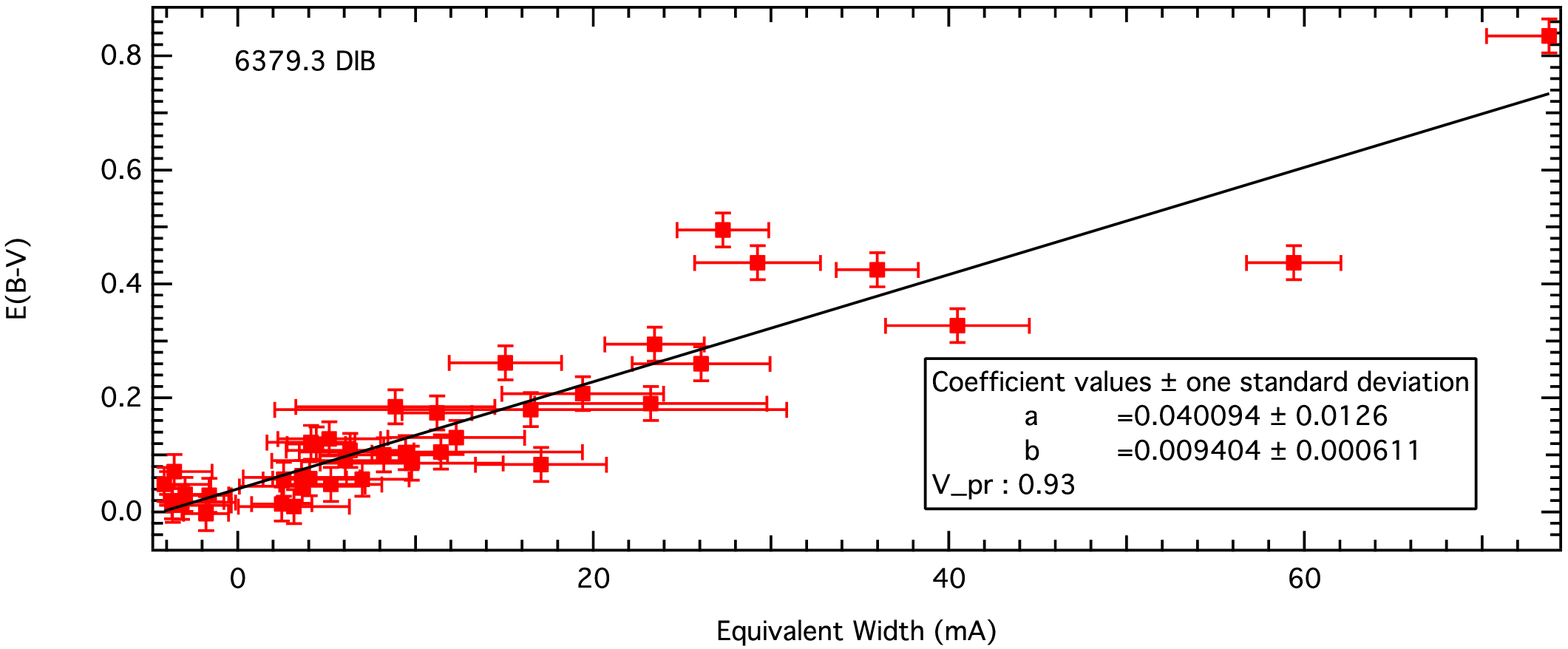}
\end{minipage}\hfill
\begin{minipage}[t]{0.5\linewidth}
\centering
	\includegraphics[width=0.8\linewidth]{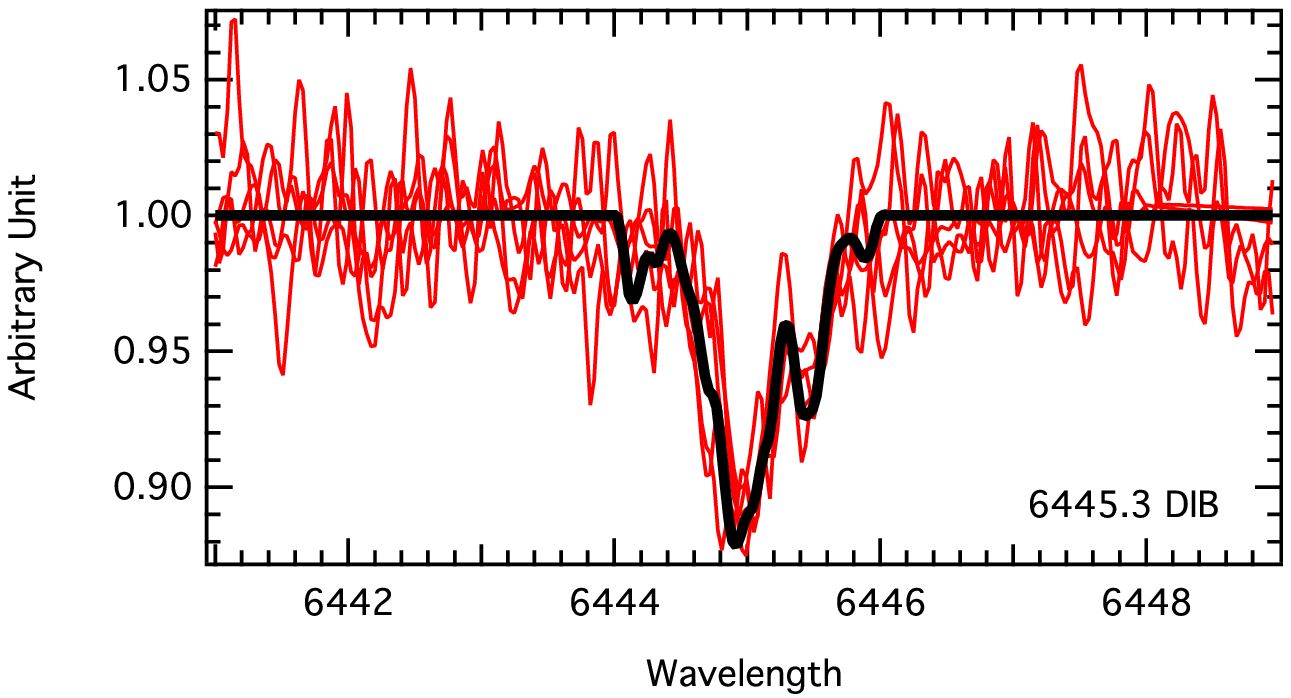} 
	\includegraphics[width=0.8\linewidth]{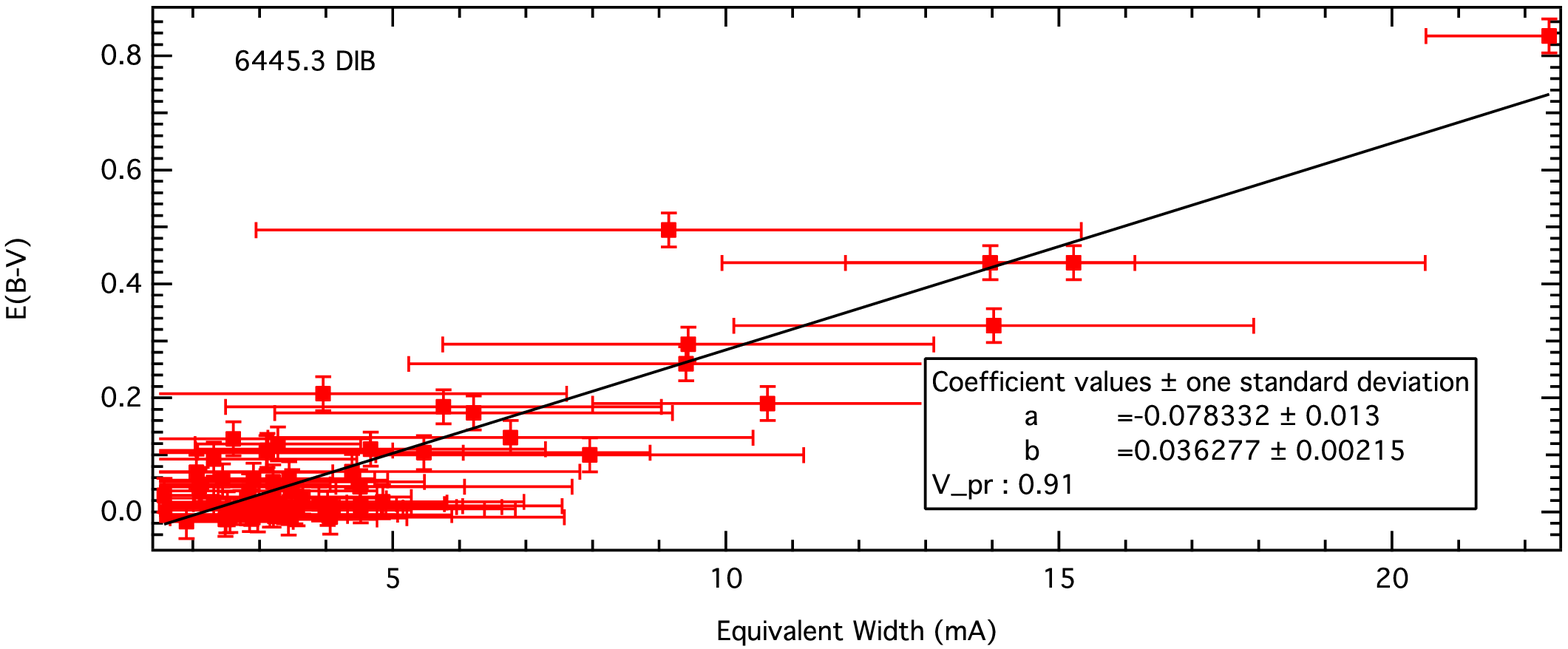}
\end{minipage}\hfill
\caption{DIB model and color excess-DIB equivalent width correlation (continued). }
\label{DIBmodel2}%
\end{figure*}

\begin{figure*}[h!b]
\centering
\begin{minipage}[t]{0.5\linewidth}
\centering
	\includegraphics[width=0.8\linewidth]{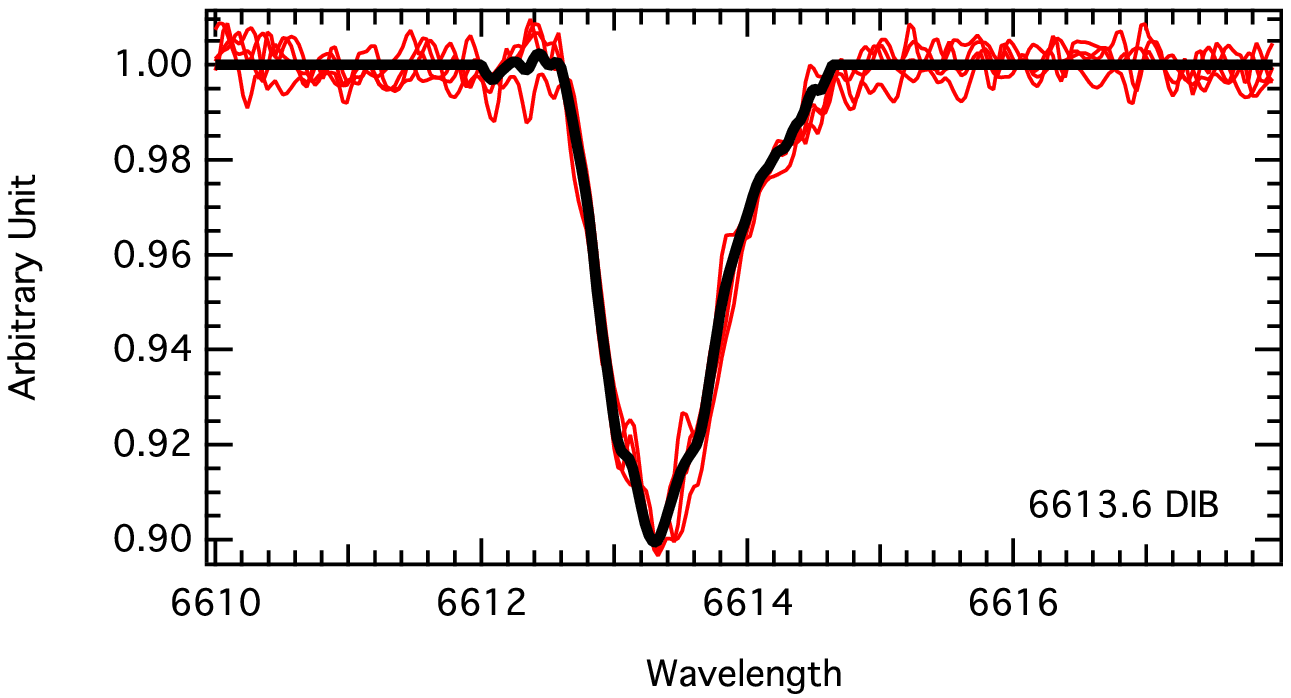} 
	\includegraphics[width=0.8\linewidth]{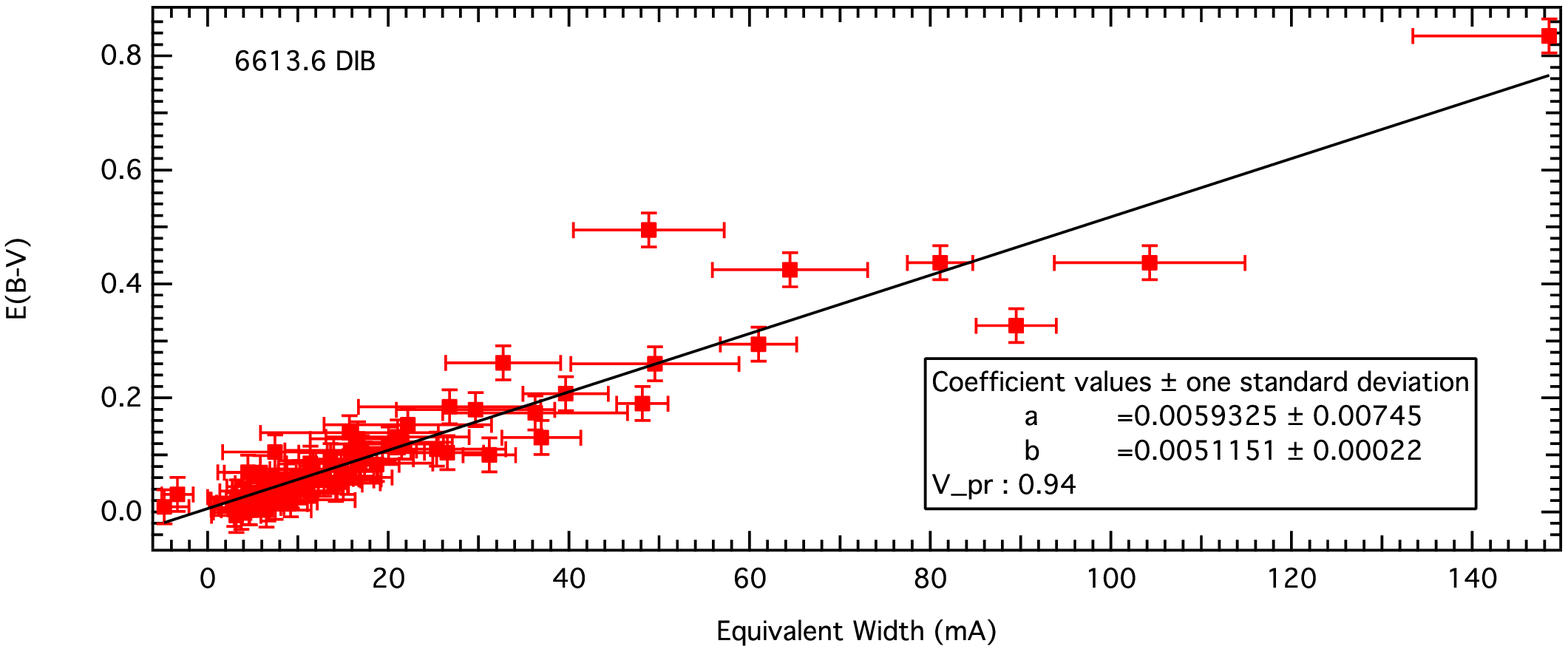}
\end{minipage}\hfill
\begin{minipage}[t]{0.5\linewidth}
\centering
	\includegraphics[width=0.8\linewidth]{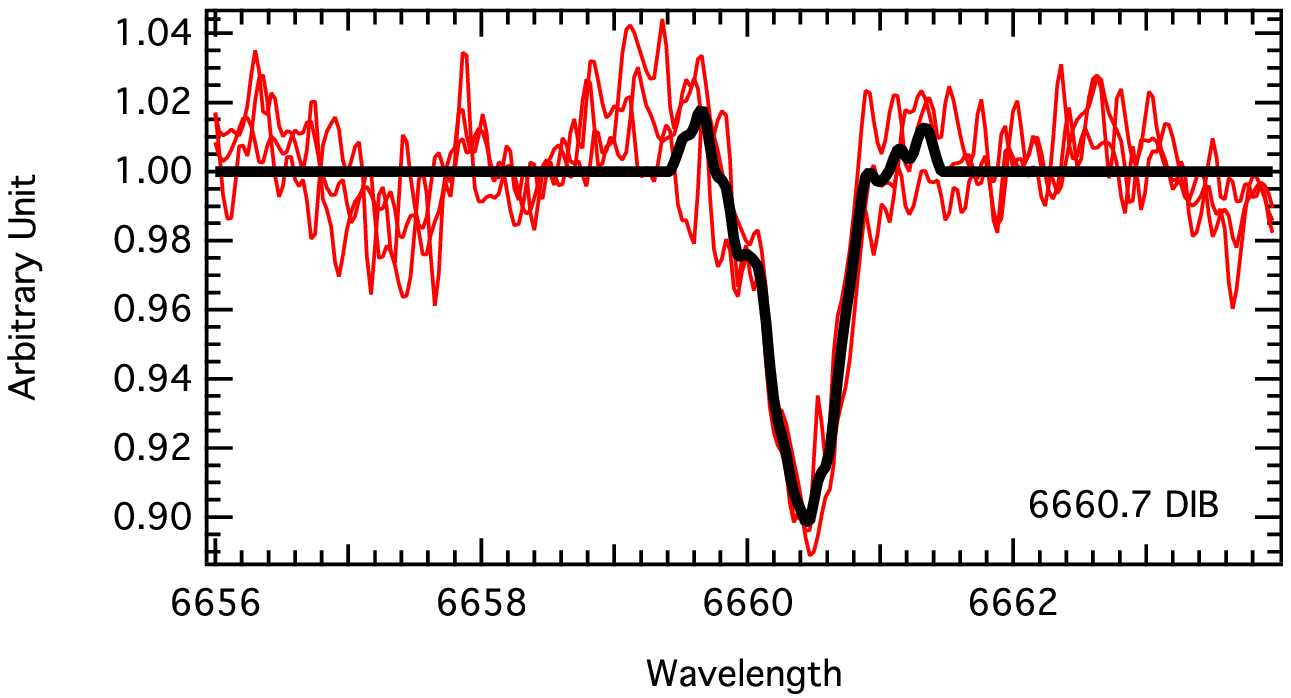} 
	\includegraphics[width=0.8\linewidth]{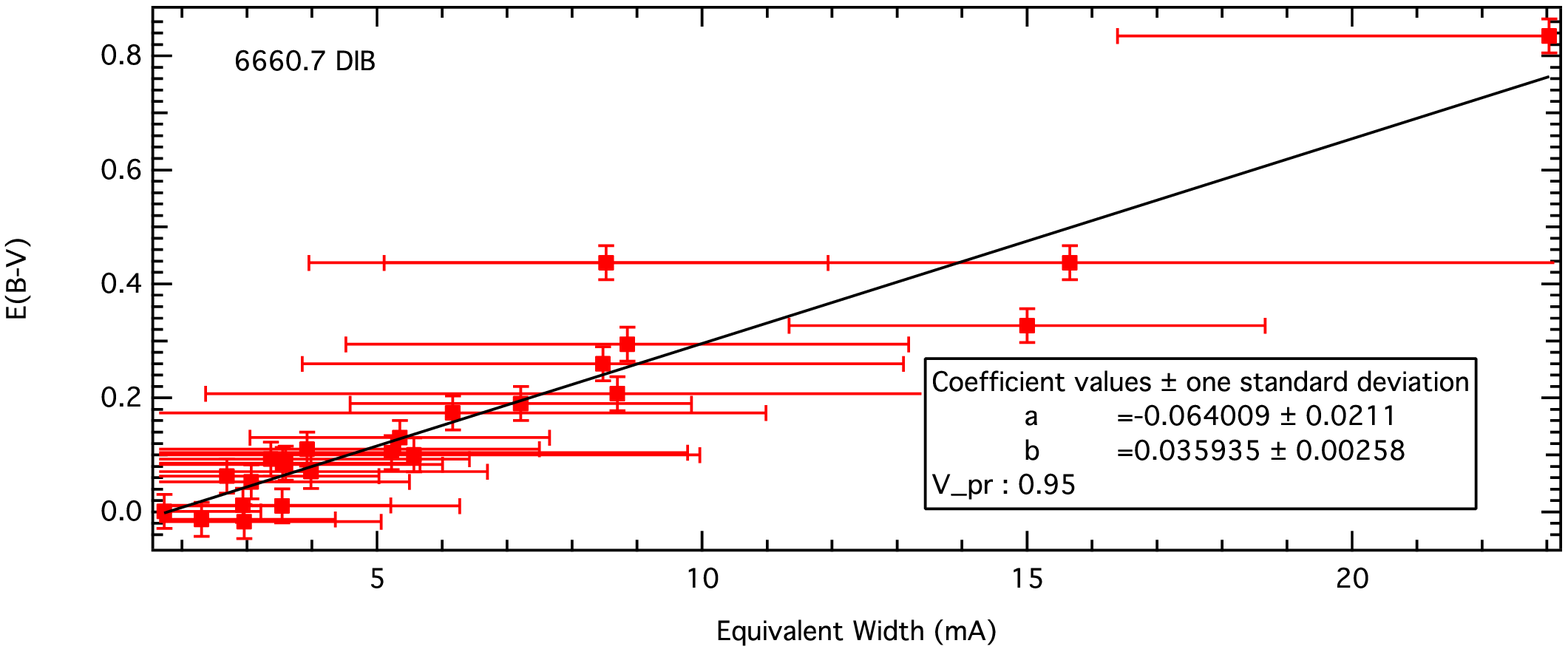}
\end{minipage}\hfill
\begin{minipage}[t]{0.5\linewidth}
\centering
	\includegraphics[width=0.8\linewidth]{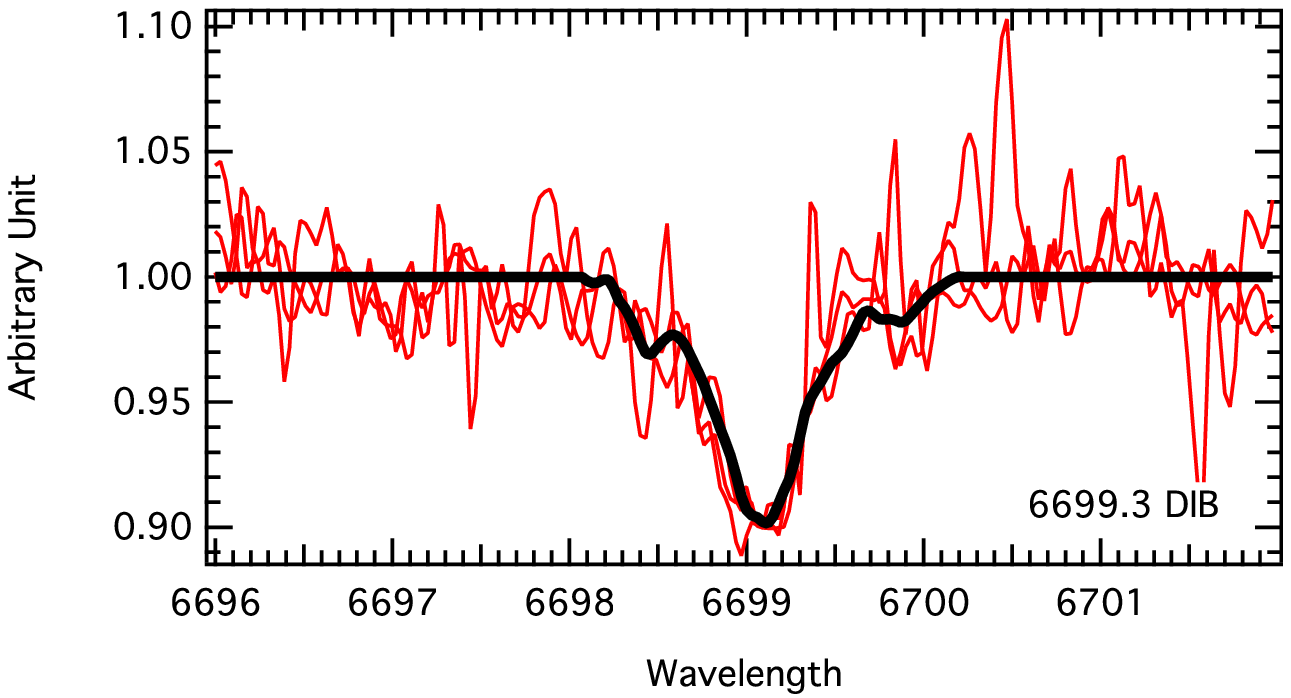} 
	\includegraphics[width=0.8\linewidth]{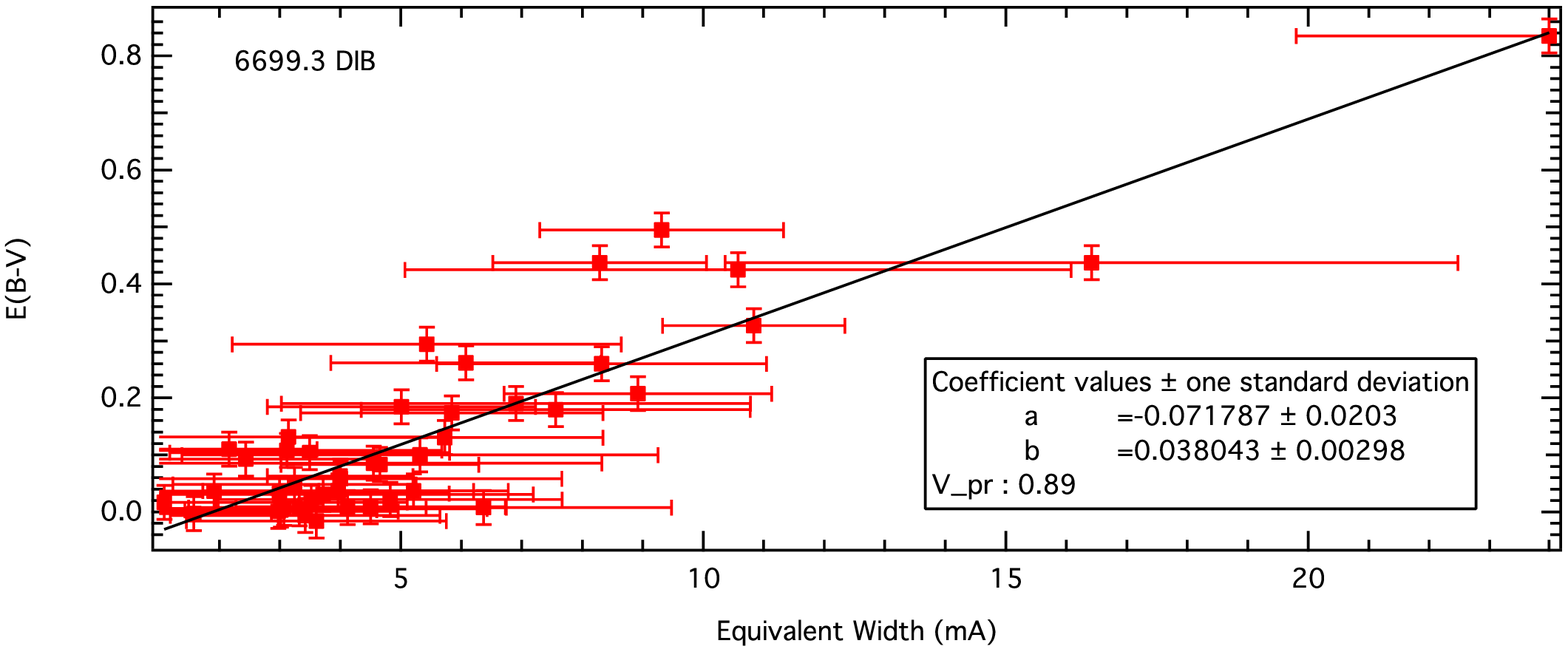}
\end{minipage}\hfill
\caption{DIB model and color excess-DIB equivalent width correlation (continued).}
\label{DIBmodel3}%
\end{figure*}

\begin{figure*}[b!]
\begin{minipage}[t]{0.5\linewidth}
\centering
	\includegraphics[width=0.65\linewidth]{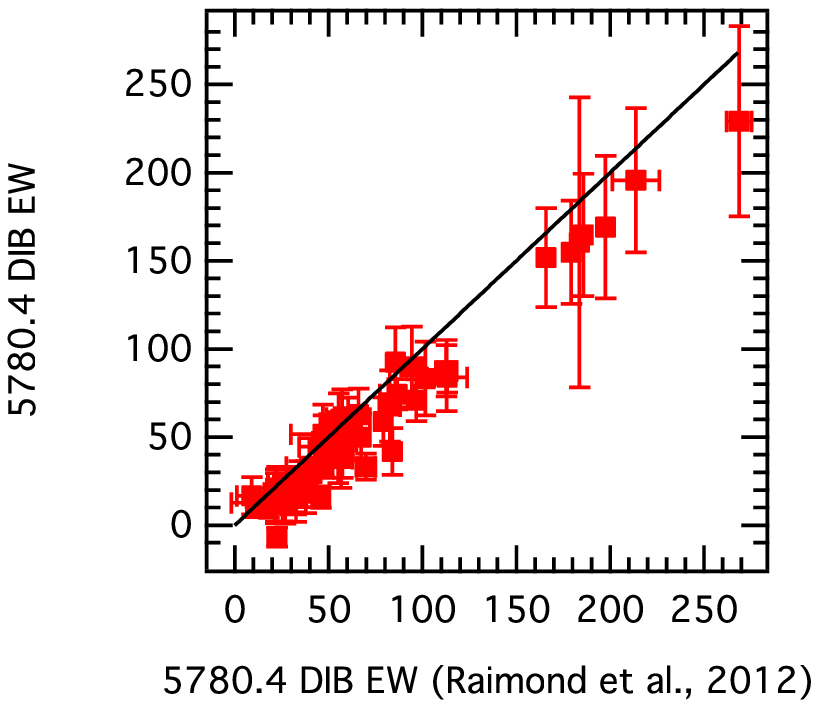}
\end{minipage}\hfill
\begin{minipage}[t]{0.5\linewidth}
\centering
	\includegraphics[width=0.65\linewidth]{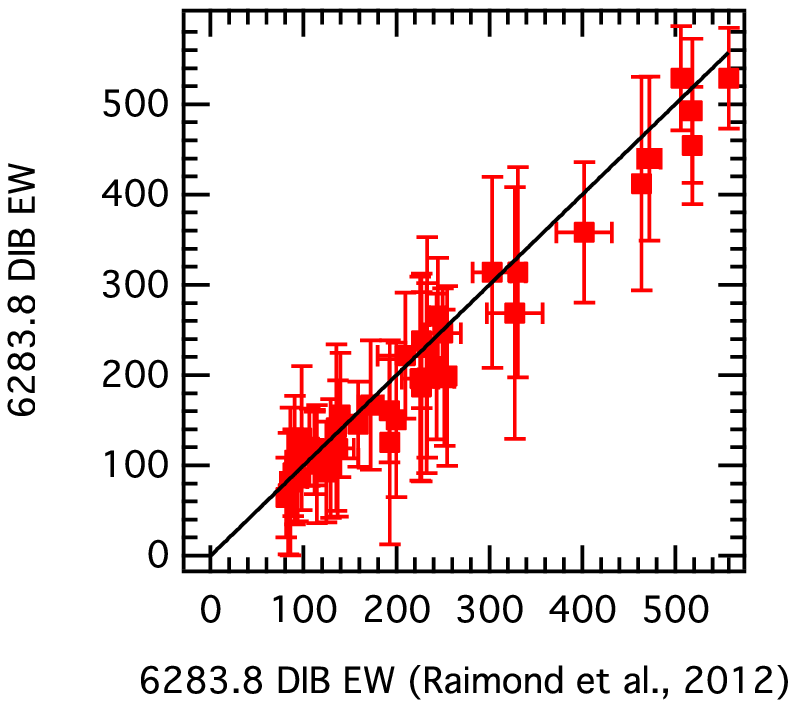}
\end{minipage}\hfill
\caption{Comparison of equivalent width value of 5780.38 DIB (left) and 6283.84  DIB (right) with \cite{raimond12}.}
              \label{compareseverin}%
    \end{figure*}

\section{Correlation with the color excess}

We have used the Geneva determinations of E(B-V) (\cite{cram99}, \cite{burcram12})  to compute, for the sixteen DIBs,  their average linear relationships with the color excess.  This was done in two ways: 
\begin{itemize}
\item by performing a simple linear fitting without taking into account individual errors on both quantities (ordinary least square, OLS). 
\item by applying the Orthogonal Distance Regression (ODR) method (or total least-squares method), taking into account individual errors on both the EWs and the E(B-V)s.  
For this method, we used the ODRPACK95 package (\cite{boggs89}) implemented in the IGOR 6.22A WaveMetrics software.
\end{itemize}

Following most of the studies, we did not impose a strict proportionality between the two quantities. 
We performed the correlation exclusively for the target stars and EWs listed in Table \ref{tabdibmeasure}, i.e. excluding binary stars or stars for which the automated fit did not converge due to the combination of noise level and DIB weakness, and also stars for which the measured EW is smaller than the estimated uncertainty. 
The Pearson correlation coefficient is deduced from the OLS method, with two values listed, the first one for the totality of the measurements, and the second after removal of two peculiar stars, when they have not been already removed from the list due to the criteria mentioned above.

The correlation parameters and coefficients  can be found in Table \ref{tablecorr} and the linear fits are displayed in Fig. \ref{DIBmodel}, \ref{DIBmodel2} and \ref{DIBmodel3}. 
We provide the first  correlation  parameters and coefficients for six DIBs, namely the 5849.81, 6089.85, 6269.85, 6445.28, 6660.71 and 6699.32 \AA\ band.  
It is interesting to note that the new correlations coefficients are in the same range as for the other DIBs in the optical, with Pearson coefficients  between 0.89 and 0.95, which confirms a significant positive correlation between color excess and EW of those DIBs. 
Since the Pearson coefficient does not take into account measurement uncertainties, it is also informative to consider the uncertainties on the slopes obtained from the OLS and ODR methods. In the case of the OLS method, uncertainties on the parameters are estimated along with the fitting process from the dispersion around the mean value. In the ODR method, the uncertainties are calculated based on all measurement uncertainties. 

We have  compared our results with relevant published values from references listed in Table \ref{tabledibparam}.
Correlation studies of the 4726.8, 4762.6, and 4963.9 \AA\ DIBs have been performed by \cite{Ordaz2011} for a large set of galactic O stars. While our correlation coefficient is similar for DIB 4726.8 (0.92 vs. 0.90), it is significantly higher for the two other DIBs. Following the same argumentation as in Raimond et al (2012), we believe this is due to the characteristics of our targets, that are late B or A stars. Their relatively low UV radiation field partly avoids strong effects on the DIB carriers in clouds that are located close to the targets, while such effects must be present in a number of their less distant O stars. This suggests that the 4726.8 DIB is less sensitive to such effects.
\cite{2011ApJ...727...33F} have established correlations with the color excess for six DIBs in common with our study, namely the 5780.4, 5797.1, 6196.0, 6204, 6283.8, and 6613.6 DIBs, using O-and early B stars from the northern hemisphere, while \cite{raimond12} have established correlations for the 5780.4 and 6283.8 DIBs using the same set of cooler stars as we are using here. In general, our results are in agreement with them, as well as with earlier measurements from \cite{herbig93}. Figure \ref{compareseverin} shows the comparison between our results and \cite{raimond12}. There are only small differences with  \cite{raimond12} for the DIB 5780.4, that are probably caused by the noise and residual contamination in the spectra during the fitting process. We find a noticeable difference in the case of the 6283.8, (a slope of 0.00076 instead of 0.00067 mag/mA) which we explain by the following bias. For this DIB, the automated sliding window method is influenced by telluric lines and gives very large uncertainties on EWs, that result in the exclusion of a large number of targets with small EWs, following our threshold rule. This introduces a bias towards higher EWs for small reddening  values and a larger slope. This implies that such a method is not fully appropriate where telluric lines are too strong and that another method should be searched for.
Similarly to the case of the 5780.4 \AA\ DIB, for which using cooler stars results in a tighter correlation, we note here that for the 5797.1, 6196.0, 6204 and 6613.6 DIBs we also obtain a significantly better correlation than the one of \cite{2011ApJ...727...33F}. This is very likely also caused by environment effects, those bands being strongly influenced by the radiation field. The same remarks apply to the correlation coefficients measured by \cite{vos11} for the upper Scorpius region that, except for DIB 5797, are lower than found here. 
\cite{josa87} have computed Pearson coefficients for three DIBS in common: 5780, 5849 and 5797 \ AA\ , using early-type target stars in reflection  nebulae.  We also computed their correlation parameters, using data from their tables. The parameters are in agreement with ours, within uncertainties. Our Pearson coefficients are higher, 0.94 vs 0.90 for DIB 5780, 0.94 
vs 0.89 for DIB 5797, and finally  0.92 
vs 0.75 for DIB 5849. The large difference for this latter band suggests that it is strongly sensitive to radiation conditions. 
Finally, \cite{cordiner11} quotes average coefficients  of 88 and 50 m\AA /mag 
for the  6379 and 6660 DIBs  resp., in their comparison with M31. Those values are similar to our results.

\begin{landscape}
\begin{table}
	\caption{DIB correlation with color excess with ordinary least square (OLS) fitting and orthogonal distance regression (ODR) fitting, where $E(B-V) = a + b*EW$.}
	\label{tablecorr}
	\begin{center}
\tiny{
		\begin{tabular}{| c | c | c | c | c | c | c | c | c | c |}
		\hline
$\lambda_c (\AA)$ & LOS & Corr. coeff. & Corr. coeff.*  &\multicolumn{4}{c|}{OLS} &\multicolumn{2}{c|}{ODR} \\\cline{5-10}
                              &        & (OLS)          & (OLS)              & a (mag) & b (mag/m\AA) & a* (mag) & b* (mag/m\AA) & a (mag) & b (mag/m\AA)\\\hline \hline

 4726.83   &       28  &     0.92 &     0.92 &   0.0479 $\pm$   0.0188 &   0.0126 $\pm$   0.0011 &   0.0446 $\pm$   0.0188 &   0.0127 $\pm$   0.0011 &   0.0130 $\pm$   0.0193 &   0.0196 $\pm$   0.0022   \\ \hline 
 4762.61   &       63  &     0.87 &     0.88 &  -0.0937 $\pm$   0.0175 &   0.0248 $\pm$   0.0018 &  -0.0945 $\pm$   0.0175 &   0.0247 $\pm$   0.0017 &  -0.1497 $\pm$   0.0420 &   0.0326 $\pm$   0.0057   \\ \hline 
 4963.88   &       23  &     0.93 &     0.92 &   0.0638 $\pm$   0.0213 &   0.0249 $\pm$   0.0022 &   0.0637 $\pm$   0.0213 &   0.0249 $\pm$   0.0023 &   0.0302 $\pm$   0.0189 &   0.0331 $\pm$   0.0042   \\ \hline 
 5780.38   &       76  &     0.94 &     0.94 &   0.0024 $\pm$   0.0071 &   0.0025 $\pm$   0.0001 &   0.0006 $\pm$   0.0071 &   0.0025 $\pm$   0.0001 &   0.0086 $\pm$   0.0054 &   0.0023 $\pm$   0.0001   \\ \hline 
 5797.06   &       32  &     0.94 &     0.94 &   0.0306 $\pm$   0.0151 &   0.0055 $\pm$   0.0004 &   0.0291 $\pm$   0.0151 &   0.0055 $\pm$   0.0004 &   0.0203 $\pm$   0.0172 &   0.0063 $\pm$   0.0005   \\  \hline 
 5849.81   &       97  &     0.92 &     0.92 &  -0.0072 $\pm$   0.0065 &   0.0117 $\pm$   0.0005 &  -0.0073 $\pm$   0.0065 &   0.0116 $\pm$   0.0005 &  -0.0163 $\pm$   0.0056 &   0.0127 $\pm$   0.0005   \\  \hline 
 6089.85   &       19  &     0.92 &     0.93 &  -0.0429 $\pm$   0.0373 &   0.0705 $\pm$   0.0075 &  -0.0566 $\pm$   0.0373 &   0.0722 $\pm$   0.0073 &  -0.1162 $\pm$   0.0752 &   0.0875 $\pm$   0.0198   \\  \hline 
 6195.98   &       76  &     0.94 &     0.95 &  -0.0269 $\pm$   0.0076 &   0.0205 $\pm$   0.0008 &  -0.0277 $\pm$   0.0076 &   0.0204 $\pm$   0.0008 &  -0.0349 $\pm$   0.0086 &   0.0217 $\pm$   0.0012   \\  \hline 
 6204        &       31  &     0.94 &     0.94 &  -0.0457 $\pm$   0.0198 &   0.0119 $\pm$   0.0008 &  -0.0475 $\pm$   0.0198 &   0.0119 $\pm$   0.0008 &  -0.1290 $\pm$   0.0767 &   0.0174 $\pm$   0.0044   \\ 
 6204$^{ci}$   &       31  &     0.93 &     0.92 &  -0.0269 $\pm$   0.0209 &   0.0084 $\pm$   0.0006 &  -0.0267 $\pm$   0.0213 &   0.0084 $\pm$   0.0007 &  -0.0801 $\pm$   0.0436 &   0.0114 $\pm$   0.0019   \\  \hline 
6269.85   &       44  &     0.89 &     0.89 &   0.0377 $\pm$   0.0141 &   0.0116 $\pm$   0.0009 &   0.0377 $\pm$   0.0141 &   0.0116 $\pm$   0.0009 &   0.0204 $\pm$   0.0104 &   0.0151 $\pm$   0.0009   \\  \hline 
 6283.84$^{**}$   &       44  &     0.83 &     0.84 &  -0.014 $\pm$   0.020 &   0.00081 $\pm$   0.00008 &  -0.010 $\pm$   0.016 &   0.00076 $\pm$   0.00007 &  -0.036 $\pm$   0.019 &   0.00089 $\pm$   0.00009   \\  \hline 
 6379.32   &       41  &     0.93 &     0.93 &   0.0401 $\pm$   0.0126 &   0.0094 $\pm$   0.0006 &   0.0383 $\pm$   0.0126 &   0.0094 $\pm$   0.0006 &   0.0359 $\pm$   0.0080 &   0.0101 $\pm$   0.0004   \\  \hline 
 6445.28   &       60  &     0.91 &     0.91 &  -0.0783 $\pm$   0.0130 &   0.0363 $\pm$   0.0022 &  -0.0783 $\pm$   0.0130 &   0.0363 $\pm$   0.0022 &  -0.0956 $\pm$   0.0185 &   0.0418 $\pm$   0.0036   \\  \hline 
 6613.62   &       69  &     0.94 &     0.95 &   0.0059 $\pm$   0.0075 &   0.0051 $\pm$   0.0002 &   0.0051 $\pm$   0.0075 &   0.0051 $\pm$   0.0002 &   0.0008 $\pm$   0.0063 &   0.0051 $\pm$   0.0002   \\  \hline 
 6660.71   &       24  &     0.95 &     0.95 &  -0.0640 $\pm$   0.0211 &   0.0359 $\pm$   0.0026 &  -0.0640 $\pm$   0.0211 &   0.0359 $\pm$   0.0026 &  -0.0953 $\pm$   0.0458 &   0.0428 $\pm$   0.0088   \\  \hline 
 6699.32   &       44  &     0.89 &     0.89 &  -0.0718 $\pm$   0.0203 &   0.0380 $\pm$   0.0030 &  -0.0737 $\pm$   0.0203 &   0.0380 $\pm$   0.0030 &  -0.1284 $\pm$   0.0333 &   0.0527 $\pm$   0.0065   \\ 

		\hline
		\end{tabular}
}
	\end{center}
\tablefoottext{*}{excluding some peculiar stars: HD147932, HD179029 }
\tablefoottext{**}{excluding HD172488}\\
\tablefoottext{ci}{based on continuum integrated EW measurement }\\
\tablefoot{\tiny{From the first to the last column, respectively: center wavelength of DIB, number of stars or line-of-sights (LOS) data used in the correlation determination, correlation coefficient (\textit{p}-value) determined by including all LOS data, correlation coefficient (\textit{p}-value) determined by removing peculiar stars, fitting parameter $a$ and $b$ from OLS method and from ODR method.}}
\end{table}

\begin{table}[h!]

	\caption{Comparison of correlations with the color excess: reference numbering from Table \ref{tabledibparam}}
	\label{tablecomparison}
\begin{tiny}
	\begin{center}
		\begin{tabular}{| c | c | c | c | c | c  | c  | c  | c  | }
		\hline
		$\lambda_c (\AA) ^*$ &  ($a$, $b$) and $p$-value obtained & (1)& (3) & (5)& (15) & (14) & (13) & (18) \\\hline
		4726.83 				&  (0.048, 0.013) 0.92	&&&  &	   				& 	 	& 0.90 &	\\
		4762.61 				&  (-0.094, 0.025) 0.87 	&& & &	   				&	 	& 0.81 &	 \\	
		4963.88 				&  (0.064, 0.025) 0.93	&& & &	   				&	 	& 0.74 &	\\
		5780.38 				&  (0.0024, 0.0025) 0.94	&0.90& (.., 0.0025) &(0,0.0021)&  (-0.006, 0.0019) 0.91	& (-0.008, 0.0020 ) 0.82 	& 		& (-0.006,0.0021) 0.79 \\
		5797.06 				&  (0.031, 0.0055) 0.94	&0.89&(.., 0.0047) &(0,0.0082)&  	   				& (-0.029, 0.0057) 0.84	&  		& (0.031,0.0063) 0.92 \\
		5849.81 				&  (-0.0072, 0.012) 0.92	&0.75&&&  	   				&		& &		 \\
		6089.85 				&  (-0.043, 0.071) 0.92	&&  &&	   				&		& 	&	\\
		6195.98 				&  (-0.027, 0.021) 0.94	&& &	 &  				& (-0.051, 0.021) 0.85	& 	&(-0.11,0.028) 0.72	\\
		6204 			 	&  (-0.046, 0.012) 0.94$^f$ $;$ 	 (-0.027, 0.0084) 0.93 $^{ci}$ &&&					&	&   (-0.072, 0.0060) 0.83  	&	& \\
		6269.85 				&  (0.038, 0.012) 0.89	&& &	&				&		& &		\\
		6283.84				&  (-0.010, 0.0076) 0.83	&&&&  (-0.005, 0.0064) 0.84	& (-0.077, 0.0096) 0.82	& 	&	 \\ 
		6379.32 				&  (0.040, 0.0094) 0.93	&&&(0,0.011)&  					&		& 	&(0.017,0.011)	0.85\\
		6445.28 				&  (-0.078, 0.036) 0.91 	&&& & 				&		& 	&	\\
		6613.62 				&  (0.006, 0.0051) 0.94	&&&(0,0.0050) & 					& (0.020, 0.0046) 0.83	& &(0.012,0.0056) 0.86		 \\
		6660.71 				&  (-0.064, 0.036) 0.95	&&&  	&				&		&	&	 \\
		6699.32 				& (-0.072, 0.038) 0.89	&&&  &					&  		& 	&	\\	
		\hline
		\end{tabular}
	\end{center}
	\tablefoottext{f}{values from fitted EW}\\
	\tablefoottext{ci}{values from continuum integrated EW}\\
\end{tiny}	
\end{table}

\end{landscape}

\section{Discussion}

We have devised and applied a simplified but automated method of DIB equivalent width measurements appropriate to early-type spectra. 
The method uses a preliminary DIB shape and fits the data to 
the product of a smooth, polynomial continuum and a DIB template profile fitted in strength and velocity shift. 
The scaling coefficient, combined with the EW of the template of reference, provides a first determination of the DIB equivalent width (the fitted EW). 
A second estimate of the equivalent width is the absorption directly measured below the normalized spectrum using the fitted continuum, independently of the DIB template (the continuum-integrated EW). 
At variance with the first, this second determination takes into account the departures from the initial shape, the first step being a way to obtain a realistic continuum at the DIB location. 

We have applied this method to a database of nearby stars high resolution spectra, for stars possessing a precise determination of he color excess. We preliminarily eliminated binary systems and corrected the spectra for weak telluric lines by means of a modeled atmospheric transmission. In the case of the strong telluric lines that contaminate the 6283.8 \AA\ DIB, we also tested the fit of the triple combination of a telluric transmission, a smooth continuum and a DIB template. We estimated the uncertainties on the DIB EWs due to the noise level and continuum placement.

For all DIBs except one, we found that the two EW determinations are in good agreement, and are compatible within our estimated uncertainties. This implies that, at least for our database,  there are only  small variations of the DIB profiles from the initial templates. 
The main reasons are the small number of absorbing clouds and the small radial velocity differences among those clouds, in comparison with the DIB intrinsic width. 
For more distant targets, departures between the two EWs may be used to initiate a second fit 
with two or more DIB profiles centered at distinct velocity shifts, in an iterative manner. 
The exception to the equality of the two determinations  is the 6203-6204 DIB, that is composed of at least two distinct bands (see e.g. \cite{porceddu91}). 

Correlations parameters with the stellar reddening are found to be in good agreement with previous determinations, again within uncertainties, and we present six new determinations. The Pearson correlation coefficients are between 0.89 
for 6269.85 and 6699.32 \AA\ to 
0.95 for 6660.71 \AA, 
in a very similar range than other DIBs in the optical domain. 
When compared with previous determinations, our Pearson coefficients for all DIBs are often significantly higher. We believe that this is due to the lack of strong stellar radiation field influences, our targets being cooler than most of those used in other studies (e.g. \cite{2011ApJ...727...33F}, \cite{Ordaz2011}). 

We note that the two DIBs that correlate the best with the reddening,  the 6196 and 6614 \AA\ bands, are also those that have high degrees of correlation with each others and other bands, such as 5544, 5809, 6699 \AA\  (see details on correlations in the recent work by \cite{xiang12} and \cite{mccall2010}).

We conclude from this study that DIBs have been correctly extracted from spectroscopic data of early-type stars in the simplified automated way we have tested.  Such treatments may be applied to build extended databases, available for correlative studies and studies of the environmental effects. The forthcoming ESA cornerstone GAIA mission will provide the distances to most targets observable with Galactic spectroscopic surveys, opening the way to three-dimensional mapping of the DIB carriers based on absorption measurements from surveys. In this perspective, the development of refined automated methods is mandatory. A condition of success for such methods is the elimination of binary stars, that must be detected prior to the DIB extraction. Here our targets are relatively nearby, and the sightline IS velocity dispersion is small in comparison with the DIB width. However, for distant targets, procedures allowing for multiple clouds with radial velocity shifts potentially broadly distributed will be necessary. They could be e.g. based on iterations for an increasing number of components, using differences between fitted DIB EWs and continuum-integrated EWs as criteria for the number of clouds, following the above remarks on those differences. 
On the other hand, if one wants to reach the best possible spatial resolution of the 3D maps, methods appropriate for cool stars will be required. Such methods are in current development, and promising results have been obtained from combinations of stellar synthetic spectra of cool stars, telluric transmission profiles and DIB absorption templates (\cite{chen12}). Again, special attention will have to be given to the case of binary systems, for which automated tools may provide erroneous results. A preliminary detection of those systems will have to precede the DIB analysis itself.

\begin{acknowledgements}
\end{acknowledgements}




\Online
\section*{Stellar data and DIB measurements}

\onllongtab{1}{

\begin{landscape}

\begin{tiny}

\begin{longtable}{| c | c | c | c | c | c | c | c | c | c | c | c | c |  }
\caption{Target stars and DIB measurements (EW in m\AA)}\\

\hline
star (HD)& \textit{l} ($^{\circ}$) & \textit{b} ($^{\circ}$) & s. type & plx (mas) & E(B-V) & EW 4726.83 & EW4762.61  & EW 4963.88 & EW 5780.38  & EW 5797.06 & EW 5849.81  & EW 6089.85  \\  
\hline
\endfirsthead
\caption{continued.}\\
\hline
star (HD) & \textit{l} ($^{\circ}$) & \textit{b} ($^{\circ}$) & s. type & plx (mas) & E(B-V) & EW 4726.83 & EW4762.61  & EW 4963.88 & EW 5780.38  & EW 5797.06 & EW 5849.81  & EW 6089.85  \\  
\hline
\endhead
\hline
\endfoot	

      480 &    319.5 &  	   -65.6	         & B5V &     3.15 $\pm$     0.72  &  -0.0009  &    &    &   &   &   &   &   \\ 
     955 &     78.9 &  	   -77.1	         & B4V &     3.24 $\pm$     1.03  &   0.0061  &    &     4.20 $\pm$     2.34  &   &    -6.69 $\pm$     6.39 &   &   &   \\ 
    1348 &    303.2 &  	   -28.7	         & B9.5IV &     1.73 $\pm$     0.60  &   0.1042  &    &    12.15 $\pm$     5.47  &     2.53 $\pm$     2.17 &    83.17 $\pm$    20.93 &    16.00 $\pm$    15.18 &   &   \\ 
    1685 &    306.9 &  	   -47.3	         & B9V &    10.65 $\pm$     0.51  &  -0.0009  &    -2.52 $\pm$     1.80  &    &   &   &   &     4.78 $\pm$     1.69 &   \\ 
    9478 &    296.8 &  	   -50.2	         & B9V &     2.00 $\pm$     0.65  &   0.0105  &    &     5.73 $\pm$     5.10  &   &   &   &   &   \\ 
   10161 &    204.4 &  	   -79.2	         & B9V &     5.64 $\pm$     0.84  &   0.0026  &    &    &   &   &   &     4.98 $\pm$     2.77 &   \\ 
   16891 &    285.4 &  	   -49.0	         & B8V &     6.68 $\pm$     0.53  &  -0.0168  &    &     3.37 $\pm$     2.27  &   &    -5.99 $\pm$     5.56 &   &   &   \\ 
   18546 &    243.7 &  	   -61.4	         & A0Vn &     9.01 $\pm$     0.64  &  -0.0027  &    &    &   &    -7.89 $\pm$     7.57 &   &   &   \\ 
   21360 &    263.1 &  	   -52.5	         & A0V &     7.49 $\pm$     0.61  &   0.0238  &    &    &   &   -10.39 $\pm$     6.25 &   &     6.14 $\pm$     2.39 &   \\ 
   24446 &    196.0 &  	   -42.3	         & B9 &     2.31 $\pm$     0.83  &   0.0220  &    &     7.34 $\pm$     3.99  &   &   &   &     4.98 $\pm$     3.33 &   \\ 
   27528 &    211.7 &  	   -40.7	         & B9V &     5.78 $\pm$     0.87  &   0.0114  &    &     7.76 $\pm$     3.85  &   &   &   &     3.86 $\pm$     2.90 &   \\ 
   29506 &    218.8 &  	   -38.2	         & B6V &     2.44 $\pm$     0.74  &   0.0070  &    &     5.98 $\pm$     2.50  &   &   &   &   &   \\ 
   32043 &    205.6 &  	   -27.6	         & B9 &     2.23 $\pm$     1.02  &   0.0556  &    &     8.53 $\pm$     7.84  &   &    15.88 $\pm$     6.22 &   &     5.42 $\pm$     1.54 &   \\ 
   33244 &    286.3 &  	   -33.4	         & B9.5V &     3.57 $\pm$     0.96  &   0.0485  &     9.12 $\pm$     5.40  &    13.60 $\pm$     3.67  &     2.91 $\pm$     1.26 &    33.28 $\pm$    16.25 &    24.90 $\pm$    11.56 &     8.80 $\pm$     1.93 &   \\ 
   35021 &    223.2 &  	   -28.9	         & B8V &     3.38 $\pm$     0.89  &   0.0052  &    &     4.27 $\pm$     2.87  &   &    -5.69 $\pm$     4.31 &   &     4.31 $\pm$     1.96 &   \\ 
   35580 &    264.2 &  	   -34.5	         & B8/B9V &     4.35 $\pm$     0.51  &   0.0061  &    &    &   &    -8.67 $\pm$     6.61 &   &     2.56 $\pm$     1.41 &   \\ 
   36965 &    233.8 &  	   -28.9	         & A0IV &     3.50 $\pm$     0.60  &   0.0167  &    &     5.07 $\pm$     5.03  &   &   &   &   &   \\ 
   37717 &    246.2 &  	   -30.5	         & B8V &     6.65 $\pm$     0.57  &   0.0220  &    &    &   &    -9.03 $\pm$     6.57 &   &   &   \\ 
   37935 &    276.5 &  	   -32.1	         & B9.5V &     3.81 $\pm$     0.47  &   0.0008  &    &    &   &    -8.43 $\pm$     5.42 &   &     3.58 $\pm$     1.35 &   \\ 
   39294 &    238.1 &  	   -26.4	         & A1V &     5.46 $\pm$     0.66  &   0.0529  &    &    &   &   &   &   &   \\ 
   40355 &    221.1 &  	   -18.7	         & B8IV &     5.43 $\pm$     0.90  &   0.0229  &    &    &   &   &   &     6.86 $\pm$     4.14 &   \\ 
   40953 &    291.1 &  	   -29.4	         & B9.5V &    12.06 $\pm$     0.46  &  -0.0062  &    &    &   &    -9.85 $\pm$     6.83 &   &     5.11 $\pm$     1.92 &     1.46 $\pm$     1.25 \\ 
   42525 &    275.8 &  	   -29.2	         & A0V &     9.82 $\pm$     0.46  &  -0.0124  &    &    &   &    -7.61 $\pm$     6.78 &   &     3.59 $\pm$     1.78 &   \\ 
   42834 &    252.7 &  	   -25.8	         & A0V &     4.92 $\pm$     0.50  &   0.0079  &     4.30 $\pm$     4.29  &    &   &    -9.24 $\pm$     6.50 &   &   &   \\ 
   44533 &    284.5 &  	   -28.5	         & B8V &     4.39 $\pm$     0.48  &   0.0627  &    &     6.73 $\pm$     4.43  &   &    19.96 $\pm$    13.31 &    14.71 $\pm$    11.77 &     8.41 $\pm$     2.19 &   \\ 
   44737 &    252.6 &  	   -23.8	         & B7V &     1.57 $\pm$     0.58  &   0.0079  &    &    &   &   &   &   &   \\ 
   45098 &    244.8 &  	   -21.1	         & B5V &     2.48 $\pm$     0.59  &   0.0370  &    &     4.35 $\pm$     2.51  &   &    17.16 $\pm$     9.82 &   &     5.33 $\pm$     1.66 &   \\ 
   45557 &    269.5 &  	   -26.6	         & A0V &    11.37 $\pm$     0.45  &  -0.0133  &    &    &   &    -8.72 $\pm$     8.01 &   &   &   \\ 
   46976 &    279.2 &  	   -27.1	         & B9V &     3.85 $\pm$     0.62  &   0.0141  &    &     3.98 $\pm$     3.45  &   &    11.07 $\pm$     9.77 &   &     5.55 $\pm$     1.24 &   \\ 
   48150 &    252.3 &  	   -20.2	         & B3V &     1.88 $\pm$     0.58  &   0.0441  &    &     6.37 $\pm$     2.75  &   &   &   &     2.93 $\pm$     2.69 &   \\ 
   48261 &    239.8 &  	   -15.5	         & B9V &     5.34 $\pm$     0.64  &  -0.0071  &    &     5.63 $\pm$     3.03  &   &   &   &     3.73 $\pm$     2.68 &     1.49 $\pm$     1.38 \\ 
   49336 &    247.1 &  	   -17.2	         & B3Vne &     2.25 $\pm$     0.55  &   0.0485  &    &    &   &   &   &   &   \\ 
   49573 &    224.1 &  	    -6.5	         & B8II/III &     2.01 $\pm$     0.79  &   0.0582  &    &     8.83 $\pm$     1.37  &    -1.48 $\pm$     1.34 &    49.96 $\pm$    13.27 &   &     3.30 $\pm$     1.56 &   \\ 
   50093 &    236.1 &  	   -11.6	         & B2III/IV &     2.71 $\pm$     0.62  &   0.0052  &     4.87 $\pm$     3.50  &    &   &   &   &   &   \\ 
   52266 &    219.1 &  	    -0.7	         & O9V &     2.06 $\pm$     0.91  &   0.2941  &    13.96 $\pm$     3.77  &    19.14 $\pm$     4.15  &     6.17 $\pm$     2.30 &   154.88 $\pm$    29.23 &    56.26 $\pm$    17.86 &    16.31 $\pm$     2.88 &     4.88 $\pm$     2.15 \\ 
   60102 &    296.8 &  	   -26.5	         & B9/B9.5V &     4.84 $\pm$     0.57  &   0.0830  &     4.28 $\pm$     4.19  &     2.35 $\pm$     2.29  &   &    27.32 $\pm$     9.20 &    23.33 $\pm$    13.37 &    10.42 $\pm$     2.40 &     2.81 $\pm$     1.05 \\ 
   60325 &    230.4 &  	     2.5	         & B2II &     2.16 $\pm$     0.80  &   0.1793  &    15.46 $\pm$     4.58  &    &   &    84.89 $\pm$    20.25 &    25.34 $\pm$    19.02 &    10.63 $\pm$     6.27 &   \\ 
   60929 &    257.1 &  	   -11.4	         & A0V &     5.62 $\pm$     0.51  &   0.0123  &    &    &   &    -6.76 $\pm$     5.36 &   &     6.18 $\pm$     2.94 &   \\ 
   61554 &    234.9 &  	     1.6	         & B6V &     3.65 $\pm$     0.74  &   0.0706  &     3.01 $\pm$     2.48  &    &   &    34.39 $\pm$     6.73 &    14.62 $\pm$     8.78 &     4.70 $\pm$     1.99 &     2.02 $\pm$     1.51 \\ 
   61672 &    242.1 &  	    -2.4	         & B6V &     3.98 $\pm$     0.69  &   0.0176  &    &     6.27 $\pm$     3.31  &   &   &   &   &   \\ 
   61831 &    252.1 &  	    -7.9	         & B3V &     5.68 $\pm$     0.46  &  -0.0089  &    &    &   &    -8.75 $\pm$     6.88 &   &   &   \\ 
   62503 &    253.2 &  	    -7.8	         & B9V &     4.87 $\pm$     0.61  &   0.0052  &    &    &   &   &   &    -2.63 $\pm$     2.45 &   \\ 
   63868 &    255.2 &  	    -7.4	         & B5V &     2.84 $\pm$     0.51  &   0.0309  &    &     7.08 $\pm$     5.60  &   &   &   &    -4.15 $\pm$     2.42 &   \\ 
   69253 &    257.8 &  	    -3.4	         & B4V &     3.40 $\pm$     0.54  &   0.0097  &    &    &   &   &   &    -3.78 $\pm$     1.85 &   \\ 
   70948 &    260.7 &  	    -3.5	         & B5V &     2.23 $\pm$     0.61  &   0.0503  &     4.98 $\pm$     1.74  &     3.88 $\pm$     3.27  &   &    11.84 $\pm$    10.69 &   &     4.93 $\pm$     2.56 &   \\ 
   71518 &    237.8 &  	    13.4	         & B2V &     3.04 $\pm$     0.75  &   0.0132  &    &    &   &   &   &    -3.22 $\pm$     2.25 &   \\ 
   72787 &    257.9 &  	     1.0	         & B2/B3V &     2.39 $\pm$     0.56  &   0.0088  &    &    &    -3.01 $\pm$     1.98 &   &   &   &   \\ 
   75009 &    264.0 &  	    -0.8	         & B8IV/V &     2.88 $\pm$     0.59  &   0.0264  &    &    &   &   &   &     5.77 $\pm$     2.77 &   \\ 
   75112 &    256.5 &  	     5.4	         & B3V &     3.64 $\pm$     0.59  &   0.0211  &    &     5.12 $\pm$     3.54  &   &    15.04 $\pm$    12.82 &   &   &   \\ 
   79752 &    245.0 &  	    22.8	         & A0V &     9.38 $\pm$     1.10  &   0.0088  &    &    &     2.64 $\pm$     1.61 &    -8.67 $\pm$     8.60 &   &     5.39 $\pm$     2.85 &   \\ 
   80590 &    260.7 &  	    10.9	         & B8V &     4.79 $\pm$     0.59  &   0.0141  &    &     7.18 $\pm$     4.98  &   &   &   &   &   \\ 
   83153 &    274.2 &  	     1.0	         & B3/B4III &     2.28 $\pm$     0.78  &   0.2075  &    15.14 $\pm$    11.62  &     6.50 $\pm$     5.21  &     5.95 $\pm$     2.08 &    67.56 $\pm$    20.13 &    34.97 $\pm$     9.10 &    12.75 $\pm$     2.42 &     4.65 $\pm$     1.77 \\ 
   84552 &    274.1 &  	     3.6	         & B6/B7V &     2.79 $\pm$     0.57  &   0.0309  &    &     7.05 $\pm$     3.47  &   &   &   &     3.29 $\pm$     1.82 &   \\ 
   85604 &    273.7 &  	     5.9	         & B8/B9V &     3.70 $\pm$     0.61  &   0.0238  &    &    &   &   &   &     5.47 $\pm$     1.84 &   \\ 
   86612 &    259.8 &  	    24.1	         & B5V &     5.19 $\pm$     0.77  &   0.0786  &     8.52 $\pm$     4.39  &    10.14 $\pm$     3.93  &   &   &   &     2.10 $\pm$     2.03 &   \\ 
   93331 &    262.2 &  	    39.3	         & B9.5V &     6.60 $\pm$     0.84  &   0.0167  &    &    &   &   &   &     6.68 $\pm$     1.19 &   \\ 
   93845 &    297.7 &  	   -19.0	         & B2.5IV &     8.97 $\pm$     0.45  &  -0.0062  &    &    &   &   &   &   &   \\ 
   95178 &    263.2 &  	    43.7	         & A0 &     4.41 $\pm$     0.97  &   0.0088  &    &    &   &   &   &     3.53 $\pm$     1.99 &   \\ 
   98161 &    283.2 &  	    21.2	         & A1V &    13.96 $\pm$     0.78  &   0.0045  &    &    &   &   -10.38 $\pm$     4.82 &   &     7.36 $\pm$     2.63 &   \\ 
   98867 &    284.6 &  	    20.6	         & B9.5V &     3.96 $\pm$     0.87  &   0.0362  &    &    &   &   &   &     5.57 $\pm$     1.47 &   \\ 
  100889 &    274.2 &  	    48.9	         & B9.5Vn &    10.70 $\pm$     0.79  &  -0.0080  &    &    &   &    -7.51 $\pm$     6.48 &   &     4.84 $\pm$     2.83 &   \\ 
  103077 &    293.1 &  	    12.3	         & B5V &     2.93 $\pm$     0.76  &   0.0441  &    &     8.81 $\pm$     6.50  &   &    18.53 $\pm$     8.12 &   &   &   \\ 
  105078 &    292.6 &  	    26.3	         & B7V &     4.48 $\pm$     0.78  &   0.0326  &    &     5.66 $\pm$     5.32  &   &   &   &   &   \\ 
  105313 &    294.9 &  	    16.6	         & B9V &     4.04 $\pm$     0.78  &   0.0556  &    &    11.81 $\pm$     6.63  &   &    29.27 $\pm$     9.15 &   &     5.43 $\pm$     4.39 &   \\ 
  106461 &    302.7 &  	   -25.6	         & B9V &     5.16 $\pm$     0.54  &  -0.0045  &    &    &     1.57 $\pm$     1.42 &   &   &     4.01 $\pm$     2.79 &   \\ 
  107931 &    298.2 &  	    15.2	         & B9V &     6.32 $\pm$     0.80  &   0.0291  &    &     3.79 $\pm$     2.74  &   &   &   &     4.23 $\pm$     2.52 &   \\ 
  108107 &    292.8 &  	    50.7	         & A1V &    15.49 $\pm$     0.75  &  -0.0137  &    &    &   &    -8.76 $\pm$     8.32 &   &     4.56 $\pm$     1.37 &   \\ 
  108267 &    298.3 &  	    17.9	         & A3V &     3.94 $\pm$     0.95  &   0.0219  &    &    &   &    11.41 $\pm$     7.02 &   &     4.66 $\pm$     4.13 &   \\ 
  108344 &    302.3 &  	   -21.0	         & B8V &     4.52 $\pm$     0.57  &   0.1307  &    &     9.41 $\pm$     4.86  &   &    60.92 $\pm$    11.26 &    18.88 $\pm$    10.15 &     7.74 $\pm$     2.78 &   \\ 
  109309 &    295.7 &  	    53.2	         & A0V &    12.47 $\pm$     0.81  &  -0.0045  &    &    &   &   &   &     4.78 $\pm$     3.43 &   \\ 
  113709 &    304.1 &  	   -10.8	         & B9V &     2.62 $\pm$     0.72  &   0.0927  &    -3.21 $\pm$     3.09  &    &   &    58.02 $\pm$    16.80 &   &     6.46 $\pm$     3.87 &     1.94 $\pm$     1.45 \\ 
  114887 &    304.9 &  	    -7.7	         & B4III &     3.45 $\pm$     0.69  &   0.1846  &    &     9.01 $\pm$     4.01  &     2.16 $\pm$     1.84 &    92.55 $\pm$    19.75 &    22.09 $\pm$     9.87 &     8.93 $\pm$     2.85 &     2.12 $\pm$     1.71 \\ 
  116226 &    308.3 &  	    14.0	         & B6IV &     1.82 $\pm$     0.71  &   0.0697  &    &     3.73 $\pm$     2.71  &    -2.60 $\pm$     1.72 &    21.20 $\pm$    10.55 &   &     3.41 $\pm$     2.37 &   \\ 
  116663 &    308.9 &  	    14.9	         & B9V &     0.98 $\pm$     1.17  &   0.0503  &     5.28 $\pm$     4.28  &    &   &    34.23 $\pm$    17.40 &    10.26 $\pm$     9.27 &     5.42 $\pm$     3.51 &   \\ 
  117484 &    310.0 &  	    15.6	         & B9V &     7.09 $\pm$     0.88  &  -0.0054  &    &    &   &   &   &   &   \\ 
  119283 &    309.6 &  	     3.0	         & B8V &     4.76 $\pm$     0.81  &   0.0856  &    &     7.80 $\pm$     3.44  &   &    41.76 $\pm$    13.24 &    11.39 $\pm$     7.48 &     5.81 $\pm$     2.56 &   \\ 
  120958 &    315.9 &  	    22.2	         & B3Vnne &     2.20 $\pm$     0.85  &   0.1051  &     7.11 $\pm$     6.46  &    &   &    16.65 $\pm$    10.64 &     6.01 $\pm$     5.43 &   &   \\ 
  124834 &    309.0 &  	   -12.2	         & B3III/IV &     3.35 $\pm$     0.62  &   0.1528  &     5.99 $\pm$     4.92  &     7.05 $\pm$     5.25  &   &    62.82 $\pm$    14.78 &    17.05 $\pm$    13.17 &    10.91 $\pm$     2.76 &     2.71 $\pm$     1.63 \\ 
  126131 &    333.1 &  	    41.3	         & A1V &     4.99 $\pm$     0.90  &   0.0688  &    &     6.76 $\pm$     5.52  &   &   &   &     3.87 $\pm$     2.70 &   \\ 
  131058 &    315.0 &  	    -6.1	         & B3Vn &     2.66 $\pm$     0.62  &   0.1104  &     3.93 $\pm$     3.23  &    10.25 $\pm$     4.35  &     2.43 $\pm$     2.08 &    70.98 $\pm$    11.89 &   &     6.70 $\pm$     1.71 &   \\ 
  132101 &    322.3 &  	     6.4	         & B5V &     2.73 $\pm$     0.86  &   0.1077  &    &     7.34 $\pm$     4.90  &   &    37.81 $\pm$    10.98 &     9.41 $\pm$     5.95 &     5.35 $\pm$     2.78 &   \\ 
  133529 &    337.3 &  	    28.0	         & B7Vn... &     5.78 $\pm$     1.01  &   0.1219  &     7.69 $\pm$     2.17  &     7.12 $\pm$     3.33  &     3.65 $\pm$     2.26 &    43.00 $\pm$    21.83 &    22.62 $\pm$    10.84 &    10.44 $\pm$     2.09 &   \\ 
  135961 &    311.6 &  	   -16.3	         & B9V &     3.17 $\pm$     0.63  &   0.0609  &    &     7.01 $\pm$     3.61  &   &    33.26 $\pm$     7.48 &   &     5.02 $\pm$     2.39 &   \\ 
  139909 &    353.4 &  	    31.8	         & B9.5V &     6.23 $\pm$     0.86  &   0.1192  &     3.08 $\pm$     2.58  &     8.42 $\pm$     2.84  &   &    58.86 $\pm$    13.70 &   &     8.60 $\pm$     2.34 &   \\ 
  140037 &    340.1 &  	    18.0	         & B5III &     3.70 $\pm$     0.96  &   0.0750  &    &     4.67 $\pm$     4.18  &   &    21.53 $\pm$    14.87 &   &   &   \\ 
  140619 &    330.2 &  	     4.6	         & B9III &     2.30 $\pm$     0.82  &   0.1042  &    &     6.44 $\pm$     4.31  &   &    51.71 $\pm$    16.68 &    13.71 $\pm$     6.43 &     4.30 $\pm$     2.34 &   \\ 
  141327 &    340.9 &  	    16.6	         & B9V &     5.00 $\pm$     0.87  &   0.0609  &    &     3.78 $\pm$     3.69  &   &    13.84 $\pm$    12.84 &    10.58 $\pm$    10.04 &     4.46 $\pm$     3.06 &   \\ 
  142315 &    349.0 &  	    23.3	         & B9V &     6.52 $\pm$     1.03  &   0.1280  &     3.49 $\pm$     1.70  &     6.15 $\pm$     3.45  &   &    47.59 $\pm$    12.73 &   &     6.90 $\pm$     1.96 &   \\ 
  143321 &    330.6 &  	     1.2	         & B5V &     4.51 $\pm$     0.90  &   0.1740  &    &     4.54 $\pm$     3.74  &     3.48 $\pm$     2.45 &    74.36 $\pm$     8.72 &    15.97 $\pm$     8.11 &     6.56 $\pm$     2.83 &   \\ 
  143326 &    313.0 &  	   -18.5	         & B8V &     3.64 $\pm$     0.70  &   0.0547  &    &     4.41 $\pm$     2.12  &   &    19.57 $\pm$    13.63 &   &     5.57 $\pm$     3.60 &   \\ 
  146029 &    352.8 &  	    20.2	         & B9V &     4.20 $\pm$     0.87  &   0.1316  &     4.08 $\pm$     2.67  &    12.74 $\pm$     3.48  &   &    83.82 $\pm$     8.68 &     9.01 $\pm$     7.18 &     7.04 $\pm$     1.82 &   \\ 
  146254 &    359.1 &  	    25.0	         & A0III &     5.31 $\pm$     0.77  &   0.1386  &    &    &   &    64.46 $\pm$    23.44 &    16.45 $\pm$    15.03 &     8.16 $\pm$     2.00 &   \\ 
  146295 &    320.2 &  	   -13.0	         & B8/B9V &     5.17 $\pm$     0.79  &   0.0450  &    &     5.92 $\pm$     4.07  &   &    23.23 $\pm$     4.79 &   &   &   \\ 
  147932 &    353.7 &  	    17.7	         & A &     7.76 $\pm$     0.96  &   0.4947  &    38.76 $\pm$    20.18  &    20.15 $\pm$     8.76  &    20.95 $\pm$     7.59 &   169.03 $\pm$    40.42 &    52.63 $\pm$     6.66 &    38.01 $\pm$     5.16 &     5.14 $\pm$     2.82 \\ 
  149425 &    342.5 &  	     4.7	         & B9V &     5.42 $\pm$     0.94  &   0.1899  &     8.04 $\pm$     3.34  &     9.04 $\pm$     7.22  &     2.83 $\pm$     2.27 &    89.85 $\pm$    22.89 &    31.36 $\pm$     9.07 &    11.26 $\pm$     2.24 &     3.39 $\pm$     2.57 \\ 
  150548 &    328.1 &  	    -9.7	         & B3V &     3.75 $\pm$     0.88  &   0.0856  &    &     8.98 $\pm$     6.19  &   &    44.43 $\pm$    17.72 &    14.76 $\pm$    10.34 &   &   \\ 
  151884 &      3.2 &  	    17.4	         & B5V &     3.73 $\pm$     0.92  &   0.4372  &    20.56 $\pm$     3.64  &    23.63 $\pm$     7.94  &     8.98 $\pm$     4.31 &   195.65 $\pm$    40.89 &    67.84 $\pm$    14.69 &    27.12 $\pm$     3.60 &     4.97 $\pm$     3.53 \\ 
  157524 &    329.0 &  	   -15.2	         & B7/B8V &     4.17 $\pm$     0.67  &   0.0441  &    &    &   &    18.31 $\pm$    10.48 &   &     4.36 $\pm$     2.17 &   \\ 
  164716 &     22.5 &  	     8.3	         & B9V &     5.62 $\pm$     0.81  &   0.2600  &    &     8.96 $\pm$     7.75  &   &    87.68 $\pm$    14.41 &    30.09 $\pm$     6.31 &    15.47 $\pm$     2.55 &     4.06 $\pm$     2.45 \\ 
  164776 &    351.8 &  	    -9.2	         & B5Vn... &     3.17 $\pm$     0.94  &   0.0574  &    &    &   &    16.71 $\pm$    14.76 &     7.18 $\pm$     6.15 &     5.13 $\pm$     3.05 &   \\ 
  165052 &      6.1 &  	    -1.5	         & O8 &     2.27 $\pm$     1.13  &   0.4249  &    13.04 $\pm$     2.89  &    17.09 $\pm$     4.55  &     7.88 $\pm$     3.80 &   160.46 $\pm$    82.25 &    54.68 $\pm$    14.79 &    29.02 $\pm$     4.33 &     5.60 $\pm$     1.65 \\ 
  165365 &      2.8 &  	    -3.7	         & B7/B8III &     2.39 $\pm$     0.89  &   0.0900  &    &    &   &    52.67 $\pm$    13.82 &   &     2.91 $\pm$     1.64 &   \\ 
  167468 &    319.3 &  	   -24.4	         & A0V &    16.59 $\pm$     0.57  &   0.0035  &    &     7.57 $\pm$     3.44  &   &   &   &     2.41 $\pm$     1.90 &   \\ 
  171722 &    324.7 &  	   -24.7	         & B9V &     5.11 $\pm$     0.70  &   0.0326  &    &    &   &    15.63 $\pm$    13.69 &   &     6.63 $\pm$     3.02 &   \\ 
  171957 &     19.0 &  	    -3.4	         & B8II/III &     4.70 $\pm$     0.94  &   0.3268  &    17.68 $\pm$     8.94  &    13.53 $\pm$     4.86  &    10.21 $\pm$     2.88 &   151.85 $\pm$    28.09 &    56.97 $\pm$    12.14 &    26.62 $\pm$     3.44 &     6.10 $\pm$     3.00 \\ 
  172016 &      0.8 &  	   -12.5	         & B9.5V &     4.43 $\pm$     0.97  &   0.0574  &    &     6.54 $\pm$     4.99  &   &    23.87 $\pm$     7.89 &   &     3.12 $\pm$     1.76 &   \\ 
  172488 &     24.0 &  	    -1.6	         & B0.5V &     3.61 $\pm$     1.16  &   0.8348  &    62.46 $\pm$     9.30  &    31.92 $\pm$    23.22  &    29.70 $\pm$     5.75 &   229.17 $\pm$    53.93 &   143.24 $\pm$    12.13 &    73.55 $\pm$     3.67 &    12.25 $\pm$     3.90 \\ 
  172882 &    313.3 &  	   -26.9	         & A0V &     3.94 $\pm$     0.64  &   0.0998  &    &    &   &    59.87 $\pm$    17.14 &   &     5.66 $\pm$     1.96 &     3.32 $\pm$     2.44 \\ 
  176853 &     24.6 &  	    -7.3	         & B2V &     3.50 $\pm$     0.81  &   0.4372  &    34.74 $\pm$     8.03  &    18.66 $\pm$     6.20  &    17.13 $\pm$     4.86 &   164.59 $\pm$    34.78 &    94.83 $\pm$    11.00 &    53.37 $\pm$     2.56 &     8.65 $\pm$     1.82 \\ 
  177756 &     30.3 &  	    -5.5	         & B9Vn &    26.05 $\pm$     0.81  &   0.0019  &    &    &   &    -8.21 $\pm$     5.97 &   &   &   \\ 
  179029 &      0.1 &  	   -20.0	         & B5V &     3.41 $\pm$     0.91  &   0.2614  &    10.39 $\pm$     4.09  &    10.07 $\pm$     3.16  &     7.81 $\pm$     4.60 &    44.14 $\pm$    20.08 &    32.41 $\pm$     9.17 &    17.63 $\pm$     1.92 &     2.60 $\pm$     1.61 \\ 
  181296 &    342.9 &  	   -26.2	         & A0Vn &    20.98 $\pm$     0.68  &  -0.0160  &    &    &   &   &   &     2.33 $\pm$     1.98 &   \\ 
  182254 &    321.0 &  	   -28.7	         & B8/B9Vn &     3.72 $\pm$     0.75  &   0.0353  &    &    &   &   &   &     4.69 $\pm$     1.75 &   \\ 
  186837 &    335.9 &  	   -30.6	         & B5V &     3.63 $\pm$     0.75  &   0.0052  &    &    &   &   &   &     3.44 $\pm$     1.93 &   \\ 
  188246 &    355.6 &  	   -29.7	         & B8/B9V &     2.02 $\pm$     0.93  &   0.0070  &    &    &   &   &   &     3.94 $\pm$     2.30 &   \\ 
  195805 &    330.6 &  	   -35.5	         & B8/B9V &     5.45 $\pm$     0.77  &   0.0088  &    &    &   &   &   &   &   \\ 
  195843 &     13.1 &  	   -34.4	         & B8V &     5.50 $\pm$     0.84  &   0.0035  &    &    &   &    -6.42 $\pm$     6.40 &   &     5.08 $\pm$     2.07 &   \\ 
  196413 &     28.5 &  	   -31.0	         & B9V &     3.15 $\pm$     1.01  &   0.0370  &    &    &   &    12.83 $\pm$     8.24 &   &     3.40 $\pm$     3.26 &   \\ 
  197630 &      2.9 &  	   -38.3	         & B8/B9V &    10.19 $\pm$     0.84  &   0.0035  &    &    &    -1.32 $\pm$     1.28 &   &   &     2.46 $\pm$     1.93 &   \\ 
  201057 &     31.2 &  	   -37.8	         & B9.5V &     7.54 $\pm$     0.90  &  -0.0027  &    &     3.85 $\pm$     2.99  &   &   &   &     4.57 $\pm$     1.56 &   \\ 
  201317 &    357.7 &  	   -42.9	         & B8V &     3.12 $\pm$     0.96  &   0.0167  &    &    &   &   &   &     4.31 $\pm$     1.81 &   \\ 
  202299 &    329.4 &  	   -39.7	         & B8V &     5.91 $\pm$     0.66  &  -0.0107  &    &    &   &   &   &     3.54 $\pm$     1.40 &   \\ 
  205348 &    322.6 &  	   -39.3	         & B8V &     5.01 $\pm$     0.63  &   0.0035  &    &    &   &   &   &   &   \\ 
  205705 &     32.3 &  	   -45.0	         & B8V &     4.96 $\pm$     0.97  &   0.0079  &    &    &     3.30 $\pm$     2.49 &   &   &     7.35 $\pm$     2.88 &   \\ 
  207158 &    338.1 &  	   -46.5	         & B9V &     4.76 $\pm$     0.71  &  -0.0054  &    &    &   &   &   &   &   \\ 
  207603 &     21.0 &  	   -50.2	         & B8/B9V &     4.92 $\pm$     0.93  &   0.0123  &    &    &   &   &   &     4.76 $\pm$     3.61 &   \\ 
  209386 &     19.4 &  	   -53.2	         & B8V &     4.01 $\pm$     0.90  &   0.0105  &    &    &   &   &   &     4.08 $\pm$     2.77 &   \\ 
  218173 &     65.5 &  	   -58.5	         & A0 &     5.96 $\pm$     0.98  &   0.0309  &    &    &   &   &   &     6.09 $\pm$     2.10 &   \\ 
  225206 &     17.9 &  	   -79.4	         & B8/B9V &     4.10 $\pm$     1.06  &  -0.0027  &    &    &   &   &   &     4.77 $\pm$     1.96 &   \\

\hline

\end{longtable}
\end{tiny}
\end{landscape}
}

\onllongtab{1}{
\begin{landscape}

\begin{tiny}
\begin{longtable}{| c | c | c | c |c | c | c | c | c | c | }
\caption{Target stars and DIB measurements (EW in m\AA)}\\
\hline
star (HD)  & EW 6195.98 & EW 6204  & EW 6269.85 & EW 6283.84  & EW 6379.32  & EW 6445.28  & EW 6613.62   & EW 6660.71  & EW 6699.32    \\
\hline
\endfirsthead
\caption{continued.}\\
\hline
star (HD) & EW 6195.98 & EW 6204  & EW 6269.85 & EW 6283.84  & EW 6379.32  & EW 6445.28  & EW 6613.62   & EW 6660.71  & EW 6699.32    \\
\hline
\endhead
\hline
\endfoot		
     480 &   &   &    -4.03 $\pm$     3.86 &    &   &   &   &    &    \\ 
     955 &   &   &    -3.69 $\pm$     3.42 &    &   &     4.02 $\pm$     2.36 &   &    &    \\ 
    1348 &     9.00 $\pm$     2.13 &   &    11.03 $\pm$     3.09 &   313.68 $\pm$   105.74  &    9.44 $\pm$     1.89  &     5.47 $\pm$     3.40 &    26.54 $\pm$     4.03 &     5.22 $\pm$     4.56  &     3.49 $\pm$     2.31  \\ 
    1685 &   &   &   &    &   &   &     3.75 $\pm$     3.09 &    &    \\ 
    9478 &   &   &   &    &   &     4.52 $\pm$     3.03 &   &     3.54 $\pm$     2.74  &    \\ 
   10161 &   &   &   &    &   &   &   &    &    \\ 
   16891 &   &   &    -3.60 $\pm$     3.55 &    &   &     1.90 $\pm$     1.61 &   &     2.95 $\pm$     2.11  &    \\ 
   18546 &   &   &   &    &   &   &   &    &     1.58 $\pm$     1.28  \\ 
   21360 &   &   &   &    &   &   &   &    &    \\ 
   24446 &     2.61 $\pm$     2.13 &   &   &    &   &   &     4.22 $\pm$     3.81 &    &    \\ 
   27528 &   &   &   &    &   -3.66 $\pm$     3.29  &   &   &     2.93 $\pm$     2.28  &    \\ 
   29506 &   &   &   &    &   &   &   &    &    \\ 
   32043 &   &   &    -2.48 $\pm$     2.00 &    &   &     2.91 $\pm$     2.01 &    11.39 $\pm$     7.04 &    &    \\ 
   33244 &     3.60 $\pm$     1.39 &     9.77 $\pm$     5.11 &   &    &    5.25 $\pm$     2.87  &   &    14.19 $\pm$     2.93 &    &    \\ 
   35021 &   &   &    -5.01 $\pm$     2.97 &    &   &   &   &    &    \\ 
   35580 &    -3.36 $\pm$     2.53 &   &   &    &   &     2.88 $\pm$     1.72 &   &    &     3.07 $\pm$     2.35  \\ 
   36965 &     2.44 $\pm$     1.24 &   &   &    &   &     2.31 $\pm$     2.00 &     4.05 $\pm$     3.06 &    &    \\ 
   37717 &   &   &   &    &   &   &   &    &    \\ 
   37935 &    -1.63 $\pm$     1.30 &   &   &    &   &   &   &     1.73 $\pm$     1.49  &     2.97 $\pm$     1.99  \\ 
   39294 &   &   &   &    &   &     3.20 $\pm$     2.27 &   &     3.06 $\pm$     2.43  &    \\ 
   40355 &   &   &   &    &   &   &   &    &    \\ 
   40953 &   &   &   &    &   &     2.56 $\pm$     1.39 &   &    &    \\ 
   42525 &   &   &    -3.31 $\pm$     2.35 &    &   &   &   &    &    \\ 
   42834 &    -2.33 $\pm$     1.47 &   &   &    &   &   &   &    &    \\ 
   44533 &     3.40 $\pm$     2.07 &   &   &    &   &   &    11.61 $\pm$     2.99 &     2.69 $\pm$     2.34  &     4.00 $\pm$     1.21  \\ 
   44737 &     2.13 $\pm$     1.65 &   &   &    &   &   &   &    &     6.37 $\pm$     3.11  \\ 
   45098 &     2.59 $\pm$     1.61 &   &    -5.09 $\pm$     4.54 &    &   &     2.84 $\pm$     1.92 &   &    &     5.21 $\pm$     1.56  \\ 
   45557 &     1.82 $\pm$     1.61 &   &   &    &   &     2.48 $\pm$     2.28 &   &     2.30 $\pm$     2.06  &    \\ 
   46976 &   &   &   &    &   &     3.89 $\pm$     1.88 &     6.74 $\pm$     4.33 &    &    \\ 
   48150 &     1.40 $\pm$     1.37 &   &   &    &   &     3.49 $\pm$     2.59 &   &    &    \\ 
   48261 &   &   &    -4.38 $\pm$     3.24 &    &   &     2.50 $\pm$     1.82 &   &    &    \\ 
   49336 &   &   &   &    &   -4.13 $\pm$     2.62  &     2.13 $\pm$     1.93 &   &    &     3.24 $\pm$     2.01  \\ 
   49573 &     6.74 $\pm$     1.88 &    11.67 $\pm$     6.86 &   &   166.96 $\pm$    71.67  &    4.05 $\pm$     2.08  &     3.45 $\pm$     1.28 &    11.62 $\pm$     4.51 &    &     4.00 $\pm$     3.66  \\ 
   50093 &   &   &   &    &   &     3.57 $\pm$     3.27 &   &    &    \\ 
   52266 &    18.90 $\pm$     1.83 &    34.16 $\pm$    29.69 &    19.26 $\pm$     2.62 &   454.23 $\pm$    64.70  &   23.46 $\pm$     2.79  &     9.44 $\pm$     3.69 &    60.98 $\pm$     4.23 &     8.85 $\pm$     4.33  &     5.43 $\pm$     3.21  \\ 
   60102 &     4.82 $\pm$     1.61 &   &   &    &   17.06 $\pm$     3.67  &   &    18.72 $\pm$     2.48 &     3.55 $\pm$     2.46  &     4.65 $\pm$     1.63  \\ 
   60325 &     9.35 $\pm$     4.19 &    18.80 $\pm$    11.95 &   &   198.93 $\pm$    99.62  &   16.48 $\pm$    14.40  &   &    29.67 $\pm$     8.77 &    &     7.56 $\pm$     3.21  \\ 
   60929 &   &   &   &    &   &   &   &    &    \\ 
   61554 &     5.72 $\pm$     1.35 &    10.75 $\pm$     6.59 &     4.11 $\pm$     2.62 &   101.52 $\pm$    65.26  &   -3.57 $\pm$     2.13  &     4.39 $\pm$     3.42 &    14.80 $\pm$     4.11 &     3.98 $\pm$     2.71  &    \\ 
   61672 &   &   &   &    &   -3.69 $\pm$     2.93  &     4.85 $\pm$     2.12 &   &    &    \\ 
   61831 &   &   &   &    &   &     4.06 $\pm$     3.52 &   &    &    \\ 
   62503 &   &   &   &    &   &   &   &    &     3.32 $\pm$     3.10  \\ 
   63868 &     3.03 $\pm$     1.40 &   &   &    56.27 $\pm$    55.80  &   &   &   &    &     3.71 $\pm$     3.47  \\ 
   69253 &   &   &    -7.89 $\pm$     6.65 &    &    3.17 $\pm$     3.12  &   &   &    &    \\ 
   70948 &   &   &   &    &   &   &   &    &    \\ 
   71518 &   &   &   &    &   &   &   &    &    \\ 
   72787 &   &   &   &    &   &     2.89 $\pm$     1.30 &   &    &    \\ 
   75009 &   &   &   &    &   &     3.65 $\pm$     1.63 &     3.23 $\pm$     3.22 &    &    \\ 
   75112 &     4.44 $\pm$     2.90 &   &   &    &   &   &     9.21 $\pm$     7.13 &    &    \\ 
   79752 &     2.67 $\pm$     2.53 &   &   &    &   &   &   &    &    \\ 
   80590 &   &   &     4.71 $\pm$     2.96 &    &    2.49 $\pm$     1.70  &     3.31 $\pm$     1.75 &   &    &    \\ 
   83153 &     9.84 $\pm$     2.46 &    16.74 $\pm$    13.96 &   &   246.74 $\pm$    49.63  &   19.40 $\pm$     4.55  &     3.95 $\pm$     3.65 &    39.65 $\pm$     4.71 &     8.70 $\pm$     6.33  &     8.92 $\pm$     2.21  \\ 
   84552 &   &   &    -2.82 $\pm$     2.54 &    &   &   &   &    &    \\ 
   85604 &     1.95 $\pm$     1.61 &   &   &    &   &     3.17 $\pm$     1.73 &   &    &    \\ 
   86612 &   &   &   &    &   &   &   &    &    \\ 
   93331 &     2.46 $\pm$     1.55 &   &   &    &   &   &     7.52 $\pm$     4.66 &    &     1.09 $\pm$     0.88  \\ 
   93845 &     2.07 $\pm$     1.92 &   &   -11.26 $\pm$     8.15 &    &   &   &     3.16 $\pm$     2.75 &    &     3.42 $\pm$     2.23  \\ 
   95178 &     1.86 $\pm$     1.72 &   &   &    &   &     3.54 $\pm$     2.42 &   &    &    \\ 
   98161 &   &   &   &    &   &   &     2.97 $\pm$     2.55 &    &    \\ 
   98867 &     3.39 $\pm$     0.90 &   &     2.84 $\pm$     2.36 &   114.94 $\pm$    46.84  &   &   &     6.22 $\pm$     4.86 &    &     1.91 $\pm$     1.81  \\ 
  100889 &     2.12 $\pm$     1.82 &   &   &    &   &   &   &    &    \\ 
  103077 &   &   &   &    82.06 $\pm$    81.76  &   &     2.10 $\pm$     2.05 &     7.71 $\pm$     5.59 &    &    \\ 
  105078 &     2.11 $\pm$     1.74 &   &   &    &   &   &   &    &    \\ 
  105313 &     4.45 $\pm$     1.63 &    13.41 $\pm$    10.98 &   &   130.31 $\pm$    79.69  &   &   &    11.49 $\pm$     5.46 &    &    \\ 
  106461 &   &   &   &    &   &   &   &    &    \\ 
  107931 &     2.68 $\pm$     1.75 &   &   &    &   -1.59 $\pm$     1.24  &     1.57 $\pm$     0.81 &   &    &    \\ 
  108107 &   &   &   &    &   &   &   &    &    \\ 
  108267 &   &   &   &    &   &   &     4.74 $\pm$     4.55 &    &     4.83 $\pm$     2.85  \\ 
  108344 &     7.40 $\pm$     2.94 &    16.25 $\pm$     7.79 &     7.93 $\pm$     2.27 &   125.35 $\pm$   112.85  &   12.30 $\pm$     3.84  &     6.77 $\pm$     3.64 &    36.94 $\pm$     4.37 &     5.35 $\pm$     2.30  &     5.72 $\pm$     2.62  \\ 
  109309 &   &   &   &    &   &     2.85 $\pm$     2.22 &   &    &    \\ 
  113709 &     7.33 $\pm$     2.86 &    16.72 $\pm$     8.94 &     6.61 $\pm$     4.78 &   221.47 $\pm$    69.77  &   &     2.31 $\pm$     2.15 &    17.39 $\pm$     5.39 &     3.37 $\pm$     3.05  &     2.44 $\pm$     2.06  \\ 
  114887 &    11.70 $\pm$     1.93 &    22.88 $\pm$     7.71 &   &   358.02 $\pm$    77.98  &    8.88 $\pm$     5.60  &     5.76 $\pm$     3.27 &    26.78 $\pm$    10.08 &    &     5.01 $\pm$     2.22  \\ 
  116226 &     4.33 $\pm$     1.84 &    12.20 $\pm$     9.10 &   &    92.99 $\pm$    55.75  &   &     2.05 $\pm$     2.05 &     4.50 $\pm$     3.35 &    &    \\ 
  116663 &     3.79 $\pm$     2.86 &   &   &    &   &   &    11.10 $\pm$     8.02 &    &    \\ 
  117484 &   &   &   &    &   &   &   &    &    \\ 
  119283 &     6.09 $\pm$     2.37 &   &   &   141.71 $\pm$    92.32  &    9.79 $\pm$     5.14  &   &    16.84 $\pm$     8.11 &     3.59 $\pm$     2.07  &    \\ 
  120958 &   &   &   &    &   11.44 $\pm$     7.96  &   &     7.47 $\pm$     5.80 &    &    \\ 
  124834 &     8.64 $\pm$     3.99 &    16.08 $\pm$    10.39 &    10.68 $\pm$     3.10 &   230.78 $\pm$   122.24  &   &   &    22.18 $\pm$     9.27 &    &    \\ 
  126131 &     3.06 $\pm$     2.24 &   &   &    &   &   &     5.83 $\pm$     3.96 &    &    \\ 
  131058 &    10.61 $\pm$     1.84 &    17.09 $\pm$    11.45 &     6.88 $\pm$     2.17 &   268.95 $\pm$   139.64  &   &     4.67 $\pm$     2.62 &    25.38 $\pm$     7.62 &     3.92 $\pm$     3.57  &     2.16 $\pm$     1.46  \\ 
  132101 &     5.21 $\pm$     3.19 &    11.40 $\pm$     9.06 &   &   155.56 $\pm$    68.91  &    6.32 $\pm$     3.59  &     3.12 $\pm$     1.88 &    15.85 $\pm$     7.30 &    &     3.12 $\pm$     2.56  \\ 
  133529 &     6.34 $\pm$     2.84 &    11.47 $\pm$     5.56 &     4.96 $\pm$     2.81 &   145.83 $\pm$    47.19  &    4.12 $\pm$     2.47  &   &    21.51 $\pm$     5.60 &    &    \\ 
  135961 &     6.73 $\pm$     2.01 &   &   &   118.84 $\pm$    75.32  &   &   &    15.51 $\pm$     4.95 &    &    \\ 
  139909 &     8.17 $\pm$     1.58 &    16.22 $\pm$    13.18 &     7.03 $\pm$     3.23 &   196.32 $\pm$   113.05  &    4.42 $\pm$     1.65  &     3.27 $\pm$     1.24 &    20.05 $\pm$     6.07 &    &    \\ 
  140037 &     5.11 $\pm$     1.80 &   &   &   107.50 $\pm$    65.95  &   &   &    11.67 $\pm$     4.76 &    &    \\ 
  140619 &     8.03 $\pm$     1.95 &    17.98 $\pm$     9.17 &   &   209.41 $\pm$    80.70  &   &     3.10 $\pm$     2.96 &    17.08 $\pm$     6.98 &    &    \\ 
  141327 &     2.83 $\pm$     2.10 &   &   &    64.49 $\pm$    44.27  &    3.60 $\pm$     3.29  &   &    10.20 $\pm$     4.73 &    &    \\ 
  142315 &     6.25 $\pm$     1.88 &    14.15 $\pm$    10.14 &     3.73 $\pm$     3.58 &   197.17 $\pm$    75.36  &    5.15 $\pm$     2.89  &     2.60 $\pm$     2.06 &    16.63 $\pm$     5.24 &    &    \\ 
  143321 &    10.23 $\pm$     1.74 &    15.82 $\pm$    12.57 &     9.84 $\pm$     2.52 &   238.07 $\pm$    74.54  &   11.21 $\pm$     1.97  &     6.21 $\pm$     2.99 &    36.28 $\pm$    10.22 &     6.16 $\pm$     4.82  &     5.84 $\pm$     2.50  \\ 
  143326 &     3.98 $\pm$     1.45 &   &     2.87 $\pm$     2.59 &    &   &     2.44 $\pm$     2.00 &     8.78 $\pm$     2.29 &    &    \\ 
  146029 &     8.82 $\pm$     2.14 &    12.59 $\pm$     9.55 &     7.84 $\pm$     4.70 &   265.30 $\pm$    64.50  &   &   &    21.04 $\pm$     7.94 &    &     3.14 $\pm$     2.50  \\ 
  146254 &     6.34 $\pm$     3.46 &    20.46 $\pm$    10.02 &     6.46 $\pm$     3.29 &   162.07 $\pm$   108.96  &   &   &    15.73 $\pm$     9.85 &    &    \\ 
  146295 &     4.19 $\pm$     2.27 &   &   &    86.21 $\pm$    48.24  &    3.64 $\pm$     3.53  &   &     9.18 $\pm$     6.41 &    &    \\ 
  147932 &    13.92 $\pm$     2.33 &    26.30 $\pm$     8.58 &    14.93 $\pm$     3.80 &   313.75 $\pm$   116.48  &   27.30 $\pm$     2.58  &     9.14 $\pm$     6.20 &    48.84 $\pm$     8.33 &    &     9.31 $\pm$     2.02  \\ 
  149425 &    11.21 $\pm$     3.16 &    16.27 $\pm$     6.76 &    15.74 $\pm$     4.97 &   160.73 $\pm$    57.02  &   23.25 $\pm$     6.54  &    10.63 $\pm$     2.63 &    48.14 $\pm$     2.83 &     7.21 $\pm$     2.62  &     6.90 $\pm$     3.88  \\ 
  150548 &     5.55 $\pm$     2.33 &     9.73 $\pm$     7.43 &   &   118.71 $\pm$    40.99  &   &   &    11.38 $\pm$     2.80 &    &     4.55 $\pm$     3.78  \\ 
  151884 &    22.28 $\pm$     4.34 &    42.76 $\pm$    15.65 &    24.57 $\pm$     4.12 &   492.60 $\pm$    79.91  &   29.24 $\pm$     3.53  &    15.22 $\pm$     5.28 &    81.12 $\pm$     3.62 &     8.53 $\pm$     3.42  &     8.29 $\pm$     1.77  \\ 
  157524 &     5.44 $\pm$     1.51 &   &   &    92.00 $\pm$    47.83  &   &     4.51 $\pm$     3.18 &     9.43 $\pm$     2.84 &    &    \\ 
  164716 &    14.12 $\pm$     1.98 &    18.03 $\pm$     7.80 &    18.76 $\pm$     6.62 &   186.86 $\pm$   104.90  &   26.08 $\pm$     3.88  &     9.40 $\pm$     4.16 &    49.52 $\pm$     9.29 &     8.48 $\pm$     4.63  &     8.32 $\pm$     2.72  \\ 
  164776 &     4.66 $\pm$     2.51 &   &   &    &    7.01 $\pm$     2.64  &   &    12.90 $\pm$     6.32 &    &    \\ 
  165052 &    17.58 $\pm$     3.97 &    39.16 $\pm$    17.13 &    24.56 $\pm$     5.13 &   528.99 $\pm$    57.47  &   35.98 $\pm$     2.32  &   &    64.46 $\pm$     8.61 &    &    10.58 $\pm$     5.51  \\ 
  165365 &     8.49 $\pm$     1.94 &    13.52 $\pm$    12.91 &     5.06 $\pm$     2.74 &   196.70 $\pm$   105.19  &    6.06 $\pm$     4.14  &   &    13.62 $\pm$     5.77 &    &    \\ 
  167468 &   &   &   &    &   &   &   &    &     3.01 $\pm$     1.59  \\ 
  171722 &     2.63 $\pm$     1.35 &   &     6.83 $\pm$     6.38 &    68.96 $\pm$    67.28  &   &   &     8.91 $\pm$     6.11 &    &     3.00 $\pm$     2.81  \\ 
  171957 &    21.57 $\pm$     2.51 &   &    39.99 $\pm$     5.01 &   411.73 $\pm$   118.41  &   40.49 $\pm$     4.04  &    14.02 $\pm$     3.90 &    89.52 $\pm$     4.44 &    15.00 $\pm$     3.66  &    10.83 $\pm$     1.51  \\ 
  172016 &     4.25 $\pm$     2.80 &   &     2.58 $\pm$     2.16 &   105.68 $\pm$    71.08  &    2.59 $\pm$     1.15  &   &    11.15 $\pm$     5.61 &    &    \\ 
  172488 &    37.96 $\pm$     3.20 &    72.93 $\pm$    24.02 &    53.42 $\pm$     4.46 &   528.94 $\pm$    55.55  &   73.78 $\pm$     3.53  &    22.36 $\pm$     1.85 &   148.51 $\pm$    15.08 &    23.03 $\pm$     6.64  &    23.98 $\pm$     4.18  \\ 
  172882 &     9.02 $\pm$     2.33 &   &    13.48 $\pm$     6.92 &   150.50 $\pm$    85.38  &    8.21 $\pm$     3.59  &     7.96 $\pm$     3.21 &    31.18 $\pm$     2.92 &     5.57 $\pm$     4.40  &     5.32 $\pm$     3.93  \\ 
  176853 &    25.07 $\pm$     3.36 &    44.93 $\pm$    17.77 &    43.95 $\pm$    12.64 &   439.72 $\pm$    90.98  &   59.40 $\pm$     2.65  &    13.97 $\pm$     2.17 &   104.30 $\pm$    10.53 &    15.65 $\pm$    11.71  &    16.42 $\pm$     6.06  \\ 
  177756 &   &   &     5.91 $\pm$     4.97 &    &   &   &   &    &    \\ 
  179029 &     9.09 $\pm$     1.59 &    21.09 $\pm$     7.41 &   &   120.25 $\pm$    42.81  &   15.06 $\pm$     3.16  &   &    32.71 $\pm$     6.37 &    &     6.07 $\pm$     2.24  \\ 
  181296 &   &   &   &    &   &   &   &    &     3.60 $\pm$     2.15  \\ 
  182254 &     2.65 $\pm$     1.72 &   &   &    46.13 $\pm$    44.85  &   &   &     7.33 $\pm$     2.98 &    &    \\ 
  186837 &     2.44 $\pm$     1.68 &   &   &    &   &     2.52 $\pm$     1.65 &   &    &    \\ 
  188246 &     2.26 $\pm$     1.25 &   &     3.54 $\pm$     3.50 &    &   &     2.81 $\pm$     2.09 &     4.67 $\pm$     3.43 &    &    \\ 
  195805 &   &   &   &    &   &     3.36 $\pm$     3.27 &    -4.78 $\pm$     2.72 &    &     4.50 $\pm$     2.23  \\ 
  195843 &     2.71 $\pm$     1.75 &   &    -2.74 $\pm$     2.24 &    &   &   &   &    &    \\ 
  196413 &     4.10 $\pm$     2.12 &   &     8.04 $\pm$     5.10 &    &   &   &     9.32 $\pm$     5.03 &    &     3.97 $\pm$     2.24  \\ 
  197630 &     2.23 $\pm$     1.25 &   &   &    &   &     3.16 $\pm$     1.67 &   &    &    \\ 
  201057 &   &   &   &    &   -1.77 $\pm$     1.26  &   &   &    &    \\ 
  201317 &   &   &   &    &   -3.14 $\pm$     3.02  &     2.99 $\pm$     2.82 &   &    &     3.52 $\pm$     1.58  \\ 
  202299 &     2.30 $\pm$     1.82 &   &     5.31 $\pm$     3.21 &    &   &     3.43 $\pm$     1.78 &   &    &    \\ 
  205348 &   &   &   &    &   &     3.21 $\pm$     2.07 &     6.51 $\pm$     4.97 &    &    \\ 
  205705 &   &   &   &    &   &     2.71 $\pm$     2.44 &     4.60 $\pm$     3.85 &    &     4.12 $\pm$     2.62  \\ 
  207158 &   &   &   &    &   &     2.97 $\pm$     2.91 &   &    &    \\ 
  207603 &   &   &   &    &   &   &   &    &    \\ 
  209386 &   &   &   &    &   &     3.23 $\pm$     2.82 &   &    &    \\ 
  218173 &     3.51 $\pm$     1.88 &   &   &    &   -2.95 $\pm$     2.45  &   &    -3.32 $\pm$     1.73 &    &    \\ 
  225206 &   &   &   &    &   &   &   &    &    \\ 

\hline
	
\end{longtable}
\end{tiny}
\end{landscape}
}



\begin{thebibliography}{}

\bibitem[Benvenuti \& Porceddu 1989]{benvenuti89} Benvenuti, P., \& Porceddu, I.\ 1989, \aap, 223, 329 

\bibitem[Boggs et al. 1989]{boggs89} Boggs, P. T., Byrd, R. H., Donaldson, J. R., Schnabel, R. B. 1989, ODRPACK, ACM Trans. Mathematical Software, 15, 348

\bibitem[Burki \& Cramer (private communication)]{burcram12} Burki, G., Cramer, N., private communication


\bibitem[Chen et al. 2012]{chen12} Chen, H.-C., Lallement, R., Babusiaux, C., et al.\ 2012, arXiv:1211.1638

\bibitem[Clough et al. 2005]{lblrtm05} Clough, S. A., M. W. Shephard, E. J. Mlawer, J. S. Delamere, M. J. Iacono, K. Cady-Pereira, S. Boukabara, and P. D. Brown, Atmospheric radiative transfer modeling: a summary of the AER codes, Short Communication, J. Quant. Spectrosc. Radiat. Transfer, 2005, 91, 233

\bibitem[Cordiner et al. 2008]{cordiner08} Cordiner, M.~A., Cox, N.~L.~J., Trundle, C., et al.\ 2008, \aap, 480, L13 

\bibitem[Cordiner et al. 2011$^a$]{cordiner11} Cordiner, M.~A., Cox, N.~L.~J., Evans, C.~J., et al.\ 2011a, \apj, 726, 39 

\bibitem[Cordiner et al. 2011$^b$]{2011AAS...21711204C} Cordiner, M., Cox, N.~L.~J., Smith, K.~T., \& Evans, C.~J.\ 2011b, Bulletin of the American Astronomical Society, 43, \#112.04 



\bibitem[Cramer 1999]{cram99} Cramer, N., 1999, New Astronomy Reviews, 43, 343

\bibitem[Destree et al. 2007]{2007ApJ...664..909D} Destree, J.~D., Snow, T.~P., \& Eriksson, K.\ 2007, \apj, 664, 909 

\bibitem[Friedman et al. 2011]{2011ApJ...727...33F} Friedman, S.~D., York, D.~G., McCall, B.~J., et al.\ 2011, \apj, 727, 33 


\bibitem[Galazutdinov et al. 2000]{gala00} Galazutdinov, G.~A., Musaev, F.~A., Kre{\l}owski, J., \& Walker, G.~A.~H.\ 2000, \pasp, 112, 648 

\bibitem[Galazutdinov et al. 2004]{gala04} Galazutdinov, G.~A., Manic{\`o}, G., Pirronello, V., \& Kre{\l}owski, J.\ 2004, \mnras, 355, 169 

\bibitem[Galazutdinov et al. 2008]{gala2008} Galazutdinov, 
G.~A., LoCurto, G., \& Kre{\l}owski, J.\ 2008, \apj, 682, 1076 

\bibitem[Gilmore et al. 2012]{2012Msngr.147...25G} Gilmore, G., Randich, S., Asplund, M., et al.\ 2012, The Messenger, 147, 25 


\bibitem[Herbig 1993]{herbig93} Herbig, G.~H.\ 1993, \apj, 407, 142 

\bibitem[Herbig 1995]{1995ARA&A..33...19H} Herbig, G.~H.\ 1995, \araa, 33, 19 


\bibitem[Hobbs et al. 2008]{hobbs2008} Hobbs, L.~M., et al.\ 2008, \apj, 680, 1256

\bibitem[Hobbs et al. 2009]{2009ApJ...705...32H} Hobbs, L.~M., York, D.~G., Thorburn, J.~A., et al.\ 2009, \apj, 705, 32 

\bibitem[Jenniskens \& Desert 1994]{JD94} Jenniskens, P., \& Desert, F.-X.\ 1994, \aaps, 106, 39 

\bibitem[Josafatsson \& Snow 1987]{josa87} Josafatsson, K., \& Snow, T.~P.\ 1987, \apj, 319, 436 

\bibitem[Krelowski \& Walker 1987]{krelo87} Krelowski, J., \& Walker, G.~A.~H.\ 1987, \apj, 312, 860

\bibitem[Krelowski \& Sneden 1994]{kre94} Krelowski, J., \& Sneden, C.\ 1994, The First Symposium on the Infrared Cirrus and Diffuse Interstellar Clouds, 58, 12 

\bibitem[Krelowski et al. 1995]{krelo95} Krelowski, J., Sneden, C., \& Hiltgen, D.\ 1995, \planss, 43, 1195


\bibitem[Kre{\l}owski et al. 1999]{krelo99} Kre{\l}owski, J., Ehrenfreund, P., Foing, B.~H., et al.\ 1999, \aap, 347, 235 


\bibitem[McCall et al. 2010]{mccall2010} McCall, B.~J., Drosback, M.~M., Thorburn, J.~A., et al.\ 2010, \apj, 708, 1628 

\bibitem[McCall \& Griffin 2013]{mccall2013} McCall, B.~J, \& Griffin, R.E.\ 2013, Proc. R. Soc. A, 469,2151


\bibitem[Moutou et al. 1999]{moutou99} Moutou, C., Kre{\l}owski, J., D'Hendecourt, L., \& Jamroszczak, J.\ 1999, \aap, 351, 680  Krelowski J. 1991, 

\bibitem[Ordaz et al. 2011]{Ordaz2011} Ordaz et al.\ 2011, arXiv:1101.3510v2



\bibitem[Porceddu et al. 1991]{porceddu91} Porceddu, I., Benvenuti, P., \& Krelowski, J.\ 1991, \aap, 248, 188 


\bibitem[Raimond et al. 2012]{raimond12} Raimond, S., Lallement, R., Vergely, J.-L., Babusiaux, C., \& Eyer, L.\ 2012, arXiv:1207.6092 

\bibitem[Rothman et al. 2009]{hitran2008} Rothman, L.~S., Gordon, I.~E., Barbe, A., et al.\ 2009, \jqsrt, 110, 533 


\bibitem[Salama et al. 1996]{salama96} Salama, F., Bakes, E.~L.~O., Allamandola, L.~J., \& Tielens, A.~G.~G.~M.\ 1996, \apj, 458, 621 

\bibitem[Sarre 2006]{sarre2006} Sarre, P.~J.\ 2006, Journal of Molecular Spectroscopy, 238, 1 



\bibitem[Sonnentrucker et  al. 1997]{sonnentrucker97} Sonnentrucker, P., Cami, J., Ehrenfreund, P., \& Foing, B.~H.\ 1997, \aap, 327, 1215 

\bibitem[Tarantola and Valette, 1982]{tarantola} Tarantola, A. \& Valette, B. 1982a, Journal of Geophysics, 50, 159

\bibitem[Thorburn et al. 2003]{thor03} Thorburn, J.~A., Hobbs, L.~M., McCall, B.~J., et al.\ 2003, \apj, 584, 339 

\bibitem[Tuairisg et al. 2000]{tuairisg00} Tuairisg, S.~{\'O}., Cami, J., Foing, B.~H., Sonnentrucker, P., \& Ehrenfreund, P.\ 2000, \aaps, 142, 225 


\bibitem[van Loon et al. 2009]{vanloon09} van Loon, J.~T., Smith, K.~T., McDonald, I., et al.\ 2009, \mnras, 399, 195 

\bibitem[Vergely et al. 2001]{vergely01} Vergely, J.-L., Freire Ferrero, R., Siebert, A., \& Valette, B.\ 2001, \aap, 366, 1016 

\bibitem[Vergely et al. 2010]{vergely010} Vergely, J.-L., Valette, B., Lallement, R., \& Raimond, S.\ 2010, \aap, 518, A31 

\bibitem[Vos et al. 2011]{vos11} Vos, D.~A.~I., Cox, N.~L.~J., Kaper, L., Spaans, M., \& Ehrenfreund, P.\ 2011, \aap, 533, A129 

\bibitem[Walker et al. 2001]{walker01} Walker, G.~A.~H., Webster, A.~S., Bohlender, D.~A., \& Kre{\l}owski, J.\ 2001, \apj, 561, 272 

\bibitem[Welsh et al. 2010]{welsh10} Welsh, B.~Y., Lallement, R., Vergely, J.-L., \& Raimond, S.\ 2010, \aap, 510, A54 

\bibitem[Welsh \& Lallement 2010]{welsh10} Welsh, B.~Y., \& Lallement, R.\ 2010, \pasp, 122, 1320


\bibitem[Weselak et al. 2010]{weselak10} Weselak, T., Galazutdinov, G.~A., Han, I., \& Kre{\l}owski, J.\ 2010, \mnras, 401, 1308 

\bibitem[Xiang et al. 2012]{xiang12} Xiang, F., Liu, Z., \& Yang, X.\ 2012, \pasj, 64, 31 


\end{thebibliography}
\end{document}